\pdfoutput=1
\documentclass[11pt,twoside,a4paper,cmspaper,final,collab]{cms-tdr}
\def\svnVersion{549d5d5}\def\svnDate{2022/07/05}\def\cmsCernNoTag{CERN-EP-2021-264}\def\cmsCernDate{\today}\def\cmsMessage{Published in the Journal of High Energy Physics as \href{http://dx.doi.org/10.1007/JHEP07(2022)081}{\doi{10.1007/JHEP07(2022)081}.}}
\begin{document}\cmsNoteHeader{EXO-20-009}

\providecommand{\cmsTable}[1]{\resizebox{\textwidth}{!}{#1}}
\newlength\cmsTabSkip\setlength{\cmsTabSkip}{1ex}
\newcommand{\cmsSmallColSkip}{\hskip 1\tabcolsep}
\newcommand{\cmsColSkip}{\hskip 3\tabcolsep}

\newcommand{\sqrts}[1][13]{\ensuremath{\sqrt{s}=#1\TeV}\xspace}
\newcommand{\TEN}[1]{\ensuremath{10^{\text{#1}}}\xspace}
\newcommand{\pp}{\ensuremath{\Pp\Pp}\xspace}
\newcommand{\keVns}{\ensuremath{\text{ke\hspace{-.08em}V}}\xspace}

\newcommand{\CP}{\ensuremath{CP}\xspace}
\newcommand{\hnl}{{\HepParticle{N}{}{}}\xspace}
\newcommand{\mhnl}{\ensuremath{m_\hnl}\xspace}
\newcommand{\tauhnl}{\ensuremath{\tau_\hnl}\xspace}
\newcommand{\mixparnosq}{\ensuremath{V_{\hnl\Pell}}\xspace}
\newcommand{\mixparenosq}{\ensuremath{V_{\hnl\Pe}}\xspace}
\newcommand{\mixparmnosq}{\ensuremath{V_{\hnl\PGm}}\xspace}
\newcommand{\mixpartnosq}{\ensuremath{V_{\hnl\PGt}}\xspace}
\newcommand{\mixpar}{\ensuremath{\abs{\mixparnosq}^2}\xspace}
\newcommand{\mixpare}{\ensuremath{\abs{\mixparenosq}^2}\xspace}
\newcommand{\mixparm}{\ensuremath{\abs{\mixparmnosq}^2}\xspace}

\newcommand{\qqtoWtoHNL}{\ensuremath{\PQq\HepAntiParticle{\PQq}{}{\prime}\to\PW\to\Pell\hnl}\xspace}
\newcommand{\Pellpr}{{\HepParticle{\Pell}{}{\prime}}\xspace}
\newcommand{\Pellprpm}{{\HepParticle{\Pell}{}{\prime\pm}}\xspace}
\newcommand{\Pellprplus}{{\HepParticle{\Pell}{}{\prime+}}\xspace}
\newcommand{\Pellprpr}{{\HepParticle{\Pell}{}{\prime\prime}}\xspace}
\newcommand{\Pellprprpm}{{\HepParticle{\Pell}{}{\prime\prime\pm}}\xspace}
\newcommand{\Pellprprmp}{{\HepParticle{\Pell}{}{\prime\prime\mp}}\xspace}
\newcommand{\PGnGellpr}{{\HepParticle{\PGn}{\!\smash{\Pellpr}}{}}\xspace}
\newcommand{\PGnGellprpr}{{\HepParticle{\PGn}{\!\smash{\Pellprpr}}{}}\xspace}
\newcommand{\PWast}{{\HepParticle{\PW}{}{\ast}}\xspace}
\newcommand{\PZast}{{\HepParticle{\PZ}{}{\ast}}\xspace}

\newcommand{\EEE}{\ensuremath{\Pe\Pe\Pe}\xspace}
\newcommand{\MMM}{\ensuremath{\PGm\PGm\PGm}\xspace}
\newcommand{\EEMss}{\ensuremath{\Pepm\Pepm\PGmmp}\xspace}
\newcommand{\MMEss}{\ensuremath{\PGmpm\PGmpm\Pemp}\xspace}

\newcommand{\Cee}{\ensuremath{\Pe\Pe}\xspace}
\newcommand{\Cmm}{\ensuremath{\PGm\PGm}\xspace}
\newcommand{\Cem}{\ensuremath{\Pe\PGm}\xspace}
\newcommand{\sigeta}{\ensuremath{\eta}\xspace}
\newcommand{\abseta}{\ensuremath{\abs{\eta}}\xspace}

\newcommand{\Wjets}{\ensuremath{\PW{+}\text{jets}}\xspace}
\newcommand{\WZ}{\ensuremath{\PW\PZ}\xspace}
\newcommand{\ZZ}{\ensuremath{\PZ\PZ}\xspace}
\newcommand{\Zgamma}{\ensuremath{\PZ\PGg}\xspace}

\newcommand{\Irel}{\ensuremath{I_{\text{rel}}}\xspace}
\newcommand{\dR}{\ensuremath{\DR}\xspace}
\newcommand{\dPhi}{\ensuremath{\Delta\phi}\xspace}
\newcommand{\ptell}{\ensuremath{\pt^{\Pell}}\xspace}
\newcommand{\Aeff}{\ensuremath{A_{\text{eff}}}\xspace}
\newcommand{\dxy}{\ensuremath{d_{xy}}\xspace}
\newcommand{\absdxy}{\ensuremath{\abs{\dxy}}\xspace}
\newcommand{\absdz}{\ensuremath{\abs{d_z}}\xspace}
\newcommand{\DeepCSV}{\ensuremath{\textsc{DeepCSV}}\xspace}

\newcommand{\lone}{\ensuremath{\Pell_1}\xspace}
\newcommand{\ltwo}{\ensuremath{\Pell_2}\xspace}
\newcommand{\lthree}{\ensuremath{\Pell_3}\xspace}
\newcommand{\ltwothree}{\ensuremath{\Pell_{2/3}}\xspace}
\newcommand{\eex}{\ensuremath{\Pe\Pe\PX}\xspace}
\newcommand{\mmx}{\ensuremath{\PGm\PGm\PX}\xspace}
\newcommand{\Ten}[2]{\ensuremath{#1\ten{#2}}\xspace}
\newcommand{\mtwol}{\ensuremath{m(\ltwo\lthree)}\xspace}
\newcommand{\DRtwol}{\ensuremath{\dR(\ltwo,\lthree)}\xspace}
\newcommand{\minDphi}{\ensuremath{\abs{\dPhi(\lone,\ltwothree)}}\xspace}
\newcommand{\mthreel}{\ensuremath{m(\lone\ltwo\lthree)}\xspace}
\newcommand{\pttwol}{\ensuremath{\pt(\ltwo\lthree)}\xspace}
\newcommand{\thetaSVll}{\ensuremath{\theta(\text{SV},\ltwo\lthree)}\xspace}
\newcommand{\SVprob}{\ensuremath{p_{\text{SV}}}\xspace}
\newcommand{\Deltwod}{\ensuremath{\Delta_{\text{2D}}}\xspace}
\newcommand{\SigTwoD}{\ensuremath{S(\Deltwod)}\xspace}
\newcommand{\MMEos}{\ensuremath{\PGmpm\PGmmp\Pepm}\xspace}
\newcommand{\EEMos}{\ensuremath{\Pepm\Pemp\PGmpm}\xspace}

\newcommand{\ellnu}{\ensuremath{\HepParticle{\Pell}{}{-}\HepAntiParticle{\PGn}{\!\Pell}{}}\xspace}
\newcommand{\LLmp}{\ensuremath{\HepParticle{\Pell}{}{-}\HepParticle{\Pell}{}{+}}\xspace}
\newcommand{\ttl}{\ensuremath{f}\xspace}
\newcommand{\ttlDB}{\ensuremath{\ttl_{\text{DB}}}\xspace}
\newcommand{\ttlSB}{\ensuremath{\ttl_{\text{SB}}}\xspace}
\newcommand{\ptc}{\ensuremath{\pt^{\text{parton}}}\xspace}
\newcommand{\rade}{\ensuremath{r_\Pe}\xspace}
\newcommand{\PiPi}{\ensuremath{\PGppm\PGpmp}\xspace}

\newcommand{\mHNLtwo}{\ensuremath{\mhnl=2\GeV}\xspace}
\newcommand{\mHNLsix}{\ensuremath{\mhnl=6\GeV}\xspace}
\newcommand{\mHNLtwelve}{\ensuremath{\mhnl=12\GeV}\xspace}
\newcommand{\mixZpEtmF}{\ensuremath{\mixpar=\Ten{0.8}{-4}}\xspace}
\newcommand{\mixOpTtmS}{\ensuremath{\mixpar=\Ten{1.3}{-6}}\xspace}
\newcommand{\mixOpZtmS}{\ensuremath{\mixpar=\Ten{1.0}{-6}}\xspace}

\newcommand{\likeli}{\ensuremath{L}\xspace}
\newcommand{\sigstr}{\ensuremath{r}\xspace}
\newcommand{\nuisan}{\ensuremath{\theta}\xspace}
\newcommand{\nuisanr}{\ensuremath{\hat\theta_r}\xspace}
\newcommand{\sigstrgl}{\ensuremath{\hat{r}}\xspace}
\newcommand{\nuisangl}{\ensuremath{\hat\theta}\xspace}

\cmsNoteHeader{EXO-20-009}
\title{Search for long-lived heavy neutral leptons with displaced vertices in proton-proton collisions at \texorpdfstring{\sqrts}{sqrt(s)=13 TeV}}
\date{\today}

\abstract{
A search for heavy neutral leptons (HNLs), the right-handed Dirac or Majorana  neutrinos, is performed in final states with three charged leptons (electrons or  muons) using proton-proton collision data collected by the CMS experiment at  \sqrts at the CERN LHC.  The data correspond to an integrated luminosity of 138\fbinv.  The HNLs could be produced through mixing with standard model neutrinos~\PGn.  For small values of the HNL mass ($<$20\GeV) and the square of the HNL-\PGn  mixing parameter (\TEN{-7}--\TEN{-2}), the decay length of these particles can  be large enough so that the secondary vertex of the HNL decay can be resolved  with the CMS silicon tracker.  The selected final state consists of one lepton emerging from the primary  proton-proton collision vertex, and two leptons forming a displaced, secondary  vertex.  No significant deviations from the standard model expectations are observed, and  constraints are obtained on the HNL mass and coupling strength parameters, excluding previously unexplored regions of parameter space in the mass range  1--20\GeV and squared mixing parameter values as low as \TEN{-7}.
}

\hypersetup{
pdfencoding=pdfdoc,
pdfauthor={CMS Collaboration},
pdftitle={Search for long-lived heavy neutral leptons with displaced vertices in proton-proton collisions at sqrt(s)=13 TeV},
pdfsubject={CMS},
pdfkeywords={CMS, heavy neutrino, displaced}}

\maketitle

\section{Introduction}
\label{sec:introduction}

The discovery of neutrino oscillations~\cite{Super-Kamiokande:1998kpq,
SNO:2002tuh, KamLAND:2002uet} has provided experimental evidence for nonzero
neutrino masses~\cite{Bilenky:2016pep}.
Constraints from cosmological considerations~\cite{RoyChoudhury:2019hls,
Ivanov:2019hqk} and several direct measurements~\cite{Formaggio:2021nfz}
indicate that the neutrino masses are extraordinarily small compared to those of
other standard model (SM) fermions.
A possible explanation is the existence of new heavy neutral leptons (HNLs) with
right-handed chirality.
The existence of these hypothetical particles would give rise to gauge-invariant
mass terms for the SM neutrinos by means of a see-saw
mechanism~\cite{Minkowski:1977sc, Yanagida:1979as, Gell-Mann:1979vob,
Glashow:1979nm, Mohapatra:1979ia, Schechter:1980gr, Shrock:1980ct, Cai:2017mow}.
The HNLs, referred to as \hnl in diagrams and equations hereafter, are singlets
under all SM gauge groups and therefore cannot interact with the SM particles
through either the electroweak or the strong interaction.
They can, however, be produced through mixing with the SM electron, muon, and
tau neutrinos~\cite{Maki:1962mu, Pontecorvo:1967fh, ParticleDataGroup:2020ssz},
with the corresponding matrix elements denoted as \mixparenosq, \mixparmnosq,
and \mixpartnosq, respectively.
Additionally, HNLs may be responsible for other unexplained cosmological
phenomena.
A stable HNL, for example, would make a viable dark matter
candidate~\cite{Dodelson:1993je, Boyarsky:2018tvu}.
Furthermore, if at least three generations of Dirac HNLs or two generations of Majorana HNLs exist, \CP~violation can occur in the HNL system and contribute to
the matter-antimatter asymmetry of the universe~\cite{Fukugita:1986hr,
Chun:2017spz}.

\begin{figure}[ht!]
\centering
\includegraphics[width=0.41\textwidth]{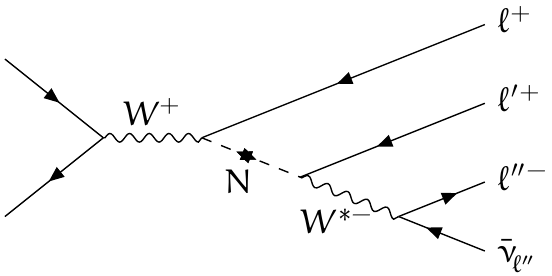}
\hspace*{0.05\textwidth}
\includegraphics[width=0.41\textwidth]{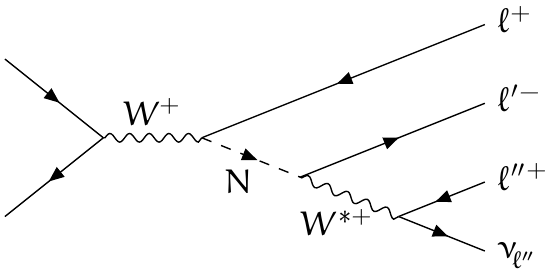}
\\[1em]
\includegraphics[width=0.41\textwidth]{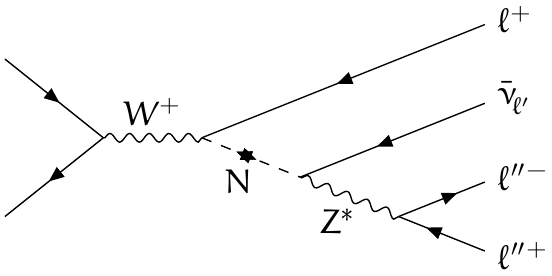}
\hspace*{0.05\textwidth}
\includegraphics[width=0.41\textwidth]{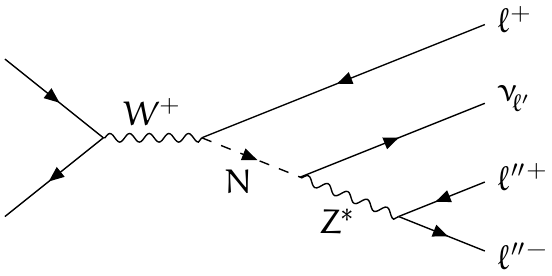}
\caption{\label{fig:introduction:feynman}
    Typical diagrams for the production and decay of an HNL through its mixing
    with an SM neutrino, leading to final states with three charged leptons and
    a neutrino.
    The HNL decay is mediated by either a \PWast (upper row) or a \PZast (lower
    row) boson.
    In the diagrams on the left, the HNL is assumed to be a Majorana neutrino,
    thus \Pell and \Pellpr in the \PWast-mediated diagram can have the same
    electric charge, and lepton number can be violated.
    In the diagrams on the right, the HNL decay conserves lepton number and can
    be either a Majorana or a Dirac particle, and, therefore, \Pell and \Pellpr
    in the \PWast-mediated diagram have always opposite charge.
    The present study considers only the case where \Pell and \Pellpr (or
    \PGnGellpr), and in case of the decay mediated by a \PZast~boson also
    \Pellprpr, are of the same flavor.
}
\end{figure}

Many experiments have searched for HNLs in the mass range from a few \keVns to
about 1\TeV~\cite{Deppisch:2015qwa, DELPHI:1996qcc, CMS:2012wqj, CMS:2015qur,
ATLAS:2015gtp, CMS:2016aro, CMS:2018iaf, CMS:2018jxx, ATLAS:2019kpx,
LHCb:2020wxx}.
The present search probes the direct production of HNLs in proton-proton (\pp)
collisions via the decay of \PW~bosons, where the SM neutrino turns into an HNL,
and the HNL subsequently decays into either a \PW~boson and a charged
lepton~\Pell, or a \PZ~boson and a neutrino~\PGn.
The resulting electroweak gauge boson can then decay leptonically, leading to a
final state with three charged leptons and a neutrino:
\begin{linenomath*}\begin{equation*}\begin{aligned}
    &\qqtoWtoHNL\to\Pell\Pellpr\PWast\to\Pell\Pellpr\Pellprpr\PGnGellprpr, \\
    &\qqtoWtoHNL\to\Pell\PGnGellpr\PZast\to\Pell\PGnGellpr\Pellprpr\Pellprpr.
\end{aligned}\end{equation*}\end{linenomath*}
The Feynman diagrams corresponding to the processes described above are shown in
Fig.~\ref{fig:introduction:feynman} for the case of a \PWp~decay.
The HNL decay width in this final state is generally dominated by the
\PWast-mediated diagrams.
If the HNL is of Majorana nature, \Pell and \Pellpr (or \Pell and \PGnGellpr)
can either have the same chirality (Fig.~\ref{fig:introduction:feynman} left) or
opposite chirality (Fig.~\ref{fig:introduction:feynman} right).
The former process is lepton number violating (LNV), while the latter is lepton
number conserving (LNC).
In the case of an HNL decay mediated by a \PWast~boson, an LNV decay
(Fig.~\ref{fig:introduction:feynman} upper left) can lead to final states with
no opposite-sign same-flavor (OSSF) lepton pairs, such as $\Pep\Pep\PGmm$ and
$\Pem\Pem\PGmp$ (referred to as \EEMss), or $\PGmp\PGmp\Pem$ and
$\PGmm\PGmm\Pep$ (referred to as \MMEss).
These final states are characterized by relatively small SM background
contributions, thus providing a pristine signal for the HNL search.
In contrast, decays mediated by a \PZast~boson
(Fig.~\ref{fig:introduction:feynman} bottom) and LNC decays
(Fig.~\ref{fig:introduction:feynman} right) are always accompanied by an OSSF
lepton pair, resulting in final states such as $\Pep\Pem\PGmp$ and
$\Pem\Pep\PGmm$ (referred to as \EEMos), or $\PGmp\PGmm\Pep$ and
$\PGmm\PGmp\Pem$ (referred to as \MMEos).

To explain neutrino observations and at the same time provide candidates for
cosmological observations, HNL models generally need to include three families
of HNLs, each of which in principle couples to leptons in all SM
generations~\cite{Asaka:2005an}.
A wide range of HNL masses and mixing parameters can give rise to realistic mass
values for the light SM neutrinos and explain the matter-antimatter asymmetry in
the universe.
For example, Ref.~\cite{Drewes:2021nqr} finds that an HNL mass scale of 10\GeV
and a range of \mixpar between \TEN{-11} and \TEN{-5} are consistent with these
requirements.
A large fraction of this phase space is testable at the CERN LHC.
For the present study, ``simplified'' models are considered where only one HNL
family couples to exactly one lepton flavor, \ie, only one of \mixparenosq,
\mixparmnosq, or \mixpartnosq is nonzero.
As a consequence, \Pell and \Pellpr (or \PGnGellpr) always belong to the same
lepton generation.
The limitations of interpretations based on such simplified models are discussed
in Refs.~\cite{Abada:2018sfh, Tastet:2021vwp, Boiarska:2021yho}.

The lifetime~\tauhnl of an HNL is strongly dependent on \mixpar and on the HNL
mass~\mhnl, and increases rapidly at small masses and low values of the mixing
parameter: $\tauhnl\propto\mhnl^{-5}\mixparnosq^{-2}$~\cite{Asaka:2005an}.
As a consequence, the kinematic properties and acceptance of HNLs with masses
below about 20\GeV are significantly affected by their long lifetimes, resulting
in HNL decay lengths ranging from several centimeters to a few meters, depending
on the coupling strength.
If the HNL has a long lifetime, its decay products (\Pellprpm, \Pellprprmp,
\PGnGellprpr; or \PGnGellpr, \Pellprprpm,
\Pellprprmp) emerge from a secondary vertex (SV) that is spatially displaced and
distinguishable from the primary vertex (PV) of the \pp interaction.
Such scenarios are referred to as ``displaced'' signatures, in contrast to
``prompt'' signatures where the SV cannot be distinguished, which is the
predominant scenario for short HNL lifetimes.
The production rates of HNLs also depend on \mhnl and
\mixpar~\cite{Degrande:2016aje, Das:2016hof}.

Previous searches for HNLs performed at the LHC used \pp collision data
corresponding to approximately 36\fbinv~\cite{CMS:2018iaf, CMS:2018jxx,
ATLAS:2019kpx}.
The ATLAS Collaboration has reported on a search for HNLs in the mass range of
4.5 to 50\GeV using prompt and displaced signatures in events with three charged
leptons~\cite{ATLAS:2019kpx}.
The search for displaced HNL decays was performed in channels with three muons,
or two muons and one electron, providing sensitivity exclusively to \mixparm.
The CMS Collaboration has performed HNL searches using prompt-lepton final
states with two same-sign leptons and one or two jets~\cite{CMS:2018jxx}, in a
mass range of 20\GeV to 1.7\TeV, as well as in final states with three charged
leptons~\cite{CMS:2018iaf}, where the HNL production was probed in a mass range
between 1\GeV and 1.2\TeV.

The present study is based on the analysis of data collected in 2016, 2017, and
2018 with the CMS detector.
The total integrated luminosity of the analyzed data set is 138\fbinv.
Events are selected with three charged leptons (electrons and muons), where one
lepton is required to originate from the primary interaction, and the other two
leptons are used to reconstruct a displaced vertex.
By optimizing the identification of leptons produced in the decay of long-lived
HNLs and by using a dedicated reconstruction of the displaced HNL vertex, the
probed phase space is extended to lower values of the HNL mass compared to the
earlier search for prompt signatures in the three-lepton
channel~\cite{CMS:2018iaf}.
Events with three same-flavor leptons (\EEE and \MMM) provide sensitivity to HNL
decays mediated through \PWast and \PZast~bosons.
Events with leptons of different flavors ($\Pe\Pe\PGm$ and $\PGm\PGm\Pe$, where
the first symbol refers to the flavor of the lepton originating from the PV)
provide sensitivity only to \PWast-mediated HNL decays, but can distinguish
between LNC and LNV decays and thus differentiate between Dirac and Majorana
signatures.
The final results of this analysis will be presented as a function of \mhnl and
\mixpar, separately for HNL couplings to electrons and muons.

Tabulated results are provided in the HEPData record for this analysis~\cite{hepdata}.

\section{The CMS detector and event reconstruction}

The central feature of the CMS apparatus is a superconducting solenoid of
6\unit{m} internal diameter, providing a magnetic field of 3.8\unit{T}.
Within the solenoid volume are a silicon pixel and strip tracker, a lead
tungstate crystal electromagnetic calorimeter (ECAL), and a brass and
scintillator hadron calorimeter (HCAL), each composed of a barrel and two endcap
sections.
Forward calorimeters extend the pseudorapidity (\sigeta) coverage provided by
the barrel and endcap detectors.
Muons are measured in gas-ionization detectors embedded in the steel flux-return
yoke outside the solenoid.
A more detailed description of the CMS detector, together with a definition of
the coordinate system used and the relevant kinematic variables, can be found in
Ref.~\cite{CMS:2008xjf}.

Events of interest are selected using a two-tiered trigger system.
The first level, composed of custom hardware processors, uses information from
the calorimeters and muon detectors to select events at a rate of around
100\unit{kHz} within a fixed latency of about 4\mus~\cite{CMS:2020cmk}.
The second level, known as the high-level trigger, consists of a farm of
processors running a version of the full event reconstruction software optimized
for fast processing, and reduces the event rate to around 1\unit{kHz} before
data storage~\cite{CMS:2016ngn}.

Charged-particle tracks are reconstructed over the full \sigeta range of the
tracker, finding charged particles with transverse momentum (\pt) as low as
0.1\GeV, or produced as far as 60\cm from the beam line~\cite{CMS:2014pgm,
CMS:DP-2017-015}.
The silicon tracker used in 2016 measured charged particles within the range
$\abseta<2.5$.
For nonisolated particles of $1<\pt<10\GeV$ and $\abseta<1.4$, the track
resolutions were typically 1.5\% in \pt and 25--90 (45--150)\mum in the
transverse (longitudinal) impact parameter~\cite{CMS:2014pgm}.
At the start of 2017, a new pixel detector was
installed~\cite{CMSTrackerGroup:2020edz}.
The upgraded tracker measured particles up to $\abseta<3.0$ with typical
resolutions of 1.5\% in \pt and 20--75\mum in the transverse impact parameter
for nonisolated particles of $1<\pt<10\GeV$~\cite{CMS:DP-2020-049}.

Tracks are used to reconstruct the vertices in each event.
The candidate vertex with the largest value of summed physics-object $\pt^2$ is
taken to be the primary \pp interaction vertex.
The physics objects used for this determination are the jets, clustered using
the jet finding algorithm~\cite{Cacciari:2008gp,Cacciari:2011ma} with the tracks
assigned to candidate vertices as inputs, and the associated missing transverse momentum, taken as the negative vector sum of the \pt of those jets.

The global event reconstruction with the particle-flow (PF)
algorithm~\cite{CMS:2017yfk} aims to reconstruct and identify each individual
particle in an event, with an optimized combination of all subdetector
information.
In this process, the identification of the particle type (photon, electron,
muon, charged or neutral hadron) plays an important role in the
determination of the particle direction and energy.
ECAL energy clusters not linked to the extrapolation of any charged-particle
trajectory to the ECAL are reconstructed as photons.
A primary charged-particle track associated with potentially many ECAL energy
clusters corresponding to this track extrapolation to the ECAL and to possible
bremsstrahlung photons emitted along the way through the tracker material is
identified as an electron.
Muons are reconstructed from tracks in the central tracker consistent with
either a track or several hits in the muon system, and associated with
calorimeter deposits compatible with the muon hypothesis.
Charged-particle tracks neither identified as electrons, nor as muons are
attributed to charged hadrons.
Finally, neutral hadrons are reconstructed from HCAL energy clusters not linked
to any charged-hadron trajectory, or from a combined ECAL and HCAL energy excess
with respect to the expected charged-hadron energy deposit.

The electron momentum is estimated by combining the energy measurement in the
ECAL with the momentum measurement in the tracker.
Within the geometric acceptance of the tracker of $\abseta<2.5$, the momentum
resolution for electrons with $\pt\approx45\GeV$ from $\PZ\to\Cee$ decays ranges
from 1.7 to 4.5\%.
It is generally better in the barrel region than in the endcaps, and also
depends on the bremsstrahlung energy emitted by the electron as it traverses
the material in front of the ECAL~\cite{CMS:2020uim}.

Muons are measured in the range $\abseta<2.4$, with detection planes made using
three technologies: drift tubes, cathode strip chambers, and resistive-plate
chambers.
The single muon trigger efficiency exceeds 90\% over the full \sigeta range, and
the efficiency to reconstruct and identify muons is greater than 96\%.
Matching muons to tracks measured in the silicon tracker results in a relative
\pt resolution, for muons with \pt up to 100\GeV, of 1 (3)\% in the barrel
(endcaps)~\cite{CMS:2018rym}.

The missing transverse momentum vector \ptvecmiss is computed as the negative
vector sum of the transverse momenta of all the PF candidates in an event, and
its magnitude is denoted as \ptmiss~\cite{CMS:2019ctu}.

\section{Event simulation}
\label{sec:simulation}

Events from Monte Carlo (MC) simulations are used to derive predictions for
signal and several of the background processes.
To reproduce the correct multiplicity of \pp interactions per bunch crossing
(pileup) observed in data, simulated minimum bias events are superimposed on the
hard collisions in simulated signal and background events.
Simulated events are reweighted to match the distribution of the number of
pileup collisions per event in data using the total inelastic \pp cross section
from Ref.~\cite{CMS:2018mlc}.
All events are processed through a simulation of the CMS detector based on
\GEANTfour~\cite{GEANT4:2002zbu}.

Signal events are generated using the \MGvATNLO~v2.4.2
generator~\cite{Alwall:2014hca, Artoisenet:2012st} at leading order (LO) in the
strong coupling constant~\alpS.
The model used to generate signal events extends the SM particle content by
introducing up to three right-handed neutrinos~\cite{Atre:2009rg, Alva:2014gxa,
Degrande:2016aje, Pascoli:2018heg}.
The production cross section is scaled to next-to-next-to-LO (NNLO) precision
using $K$-factors derived from the comparison of SM samples of \PW~boson
production generated at LO precision, using exactly the same settings as the
signal samples, to the cross section calculated at NNLO with the \FEWZ3.1
program~\cite{Melnikov:2006kv, Gavin:2010az, Gavin:2012sy, Li:2012wna}.

In the simulation of signal events, we consider HNL masses~\mhnl in the range
1--20\GeV.
The HNLs are assumed to couple exclusively to one of the three SM neutrino
families with different mixing probabilities~\mixpar in the range
\TEN{-8} to \TEN{-1}, depending on the HNL mass.
The values of \mhnl and \mixpar not only determine the HNL production cross
section, but also its mean lifetime and, consequently, acceptance and
reconstruction efficiency.
For a fixed value of \mhnl, therefore, a simple cross section rescaling is not
sufficient to correctly reproduce the behavior of other HNLs with same mass and
different \mixpar.
To emulate an HNL signal sample with a specific value of \mixpar, we thus apply
per-event weights to the events from all signal samples with the same mass such
that the HNL lifetime distribution (taken before the parton shower and detector
simulation) matches the predicted distribution for the chosen \mixpar value, and
additionally we apply a global weight to match the expected cross section.

The signal MC samples are generated assuming Majorana HNLs with both LNC and LNV
decays.
Dirac HNL scenarios for any (\mhnl,\mixpar) values are emulated from the
Majorana HNL samples, using the fact that the HNL production cross section is
the same but the HNL lifetime is twice as large for a Dirac HNL without the LNV
decay channels.
To that end, all Majorana HNL events are reweighted to produce the expected
Dirac HNL lifetime, and all LNV events are classified as LNC events assuming
equal selection efficiencies for LNV and LNC events.

{\tolerance=900
Contributions from background processes are estimated mostly from control
samples in data.
However, MC event simulation is still used to validate the data-driven methods,
as well as to derive predictions for minor background processes contributing to
the signal regions.
Drell--Yan (DY) dilepton production in association with up to four additional
jets is simulated at next-to-LO (NLO) in perturbative quantum chromodynamics
(QCD) with \MGvATNLO and the FxFx algorithm~\cite{Frederix:2012ps} for matching
jets from matrix element calculations and parton showers.
Simulation of \PW~boson production in association with jets (\Wjets) is carried
out at LO precision with the \MGvATNLO generator using the MLM matching
scheme~\cite{Alwall:2007fs}.
Top quark and diboson ($\PW\PW$, \WZ, \ZZ) processes with additional jets are
generated with the \POWHEG~v2~\cite{Nason:2004rx, Frixione:2007nw,
Frixione:2007vw, Alioli:2009je, Alioli:2010xd, Re:2010bp, Melia:2011tj,
Nason:2013ydw} and \MGvATNLO generators at NLO precision.
The production of three vector bosons (\PW and \PZ~bosons) is simulated with
\MGvATNLO at NLO precision.
\par}

The NNPDF3.0 and 3.1~\cite{NNPDF:2014otw, NNPDF:2017mvq} parton
distribution functions (PDFs) are used for simulating the hard process.
The \PYTHIA~v8.226 and v8.230 event generator~\cite{Sjostrand:2014zea} is used
to model parton shower effects using the CUETP8M2T4 and CUETP8M1 underlying
event tunes~\cite{CMS:2016kle} for the simulated samples produced for the
analysis of the 2016 data set, while the CP5 tune~\cite{CMS:2019csb} is used in
the production of the corresponding samples simulated with the 2017 and 2018
detector conditions.

\section{Selection of leptons and jets}
\label{sec:leptonsjets}

Signal events consistent with the HNL decay topology are expected to contain one
lepton originating from the PV (prompt lepton), as well as two (displaced)
leptons whose tracks reconstruct an SV.
The analysis makes use of single-electron and single-muon triggers to select
events.
The single-electron triggers require the presence of at least one isolated
electron with $\pt>27$ (2016) or 32\GeV (2017--2018 data-taking periods),
while the single-muon triggers select events with at least one isolated muon
with $\pt>24\GeV$.
An additional set of nonisolated triggers at reduced rate with looser \pt
requirements on leptons is used for the estimation of backgrounds from
misidentified leptons.

Signal leptons are required to be isolated from any hadronic activity in the
event.
An isolation variable~\Irel is computed as the scalar \pt sum of charged hadrons
originating from the PV, neutral hadrons, and photons, all within a cone of
$\dR=\sqrt{\smash[b]{\Delta\eta\,^2 +\dPhi\,^2}}<0.3$ around the lepton
candidate direction at the vertex, divided by the transverse momentum \ptell of
the lepton candidate:
\begin{linenomath}\begin{equation*}
    \Irel=\frac{1}{\ptell}\left(\sum_{\text{ch. hadr.}}\pt^{\text{PV}}+
    \max{\left[0,\sum_{\text{neu. hadr.}}\pt+\sum_{\text{pho.}}\pt
    -\rho\Aeff\right]}\right).
\end{equation*}\end{linenomath}
The term $\rho\Aeff$ is used to mitigate the contribution of the pileup to the
isolation variable: the average \pt flow density $\rho$ is calculated in each
event using a ``jet area'' method~\cite{Cacciari:2007fd}, and the effective area
\Aeff is the geometric area of the isolation cone multiplied by an
\sigeta-dependent correction factor that accounts for the residual dependence of
the isolation on pileup.

Electron reconstruction is based on the combination of ECAL information and a
Gaussian sum filter tracking algorithm that accounts for possible bremsstrahlung
from the electron~\cite{CMS:2020uim}.
Electrons are required to have $\abseta<2.5$ to be within the geometric acceptance of the CMS tracking system.
To select signal electrons and reduce the rate of misidentified electrons, we
impose identification criteria based on the following properties of the
reconstructed electron: the electromagnetic shower shape, calorimetric energy
ratios, track-cluster matching, track quality, and track impact parameters with
respect to the PV.

Prompt electrons are identified using a multivariate (MVA)
discriminant~\cite{CMS:2020uim} with a selection efficiency of about 90\%, with
$\pt>30$ (32)\GeV in the 2016 (2017--2018) data set.
Additional selection requirements are imposed on the transverse and
longitudinal impact parameters with respect to the PV of the associated track,
$\absdxy<0.05\cm$ and $\absdz<0.1\cm$.
Prompt electrons are required to be isolated with $\Irel<0.1$, and must be
matched to the corresponding trigger objects.
Prompt electrons passing this selection will be referred to as ``tight prompt
electrons'' hereafter.

Displaced electrons are identified using the standard set of sequential
requirements used in prompt electron identification~\cite{CMS:2020uim}, but
without requiring a veto on photon conversions, as well as not including the
requirement on the maximum number of missing inner hits, since it significantly
affects the identification efficiency for electrons not emerging from the PV.
``Tight displaced electrons'' are required to have $\pt>7\GeV$ and to originate
from a nonprimary vertex by requiring $\absdxy>0.01\cm$ (no \absdz requirement
is imposed).
In addition, these electrons must be isolated ($\Irel<0.2$) and satisfy the
standard prompt electron identification criteria.
Samples enriched in misidentified electrons are selected by applying the tight
displaced electron identification criteria with a looser requirement on the
relative isolation ($\Irel<2$).
These electrons are referred to as ``loose displaced electrons''.

Muons are reconstructed by combining the detector information of the tracker and
of the muon spectrometer~\cite{CMS:2018rym}.
The geometric compatibility between these separate measurements is used in the
further selection of muons.
Muons are required to have $\abseta<2.4$ to fall inside the geometric acceptance
of the muon detector.
All muons considered for the analysis are identified by loose criteria on the
matching between inner tracker tracks and muon spectrometer measurements.
Additional loose selection criteria are formulated based on the impact
parameters of the tracks with respect to the PV.

Prompt muons must satisfy identification criteria based on the track quality and
the matching of the inner tracker track with the measurements in the muon
detectors, have $\pt>25\GeV$, and be matched to the corresponding trigger
objects.
Prompt muons must be isolated ($\Irel<0.1$) and associated with the PV by
satisfying $\absdxy<0.05\cm$ and $\absdz<0.1\cm$.
The muon track in the tracking detector is required to use hits from more than
80\% of the layers it traverses.
Prompt muons passing this selection will be referred to as ``tight prompt
muons''.

Displaced muons must satisfy a set of standard identification criteria, removing
some of the requirements that may reduce the efficiency for reconstructed muons
emerging from an SV.
``Tight displaced muons'' are required to have $\pt>5\GeV$, not be associated
with the PV ($\absdxy>0.01\cm$ and $\absdz<10\cm$), and satisfy a looser
isolation requirement ($\Irel<0.2$).
In contrast to the prompt muon identification, no selection is applied on the
fraction of hits in the tracking detector, nor on the quality of the matching
between the tracks reconstructed in the tracking detector and in the muon
system.
``Loose displaced muons'' are selected with the same identification criteria as
used to define tight displaced muons, but with a looser requirement on the
isolation ($\Irel<2$).

We evaluate the efficiencies to reconstruct displaced leptons from an HNL decay
in simulated signal samples for typical model parameters.
The average efficiency to reconstruct a loose displaced electron with a vertex
displacement of 10 (25)\cm in the transverse plane is found to be 20--40
(15--20)\%, while the average efficiency for those loose displaced electrons to
also satisfy the tight identification requirements is 60--80 (50--60)\%.
The average efficiency to reconstruct a loose displaced muon with a vertex
displacement of 10 (50)\cm in the transverse plane is 85--90 (40--50)\%, while
the average efficiency for those loose displaced muons to also satisfy the tight
identification requirements is higher than 95 (80)\%.
While the selection criteria for displaced electrons and muons are slightly
different, the large difference in the observed efficiencies is primarily due to
the better resolution and higher efficiency in the basic reconstruction of muons
compared to electrons~\cite{CMS:2020uim, CMS:2018rym}.

Particles reconstructed by the PF algorithm~\cite{CMS:2017yfk} are clustered
into jets using the anti-\kt algorithm~\cite{Cacciari:2008gp} with a value of
0.4 for the distance parameter, as implemented in the
\FASTJET~\cite{Cacciari:2011ma} package.
Jets are required to pass several quality criteria, designed to remove those jet
candidates that are likely to originate from anomalous energy deposits.
To account for the effect of pileup, nonuniformities of the detector response,
and residual differences between jets in data and in simulation, jet energy
corrections are applied~\cite{CMS:2016lmd}, and the \ptvecmiss is modified
accordingly.
Jets selected for analysis are required to have $\pt>25\GeV$ and $\abseta<2.4$.

Jets originating from the decay of \PQb~hadrons are identified using the
\DeepCSV \PQb~tagging algorithm~\cite{CMS:2017wtu}.
Jets are considered \PQb~tagged, and referred to as ``\PQb~jets'' hereafter, if
they satisfy the loose selection on the \DeepCSV tagger, corresponding to an
efficiency of about 84\% in selecting jets containing \PQb~hadron decays, with
misidentification probabilities of about 11 and 41\% for light and \PQc~jets,
respectively.

\section{Event selection and search strategy}

Signal events are characterized by the presence of a prompt lepton (referred to
as \lone), two displaced leptons (\ltwo and \lthree), and a neutrino.
They are selected with the identification criteria discussed in
Section~\ref{sec:leptonsjets}.
In the following, ``lepton'' will refer to an electron or muon, and \ltwo
(\lthree) will refer to the lepton in the pair with the higher (lower) \pt.
Events are split into final states in which \lone and at least one of \ltwo or
\lthree are electrons (\eex) or muons (\mmx), as expected for LNC HNL decays.
Events with \eex final states are sensitive to \mixpare, while \mmx events are
sensitive to \mixparm.

\begin{figure}[t!]
\centering
\includegraphics[width=.31\textwidth]{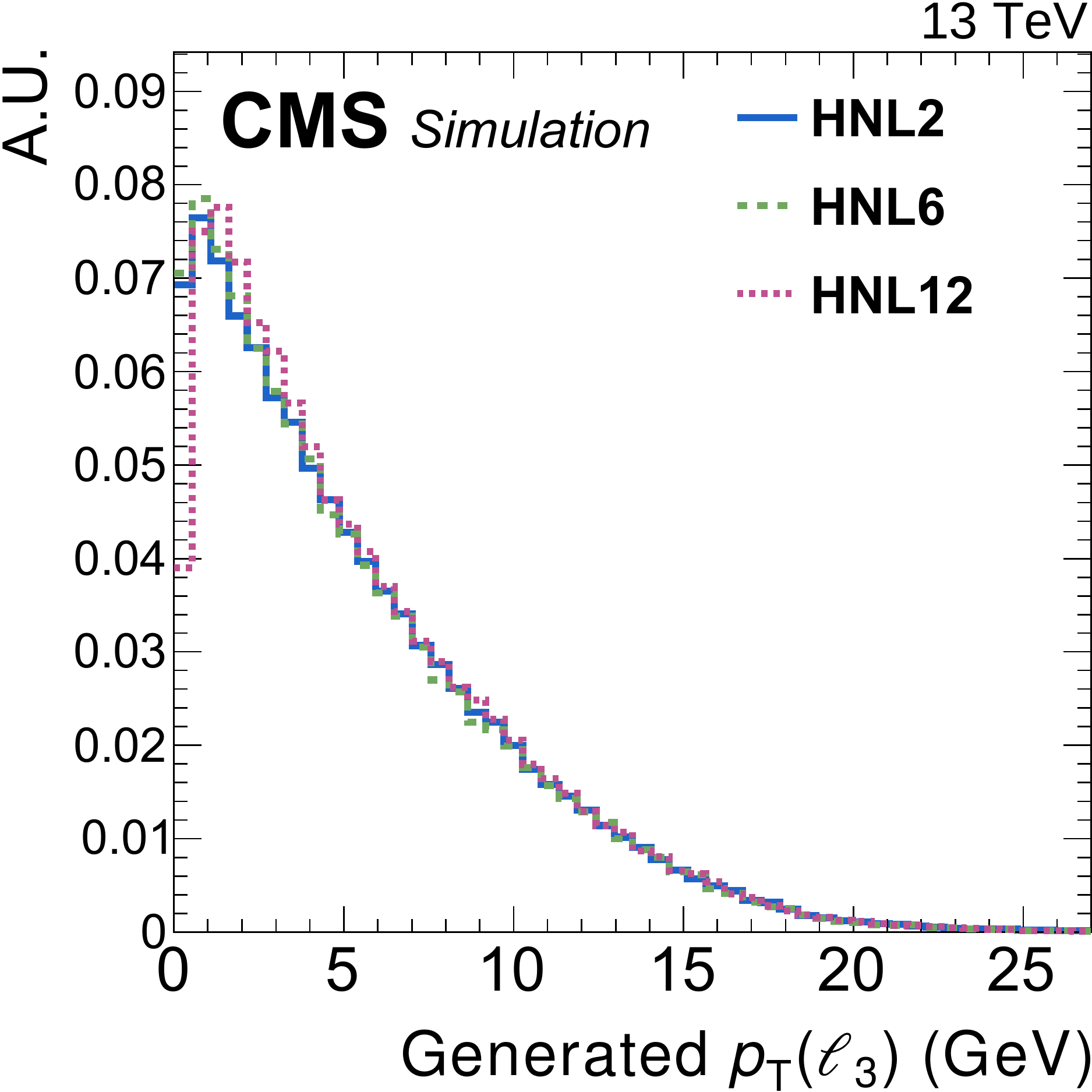}
\hfill
\includegraphics[width=.31\textwidth]{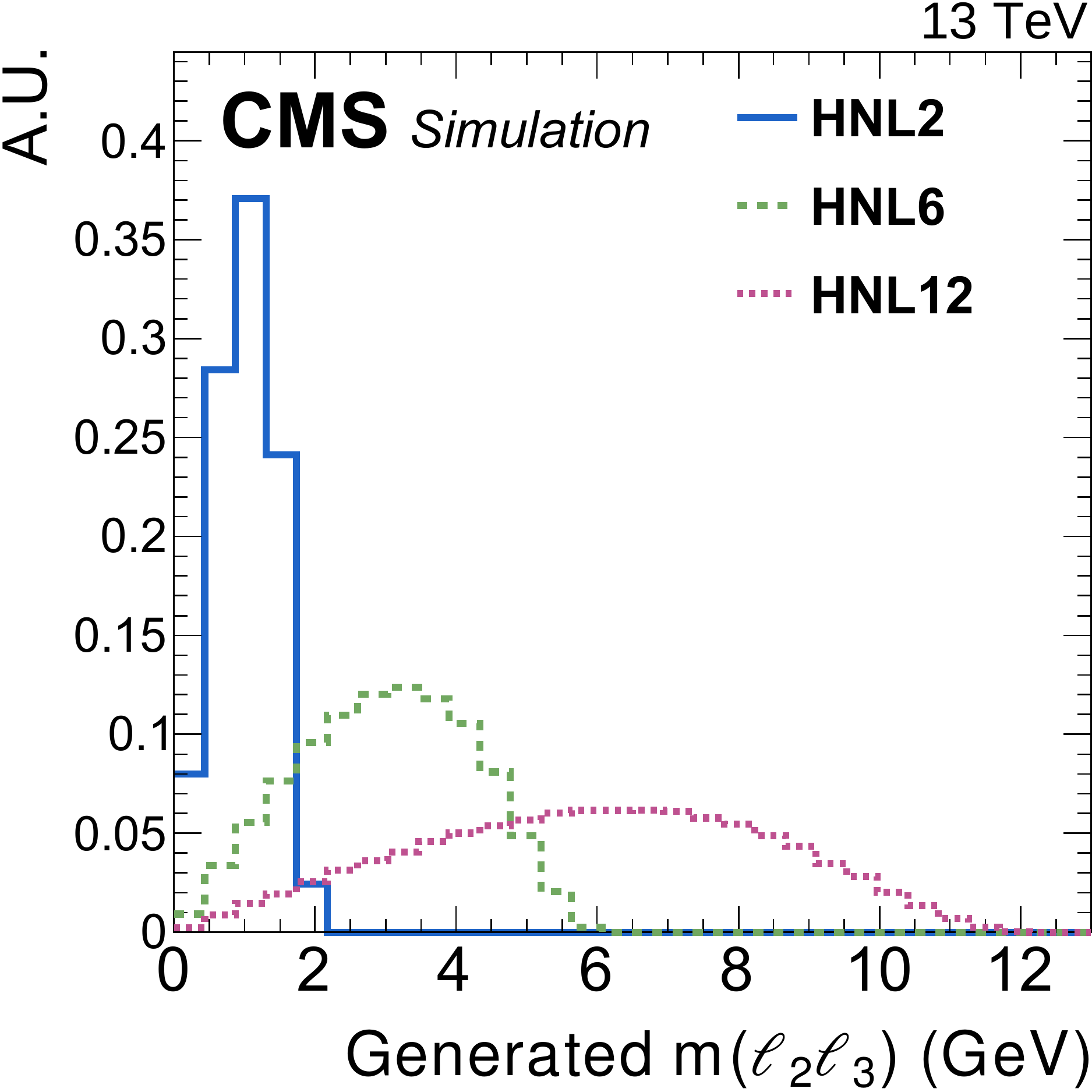}
\hfill
\includegraphics[width=.31\textwidth]{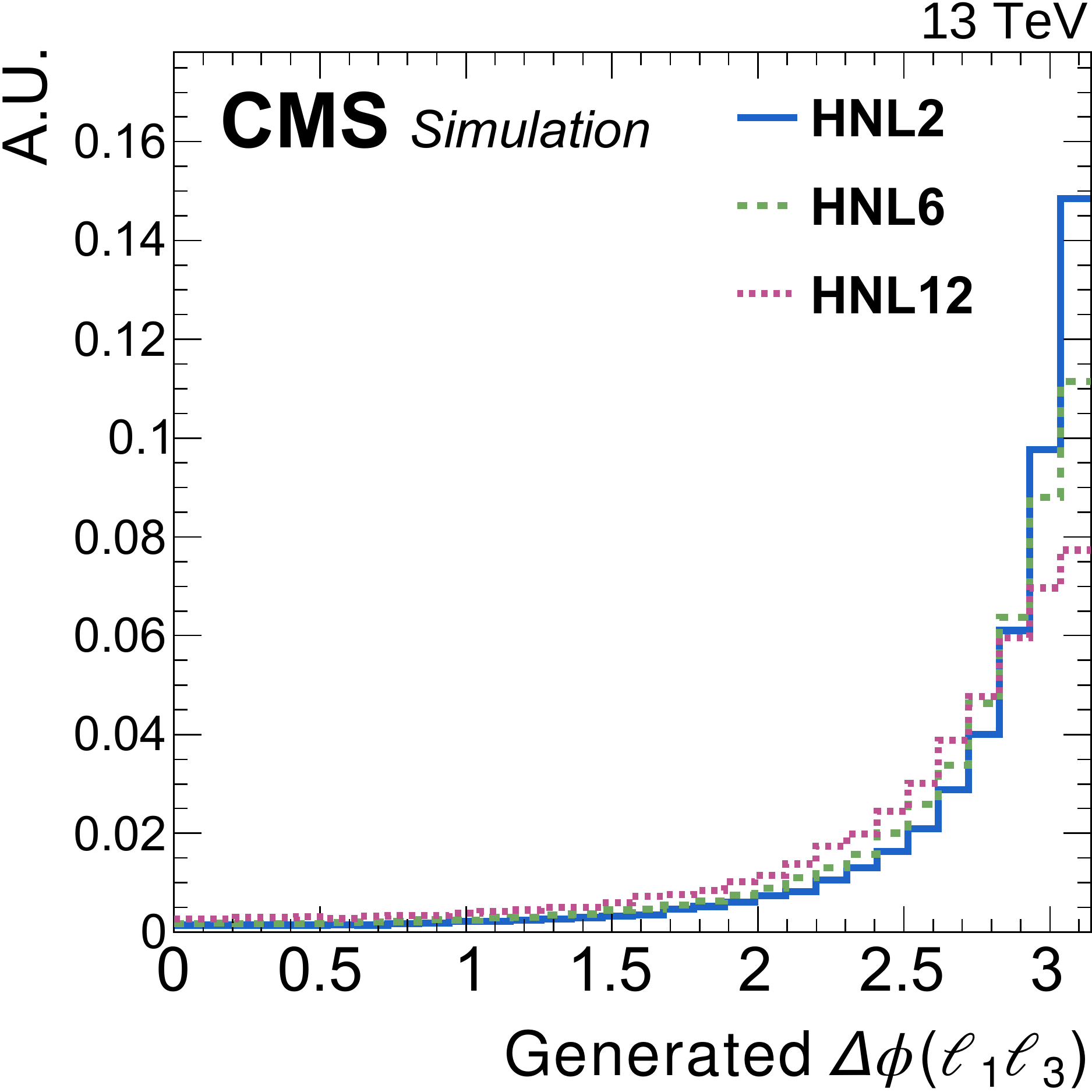}
\caption{\label{fig:selection:signal}
    Distribution of kinematic properties of the generated leptons for simulated HNL signal events: the \pt of \lthree
    (left), the \mtwol variable (center), and the angular separation between
    \lone and \lthree (right).
    Predictions are shown for several benchmark hypotheses for Majorana HNL
    production: \mHNLtwo and $\mixpar=\Ten{1.2}{-4}$ (HNL2), \mHNLsix and
    $\mixpar=\Ten{4.1}{-6}$ (HNL6), \mHNLtwelve and $\mixpar=\Ten{1.0}{-5}$
    (HNL12).
}
\end{figure}

Given the low HNL masses considered in this analysis ($\mhnl<20\GeV$), \lone has
the typical \pt spectrum expected for \PW~boson decays, with a Jacobian peak
around 40\GeV, while \ltwo and \lthree have rather soft \pt spectra, an
invariant mass~\mtwol smaller than \mhnl, and a small opening angle.
In the absence of significant hadronic activity, \lone and the HNL are typically
separated by a large azimuthal angle.
In Fig.~\ref{fig:selection:signal}, we show properties of the generated leptons
in simulated event samples (\ie, for leptons after the simulation of additional
radiation but before detector simulation), for a selection of HNL masses and
couplings.
These features, along with the possible displacement of \ltwo and \lthree, can
be used to identify the two leptons coming from the HNL decay.

The requirement $\absdxy>0.01\cm$ is imposed on \ltwo and \lthree, since they
are expected to originate from the SV.
Note that given the rapid variation of the expected HNL displacement as a
function of \mhnl and \mixpar, this initial displacement requirement is kept
loose, and does not fully resolve possible ambiguities between prompt and
displaced reconstructed leptons.

Among all the leptons satisfying the prompt selection, the one with the highest
\pt is chosen as \lone.
Among all the selected displaced leptons, the two leptons of any flavor with the
highest dilepton \pt and opposite sign are selected as \ltwo and \lthree
($\Pepm\Pemp$, $\Pepm\PGmmp$, $\PGmpm\PGmmp$).
If there are no opposite-sign displaced lepton pairs, the event is rejected.
This selection requirement reduces background processes with misidentified
leptons, while retaining almost full efficiency for the signal.

The decay vertex of the HNL is reconstructed from the tracks associated with
\ltwo and \lthree.
This is done by fitting the two tracks to a common point with a Kalman-filter
approach~\cite{Fruhwirth:1987fm}.
The fitted SV provides an estimate for the position of the production vertex of
\ltwo and \lthree.

After selecting the events with one lepton from the PV and two leptons forming
an SV, the kinematic properties of the low-mass HNL events are investigated and
used to separate the signal from the SM backgrounds.
Typical backgrounds arise from SM processes containing either one or two
misidentified leptons, \eg, \ttbar, single top quark, DY, and \Wjets production,
or containing a real photon that converts into a lepton pair---mostly
electrons---in the detector material, such as $\PW\PGg$ and $\PZ\PGg$.

Compared to these background processes, \ltwo and \lthree in signal events are
expected to have a small opening angle, given the small mass and relatively
large momentum of the HNL.
The \dR~separation between \ltwo and \lthree, \DRtwol, discriminates the signal
from various background processes, especially background events with one
misidentified lepton.
In order to retain high signal efficiency, a relatively loose requirement of
$\DRtwol<1$ is applied.

In the absence of energetic jets in signal events, \lone is expected to recoil
at a large angle in the transverse plane from the HNL, and thus from \ltwo and
\lthree.
An additional selection criterion is applied to the azimuthal separations
between \lone and each of the displaced leptons \ltwothree to be $\minDphi>1$.

The invariant mass of the three charged leptons, \mthreel, is limited by the
mass of the on-shell \PW~boson.
Given the relatively low momentum carried away by the neutrino, \mthreel peaks
just below the \PW~boson mass, with a steep falloff above 80\GeV and a larger
tail at lower masses.
We select events with \mthreel values between 50 and 80\GeV.
This requirement proves to be particularly efficient against the \Zgamma
background, where the photon radiated by one of the leptons from the \PZ~boson
decay undergoes an asymmetric conversion: one of the leptons receives most of
the photon momentum, while the other is too soft to be detected or identified.
Events with a \PQb~jet with $\pt>25\GeV$ are rejected, in order to substantially
reduce the background from processes with top quarks.
The transverse momentum of the $\ltwo\lthree$ dilepton system, \pttwol, is
required to be larger than 15\GeV, since the low-\pt region is dominated by
background processes.
The decay products of the HNL are emitted at small opening angles, and thus the
vector sum of the \ltwo and \lthree momenta should be aligned with the HNL
flight direction.
We assume that the vector from the PV to the SV corresponds to the HNL flight
direction, and calculate the angle between this direction and the dilepton
momentum, \thetaSVll.
Events are required to satisfy $\cos\thetaSVll>0.99$.

The quality of the reconstructed SV necessarily correlates with the precision on
the track parameters of \ltwo and \lthree, as well as the spatial separation
between their trajectories at the intersection point.
The SV quality is estimated as a probability~\SVprob based on the $\chi^2$~fit
from the kinematic vertex fitter.
A loose requirement of $\SVprob>0.001$ is applied to remove events that are
clearly not compatible with a reconstructed SV.
Additionally, we calculate the significance~\SigTwoD, which is defined as the
ratio of the transverse PV-SV distance~\Deltwod and its uncertainty.
From simulated event samples, it is found that the distribution of~\SigTwoD in
signal events is approximately constant for a wide range of mass and coupling
values, while background processes typically result in small values of~\SigTwoD.
Thus, a tight requirement of $\SigTwoD>20$ is applied to increase the
signal-over-background ratio in the signal selection.

In the kinematic region of interest there are several background contributions
arising from SM processes that include distinct resonances in the dilepton
invariant mass spectrum.
Events with an OSSF (\ltwo,\lthree) pair are removed when $\Deltwod<1.5\cm$ and
the invariant mass~\mtwol is consistent with the mass of the \PGo, \PGf, \PJGy,
or \Pgy mesons.
The contribution from these processes to the larger displacement region is
negligible.
In the case where \lone forms an OSSF pair with one of \ltwo or \lthree, we
remove events if the invariant mass of that pair is consistent with either the
mass of one of the resonances mentioned above, or with one of the higher-mass
resonances \PgUa, \PgUb, \PgUc, and \PZ.

\begin{table}[ht!]
\centering
\renewcommand{\arraystretch}{1.2}
\topcaption{\label{tab:selection:baseline}
    Baseline selection criteria (left) and dilepton invariant mass vetoes
    (right) applied in the analysis.
    The width of the vetoed range for the meson resonances reflects the
    experimental resolution of the dilepton invariant mass reconstruction.
}
\begin{tabular}[t]{c@{\hspace{0.08\textwidth}}c}
\begin{tabular}[t]{lc@{\,}l}
    \multicolumn{3}{l}{Selection criteria} \\
    \hline
    \DRtwol & $<$ & 1 \\
    \minDphi & $>$ & 1 \\
    \mthreel & $\in$ & [50,80]\GeV \\
    number of \PQb jets & $=$ & 0 \\
    \pttwol & $>$ & 15\GeV \\
    $\cos\thetaSVll$ & $>$ & 0.99 \\
    \SVprob & $>$ & 0.001 \\
    \SigTwoD & $>$ & 20 \\
    $m(\Pell\Pell)$ & $\notin$ & vetoed ranges
\end{tabular} & \begin{tabular}[t]{ccr@{${}\pm{}$}l}
    OSSF pair & Resonance & \multicolumn{2}{l}{Vetoed range (\GeVns{})} \\
    \hline
    any & \PGo & 0.78 & 0.08 \\
    & \PGf & 1.02 & 0.08 \\
    & \PJGy & 3.10 & 0.08 \\
    & \Pgy & 3.69 & 0.08 \\[\cmsTabSkip]
    $\lone\ltwothree$ & \PgUa & 9.46 & 0.08 \\
    & \PgUb & 10.02 & 0.08 \\
    & \PgUc & 10.36 & 0.08 \\
    & \PZ & 91.2 & 10.0 \\
\end{tabular}
\end{tabular}
\end{table}

The requirements introduced above represent the baseline selection applied in
the analysis and are summarized in Table~\ref{tab:selection:baseline}.
The selection efficiency for signal events is strongly dependent on the HNL mass
and mixing parameters, as well as on the selected final state.
The average product of acceptance and efficiency is approximately 0.1\% for
$\mhnl=2\GeV$ and $\mixpar=\Ten{8}{-5}$, and close to 2\% for $\mhnl=12\GeV$ and
$\mixpar=\TEN{-6}$.

In regions of large displacement, the sensitivity of the analysis greatly
benefits from small contributions from the displaced lepton background and from
the absence of prompt lepton background contributions.
In contrast, signal events corresponding to smaller values of \tauhnl populate
regions of smaller displacement, where both the prompt and displaced lepton
contributions are still significant.
No explicit selection is applied on the SV displacement in order for this search
to be sensitive to the full range of \tauhnl values that we consider.
The sensitivity of the present search is optimized in bins of the HNL
displacement and \mtwol.

The residual background contributions are dominated by DY and top quark
production and processes with photon conversions, and have predominantly small
\Deltwod.
In contrast, HNL signal events have typically larger \Deltwod, depending on
\mhnl and \mixpar.
As a result, \Deltwod provides the best discrimination between signal and
background.

Event categories are defined based on the value of \Deltwod, with optimized
sensitivities for various signal scenarios.
Selected events are additionally split into categories corresponding to
three different final states per flavor of the prompt lepton.
The mixed-flavor final states are split into categories where the prompt and
displaced lepton of the same flavor have the same (\MMEss and \EEMss) or the
opposite (\MMEos and \EEMos) charge, such that the former (latter) categories
probe the LNV (LNC) hypothesis.
For the same-flavor final states (\MMM and \EEE, where the two displaced leptons
are also always of opposite charge), no distinction between LNC and LNV events
is possible, so no additional categorization is performed.
The categories are associated with different background rates and process
composition, enhancing the sensitivity of the search.

The top quark background processes, where \ltwo and \lthree are typically
produced in the semileptonic decay of \PQb~hadrons, populate the region
$\mtwol<4\GeV$, and are nearly absent at higher dilepton masses.
The data are therefore further split into two \mtwol categories of $<$4 and
$>$4\GeV.
In the high-mass region, only small background contributions are expected
originating mainly from DY and \Wjets production, and thus fewer \Deltwod
categories are used for $\mtwol>4\GeV$.
The event categorization in \mtwol and \Deltwod is displayed in
Table~\ref{tab:selection:binning}.

\begin{table}[ht!]
\centering
\renewcommand{\arraystretch}{1.2}
\topcaption{\label{tab:selection:binning}
    Definition of kinematic regions in terms of dilepton invariant
    mass~\mtwol and SV displacement~\Deltwod.
}
\begin{tabular}{lc@{\cmsSmallColSkip}c@{\cmsSmallColSkip}c@{\cmsSmallColSkip}c@{\cmsColSkip}c@{\cmsSmallColSkip}c}
    \mtwol (\GeVns{}) & \multicolumn{4}{c}{$<$4} & \multicolumn{2}{c}{$>$4} \\ \hline
    \Deltwod (cm) & $<$0.5 & 0.5--1.5 & 1.5--4 & $>$4 & $<$0.5 & $>$0.5 \\
\end{tabular}
\end{table}

\section{Background estimation}
\label{sec:backgrounds}

The dominant source of background for the final states considered in this search
consists of events with misidentified hadrons, muons from pion or kaon decays,
and leptons from heavy-flavor hadron decays.
We refer to such leptons as ``background leptons''.
Events with background leptons originate primarily from top quark, DY, and
\Wjets production, where the background leptons are typically selected as
nonprompt leptons.
The contribution of events with displaced background leptons is estimated with a
``tight-to-loose'' method using control samples in data.
In contrast, the contribution of events with prompt background leptons, \eg,
from SM events composed uniquely of jets produced through the strong interaction, referred to as QCD multijet events, is found to be negligible, because of the
much smaller probability for background leptons to satisfy the tight prompt
selection criteria.

Processes that include \PW or \PZ~bosons can contribute to trilepton final
states with at least one displaced electron if a photon radiated from the
initial or final state converts into an electron-positron pair in the material
of the beam pipe or the detector.
This background gives only a minor contribution, and is estimated from control
samples in data.

Processes with multiple bosons (\PW, \PZ, \PGg, \PH) or top quarks can
contribute to final states with three or more prompt leptons.
The dominant contribution arises from \WZ and \ZZ diboson production, while
processes involving more than two bosons or top quarks have very small
production rates and are further suppressed by the lepton and \PQb~jet vetoes.
The background contribution from all these processes is almost negligible, and
is estimated from simulated event samples.

We use data control samples to estimate two classes of background lepton events.
Single-background (SB) leptons are defined as single reconstructed leptons
produced via one of the aforementioned mechanisms.
Multiple SB leptons can be found in an event, produced in the decays of
different hadrons.
In this case, the probabilities of selecting background leptons can be treated
as uncorrelated.
Double-background (DB) leptons represent pairs of reconstructed leptons produced
in the decay chain of the same hadron (\eg, in
$\PQb\to\ellnu\PQc\to\ellnu\Pellprplus\PGnGellpr\PQs$) or a quarkonium state
(\eg, in $\PJGy\to\LLmp$).
In such decays, the two leptons are not independent and their selection
probabilities are correlated.
In background events with background leptons, the main contribution consists of
DB leptons, with subdominant contributions arising from SB~lepton events.

In the ``tight-to-loose'' method, the probability~\ttl (or ``tight-to-loose
ratio'') for displaced leptons passing a ``loose'' isolation criterion of
$\Irel<2$ to also pass the ``tight'' criterion of $\Irel<0.2$ (which is part of
the tight lepton selection described in Section~\ref{sec:leptonsjets}) is
measured in a ``measurement region'' selected with requirements orthogonal to
those of the baseline event selection.
A weight based on this probability is then applied to the number of events found
in an ``application region'', to estimate the background contribution in the
signal region.
Events in the application region are selected with requirements identical to the
baseline selection except for having at least one ``loose-not-tight'' displaced
lepton (\ie, with $0.2<\Irel<2$).
This method is applied separately for the estimation of SB and DB lepton events,
with separate measurement and application regions in the two cases.

Background leptons are typically found in the proximity of a jet.
Each loose displaced lepton is therefore associated with the jet of $\pt>10\GeV$
closest in \dR, if such a jet exists within $\dR(\Pell,\text{jet})<0.4$.
If two loose displaced leptons are either associated with the same jet or have
$\DRtwol<0.45$, they are considered DB leptons and treated as a single dilepton
system.
In all other cases, \ie, if they have $\DRtwol>0.45$, and are either associated
with different jets or at least one is not associated with a jet, the loose
displaced leptons are considered SB leptons and treated independently.
To reduce the dependency of \ttl on the flavor and momentum of the initial
parton from which the background leptons originate, the parton momentum is
reconstructed from the background leptons and the associated jets, and \ttl is
measured as a function of the reconstructed parton momentum.

For DB leptons, one common parton momentum is reconstructed as the sum of the
two displaced lepton momenta and the recalibrated momenta of associated jets.
The jet momenta are recalibrated by subtracting the associated lepton momenta
from the raw jet momentum, and applying the appropriate jet energy corrections.
The tight-to-loose ratio for DB leptons, \ttlDB, is parametrized by the \pt of
the reconstructed parton.
The measurement region for \ttlDB is defined by the baseline event selection
summarized in Table~\ref{tab:selection:baseline} except that we also require the
presence of at least one \PQb~jet and that the two loose displaced leptons pass
the DB lepton selection.
The \PQb~jet requirement enriches the selection in top quark events, which are
the main source of DB leptons.
We measure the \ttlDB values to be in the range 4--60\%, depending on the parton
\pt.
The application region for the DB lepton estimation is also defined by the
baseline event selection except for the requirements that at least one displaced
lepton is loose-not-tight and that the two displaced leptons pass the DB lepton
selection.

For SB leptons, the transverse momentum of the initial parton, \ptc, is
reconstructed from the lepton~\pt and isolation as:
\begin{linenomath}\begin{equation*}
    \ptc=\ptell\Big(1+\min\big(0,\Irel-0.2\big)\Big),
\end{equation*}\end{linenomath}
where 0.2 is the maximum \Irel value allowed for tight displaced leptons.
This definition accounts for the hadronic activity in the proximity of the
reconstructed lepton, and does not depend on the presence of a jet close to the
lepton.
The tight-to-loose ratio for SB leptons, \ttlSB, is parametrized as a function
of \ptc and of the lepton \sigeta.
To construct the \ttlSB measurement region we use prescaled single-lepton
triggers to select an event sample enriched in multijet events.
Events in the measurement region are required to contain exactly one loose
displaced lepton, and at least one jet separated from the lepton by $\dR>1$.
To reduce the contribution from events with prompt leptons, mainly arising from
\Wjets and DY production, we require $\ptmiss>20\GeV$ and
$\mT=\sqrt{\smash[b]{2\ptmiss\pt(\Pell)(1-\cos\dPhi)}}<20\GeV$, where \dPhi is
the azimuthal angle difference between the lepton and \ptvecmiss.
The residual contribution from prompt lepton events is subtracted using
simulated event samples for \Wjets, DY, and \ttbar production, with yields
normalized to the observed yields in an event selection defined by
$\ptmiss>20\GeV$ and $70<\mT<120\GeV$.
We obtain \ttlSB values of 10--60 (0.1--50)\% for electrons (muons), depending
on the parton \pt and \abseta.
The application region for the SB lepton estimation is defined by the baseline
event selection except for the requirements that at least one displaced lepton
is loose-not-tight and that the two displaced leptons do not pass the DB lepton
selection.

\begin{figure}[!p]
\centering
\includegraphics[width=.85\textwidth]{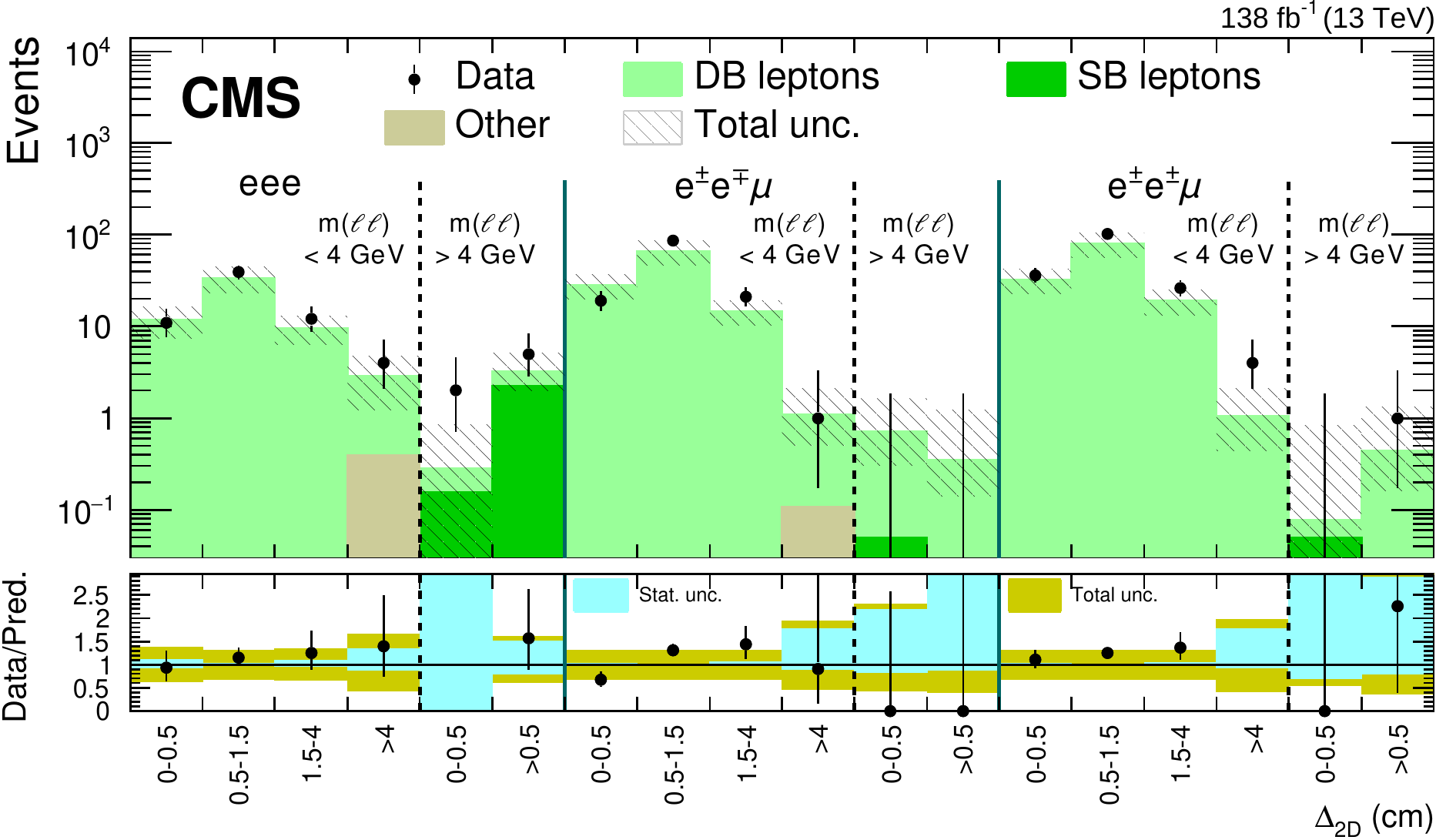} \\[1ex]
\includegraphics[width=.85\textwidth]{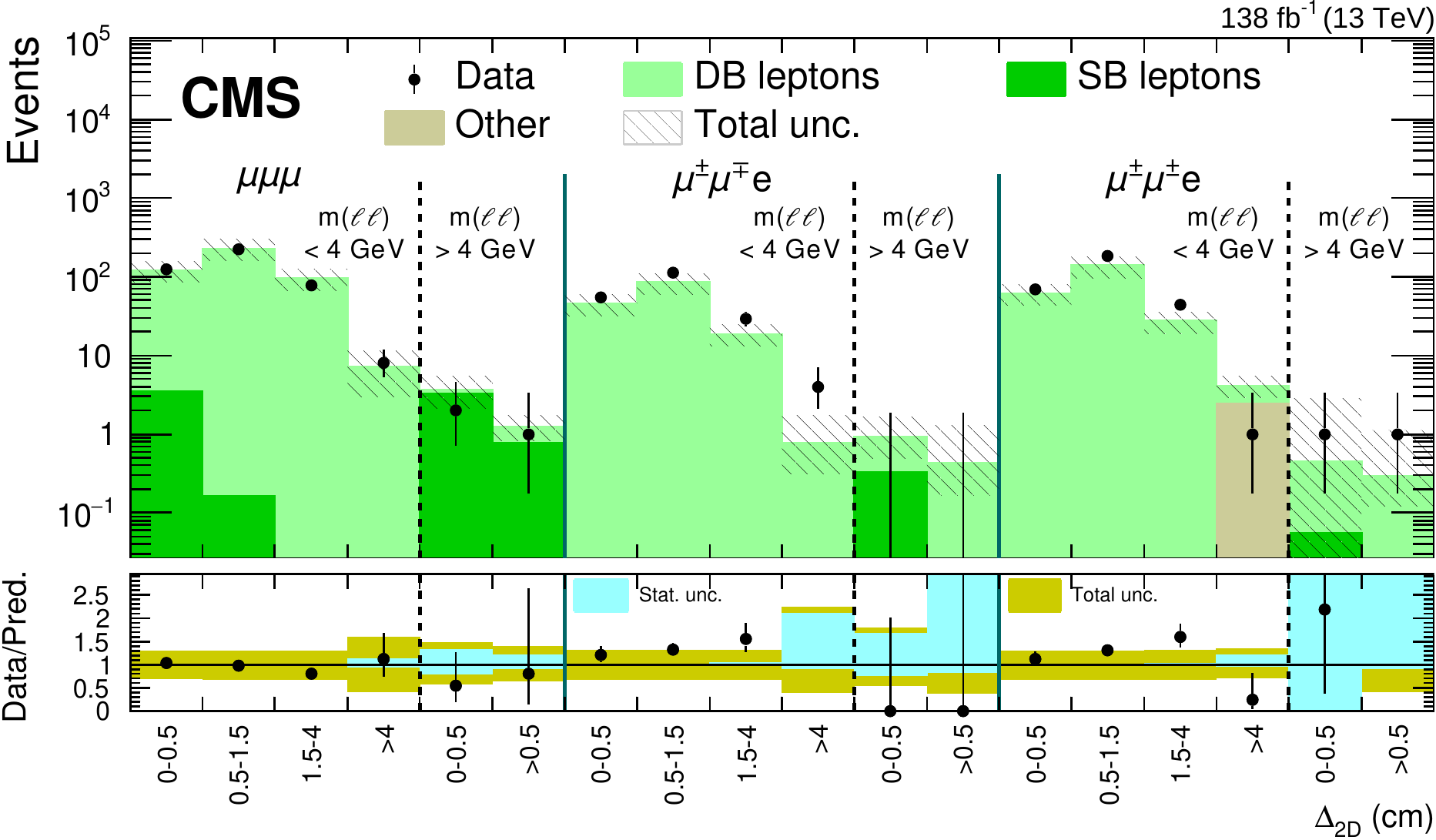}
\caption{\label{fig:backgrounds:controlregions}
    Comparison between the number of observed events in data and the background
    predictions (shaded histograms, stacked) in the background validation region
    for \eex (upper) and \mmx (lower) final states.
    The hashed band indicates the total systematic and statistical uncertainty
    in the background prediction.
    The lower panels indicate the ratio between data and prediction, and missing
    points indicate that the ratio lies outside the axis range.
    The uncertainty band assigned to the background prediction includes
    statistical and systematic contributions.
    Small contributions from background processes that are estimated from
    simulation are collectively referred to as ``Other''.
}
\end{figure}

The accuracy of the background estimates obtained with the SB and DB lepton
methods is derived from closure tests performed in data.
A background validation region is formed by selecting events that satisfy the
baseline event selection summarized in Table~\ref{tab:selection:baseline},
except for either fulfilling $\mthreel\notin[50,80]\GeV$ or having at least one
\PQb~jet.
The contribution of background processes in the background validation region is
estimated with the tight-to-loose method, \ie, by applying the measured values
of \ttl to events that pass the selection criteria of the background validation
region except for having at least one loose-not-tight displaced lepton, and
compared to the observed event yields.
Using the kinematic regions defined in Table~\ref{tab:selection:binning}, the
comparison is shown in Fig.~\ref{fig:backgrounds:controlregions}.
We find excellent agreement within the statistical and systematic uncertainties
discussed in Section~\ref{sec:systematic}.

Signal contamination in the measurement or application regions used for the SB
and DB lepton background contributions would result in an enlarged background
prediction and could thus hide the signal contribution in the signal region.
We have evaluated this for several HNL signal scenarios that are not excluded by
previous searches and find that the signal contribution in the measurement or
application regions, when propagated to the signal region in the same way as the
background prediction, is negligible compared to the direct HNL contribution in
the signal region.

\section{Validation of identification techniques}

The contributions from the main background processes in this search are
estimated directly from data.
The efficiencies for reconstruction, identification, and isolation of displaced
leptons in signal events are also validated in dedicated control regions in
data.

The performance of the displaced electron reconstruction, identification, and
isolation has been studied in data and MC simulation using a \Zgamma
conversion-enhanced selection.
The conversion product is taken as a proxy for the displaced lepton.
The asymmetric photon conversions include
$\PZ\to\LLmp\PGg\to\LLmp\Pepm(\Pemp)$ events, where (\Pemp) represents a low-\pt
electron that may fail to be reconstructed in the detector.
The other three leptons have an invariant mass close to the \PZ~boson mass.
Events are selected with two oppositely charged prompt tight electrons with
$\pt>35$ and 7\GeV, respectively, or two oppositely charged prompt tight muons
with $\pt>28$ and 5\GeV, respectively.
Additionally, one or two loose electrons are required.
The invariant mass of the tight OSSF lepton pair together with one of the
additional electrons must be within 10\GeV of the \PZ~boson mass.

While conversion events with four leptons allow for a reconstruction of the SV,
and thus directly give a \Deltwod value, the size of the selected event sample
is very small and does not allow for a precise determination of correction
factors.
The yield of conversion events with three leptons, in contrast, is much larger,
but the SV cannot be reconstructed and no \Deltwod value can be calculated.
As a proxy for the actual displacement of the converted photon with respect to
the PV, we define the ``displacement variable'':
\begin{linenomath}\begin{equation*}
    d=\sqrt{2\rade\dxy+\dxy^2}\,,
\end{equation*}\end{linenomath}
where \rade is the radius of curvature of the path of the displaced electron in
the magnetic field, and \dxy is the transverse impact parameter of the
associated track.

\begin{figure}[ht!]
\centering
\includegraphics[width=0.48\textwidth]{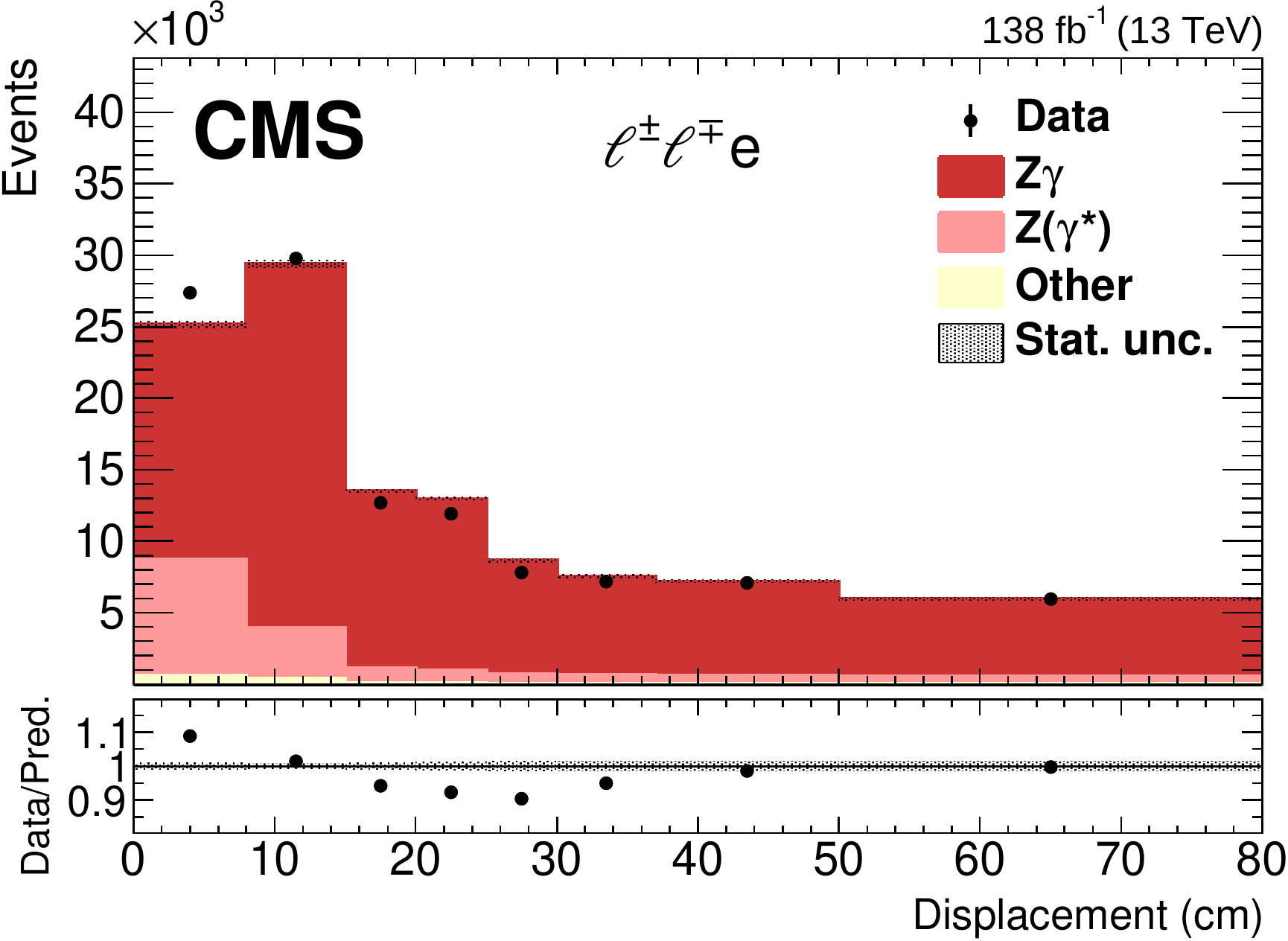}
\hfill
\includegraphics[width=0.48\textwidth]{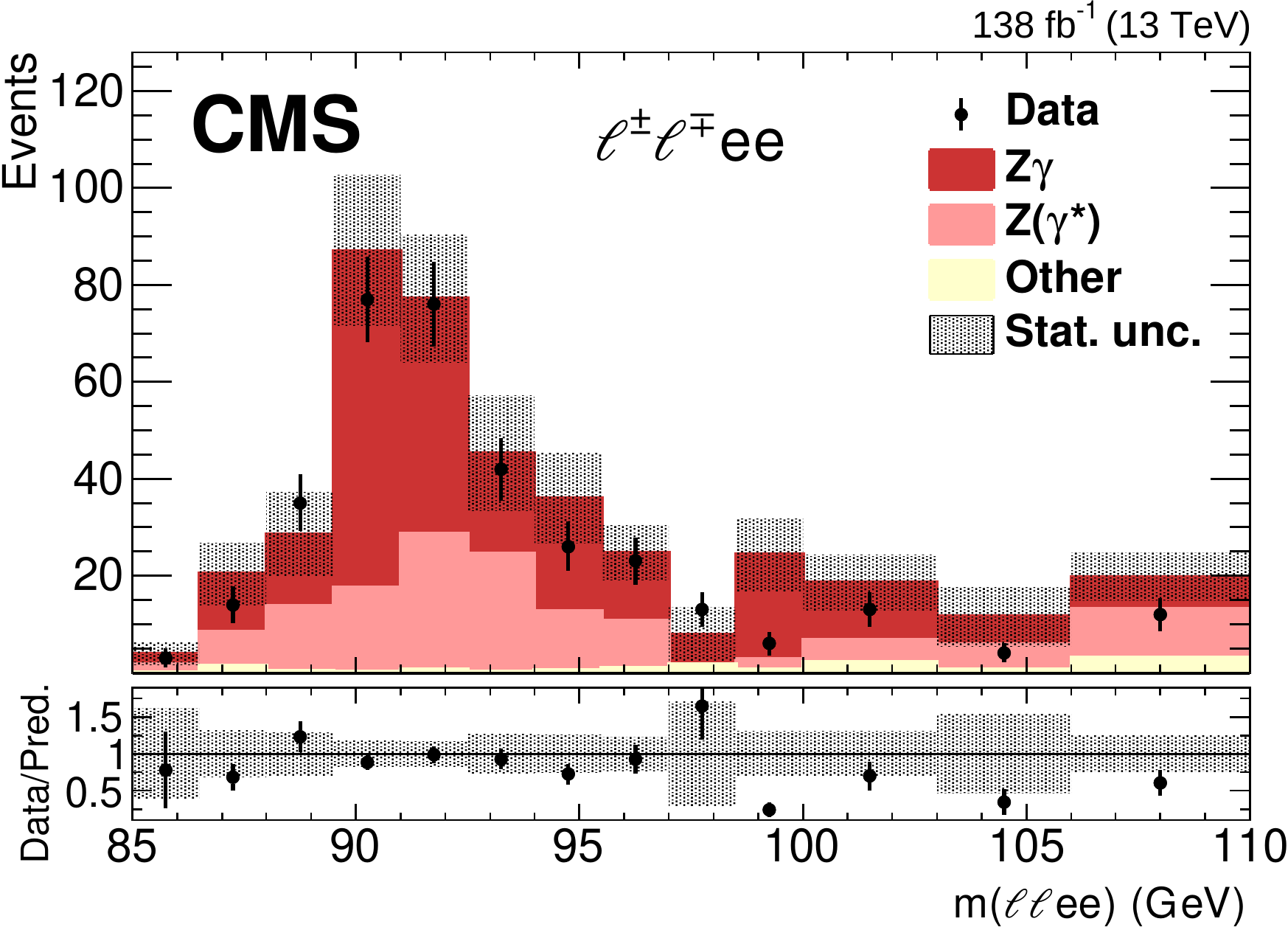} \\
\includegraphics[width=0.48\textwidth]{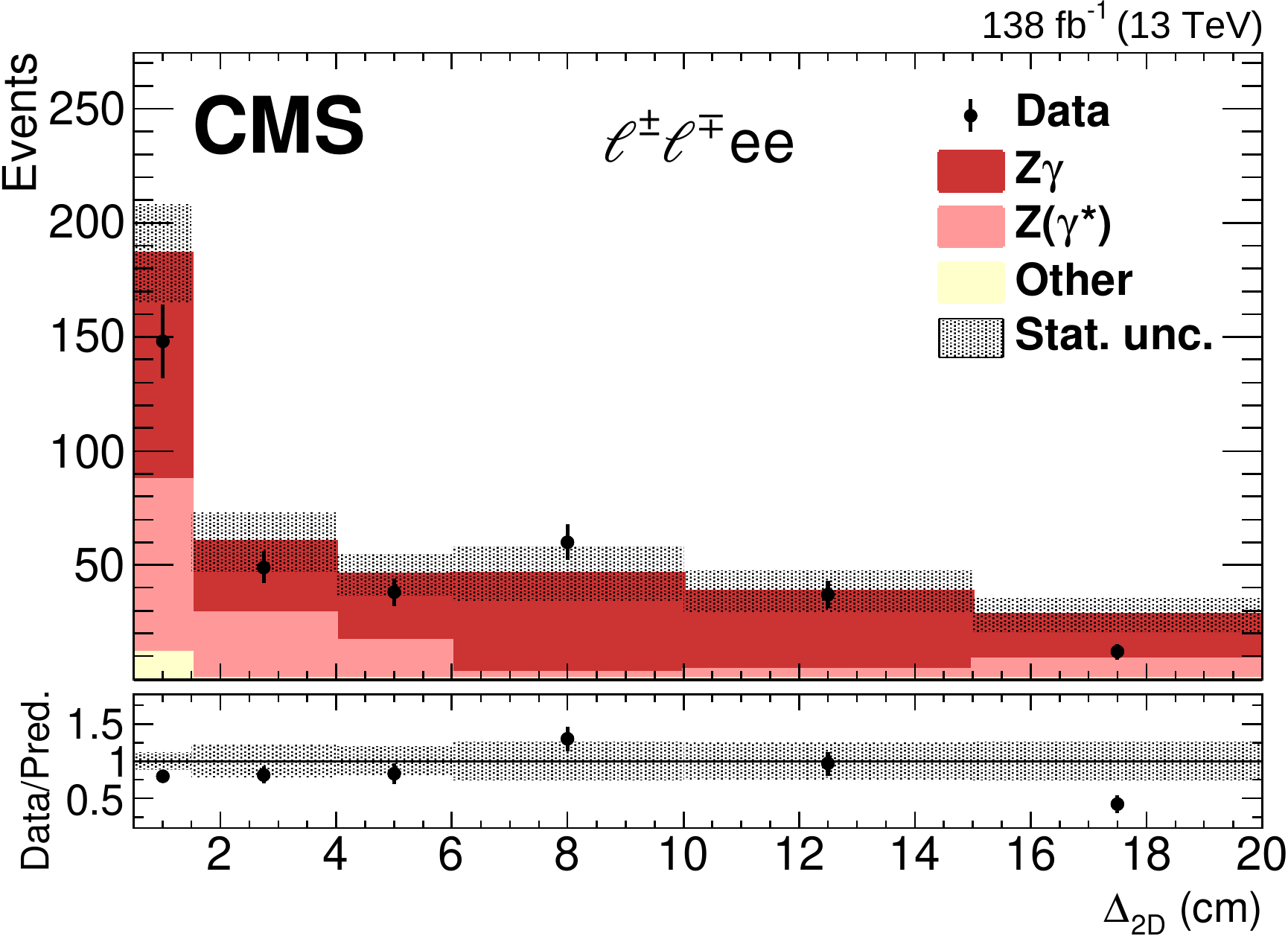}
\caption{\label{fig:validation:conversion}
    Comparison between the observed number of events in data and simulation for
    converted photons.
    Events are selected in the final states with three (upper left) or four
    (upper right, lower) leptons, with one (or two) of the leptons identified as
    displaced electron(s).
    The distributions are shown for the displaced electron displacement $d$
    (upper left) and reconstructed invariant mass of four leptons (upper right).
    Additionally, the \Deltwod variable is presented (lower).
    The simulated events correspond to the processes with external conversions,
    \Zgamma; internally converted photons, $\PZ(\PGg^{\ast})$; and other
    processes with the production of vector bosons and top quarks.
    In the distributions for the four-lepton final state (upper right, lower),
    the scale factor derived from the distribution of $d$ in three-lepton events
    (upper left) has been applied to the simulated events.
    The hashed band represents the statistical uncertainty in the simulation.
    The lower panels indicate the ratio between data and prediction.
}
\end{figure}

In Fig.~\ref{fig:validation:conversion} (upper left), the distribution of $d$ is
shown for conversion events with three leptons.
The measured data-to-simulation ratios in bins of $d$ are used as scale factors
per electron to apply to the HNL signal events.
These scale factors deviate from unity by up to 10\% per electron, depending on
the displacement value.
The four-lepton invariant mass and the \Deltwod distribution for
conversion events with four leptons are shown in
Fig.~\ref{fig:validation:conversion} as well, where the scale factors derived
from the events with three leptons have been applied to the simulated events.
The agreement within uncertainties between the observed and predicted yields
validates the correction procedure.

For displaced muons, the efficiencies in the reconstruction, identification, and
isolation are studied separately.
The general muon reconstruction efficiency is known to be very high and to agree
between data and simulation~\cite{CMS:2018rym}, and thus only the
displaced-track reconstruction efficiency is evaluated below.
Displaced-muon identification efficiencies without the isolation requirement
applied are validated using $\PBpm\to\PJGy\PKpm\to\MM\PKpm$ events, where the
two muons from the \PJGy~meson decay are studied with a ``tag-and-probe''
method~\cite{CMS:2018rym}.
The measured efficiencies are found to be very close to 100\% and the agreement
between data and simulation is within the statistical uncertainties.
Displaced-muon isolation efficiencies are validated using $\PZ\to\MM$ events,
with a tag-and-probe method as well, and no disagreement between data and
simulation is observed.

\begin{figure}[ht!]
\centering
\includegraphics[width=0.42\textwidth]{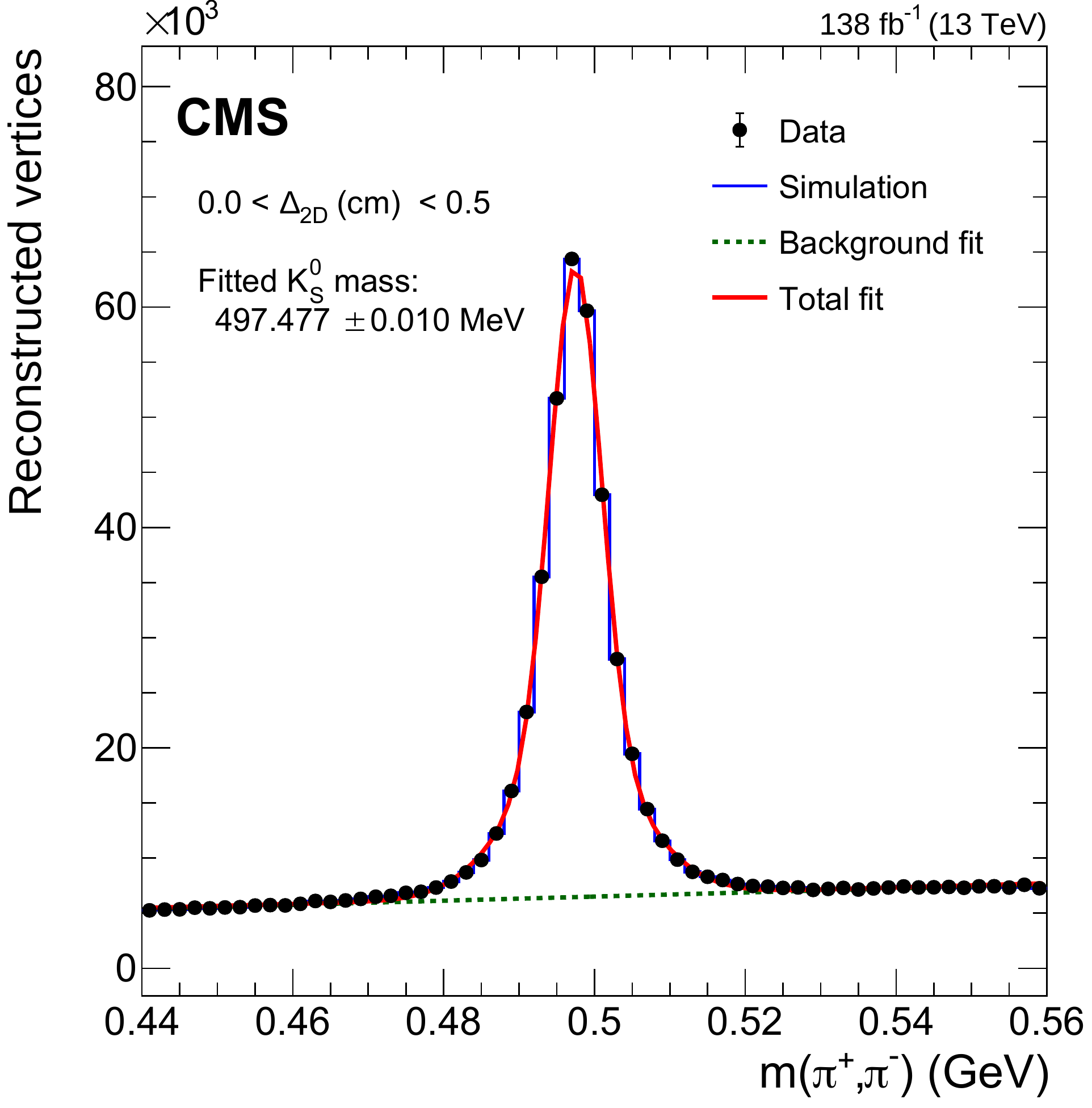}
\hspace{0.04\textwidth}
\includegraphics[width=0.42\textwidth]{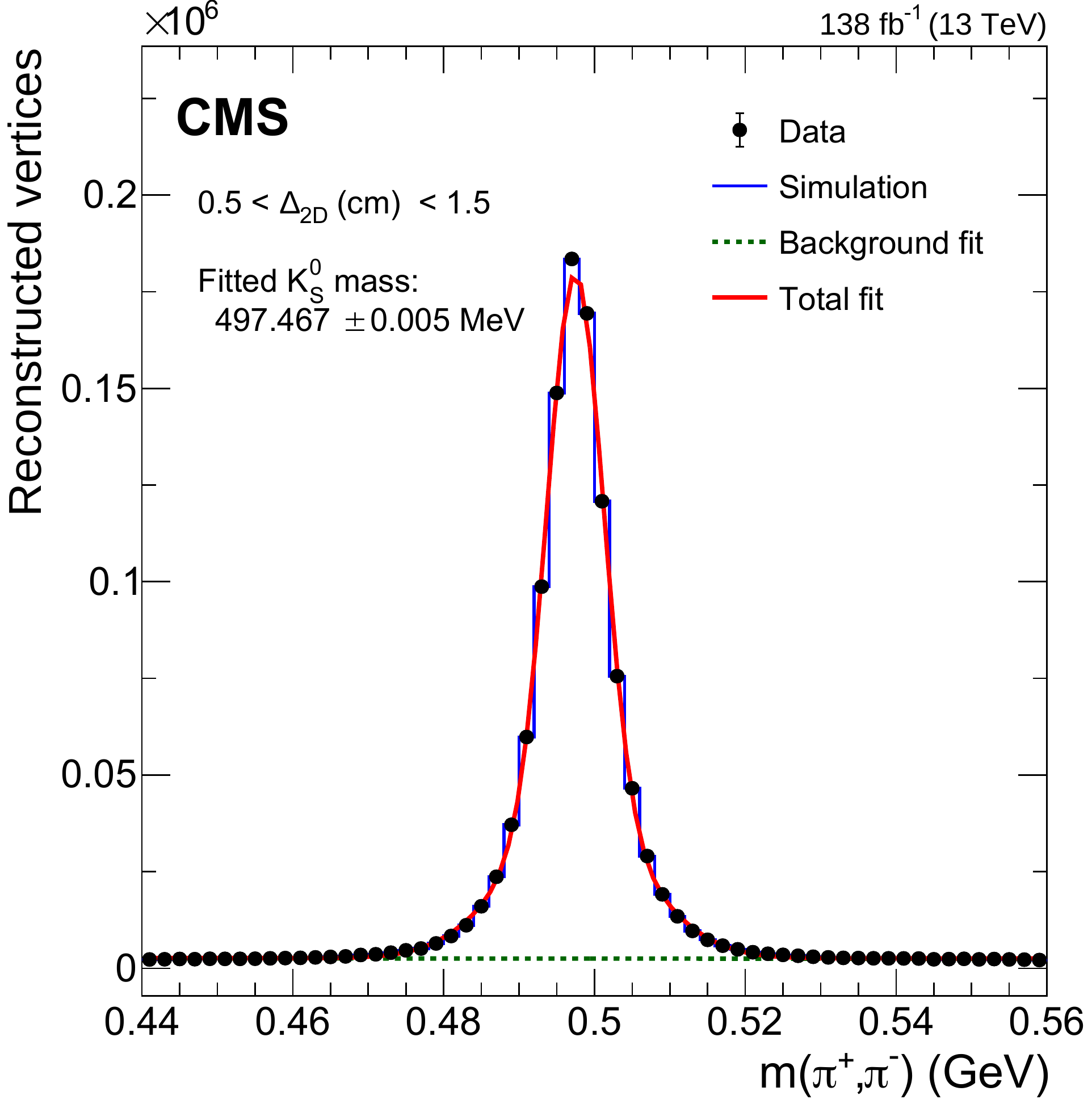} \\
\includegraphics[width=0.42\textwidth]{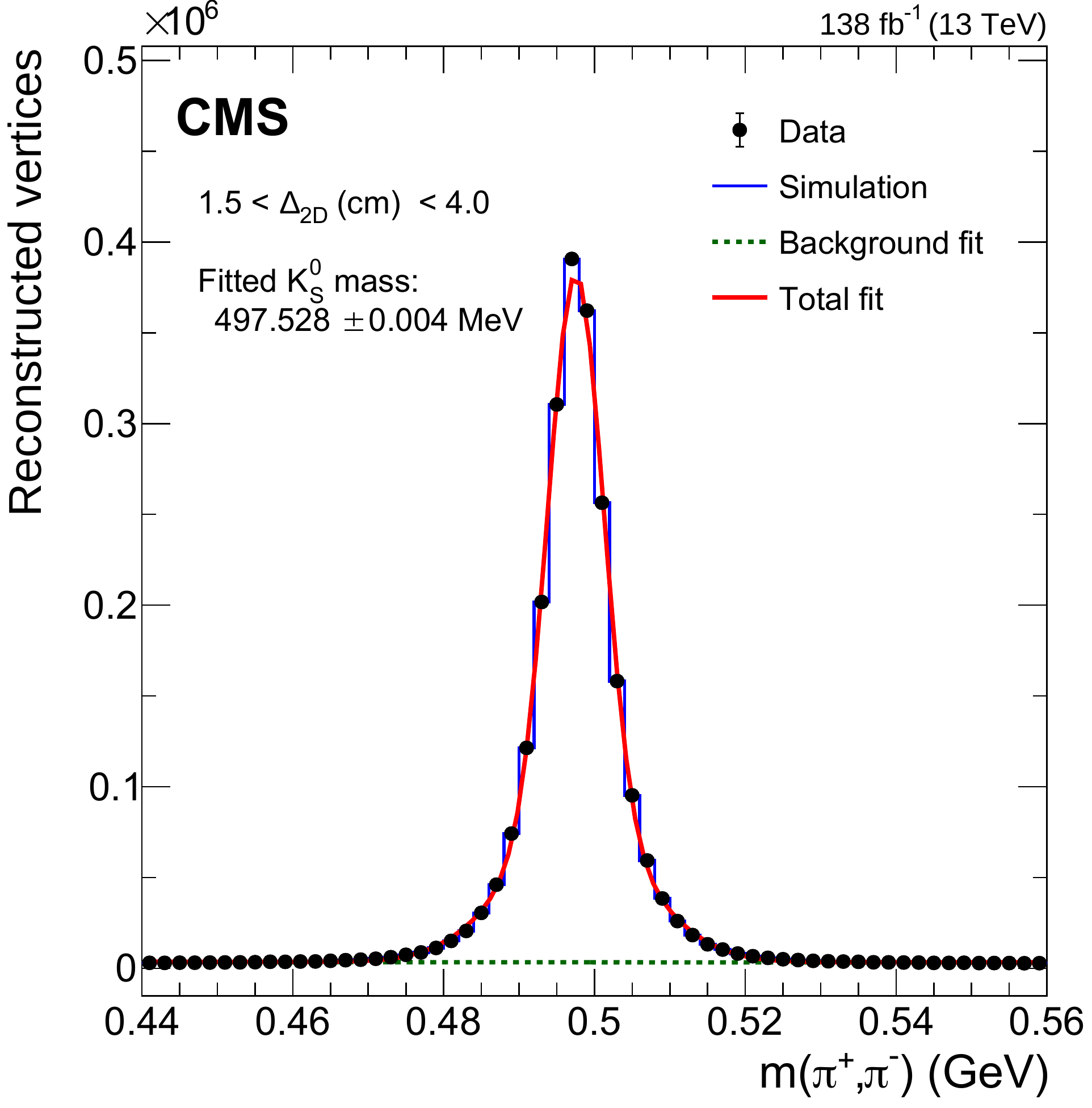}
\hspace{0.04\textwidth}
\includegraphics[width=0.42\textwidth]{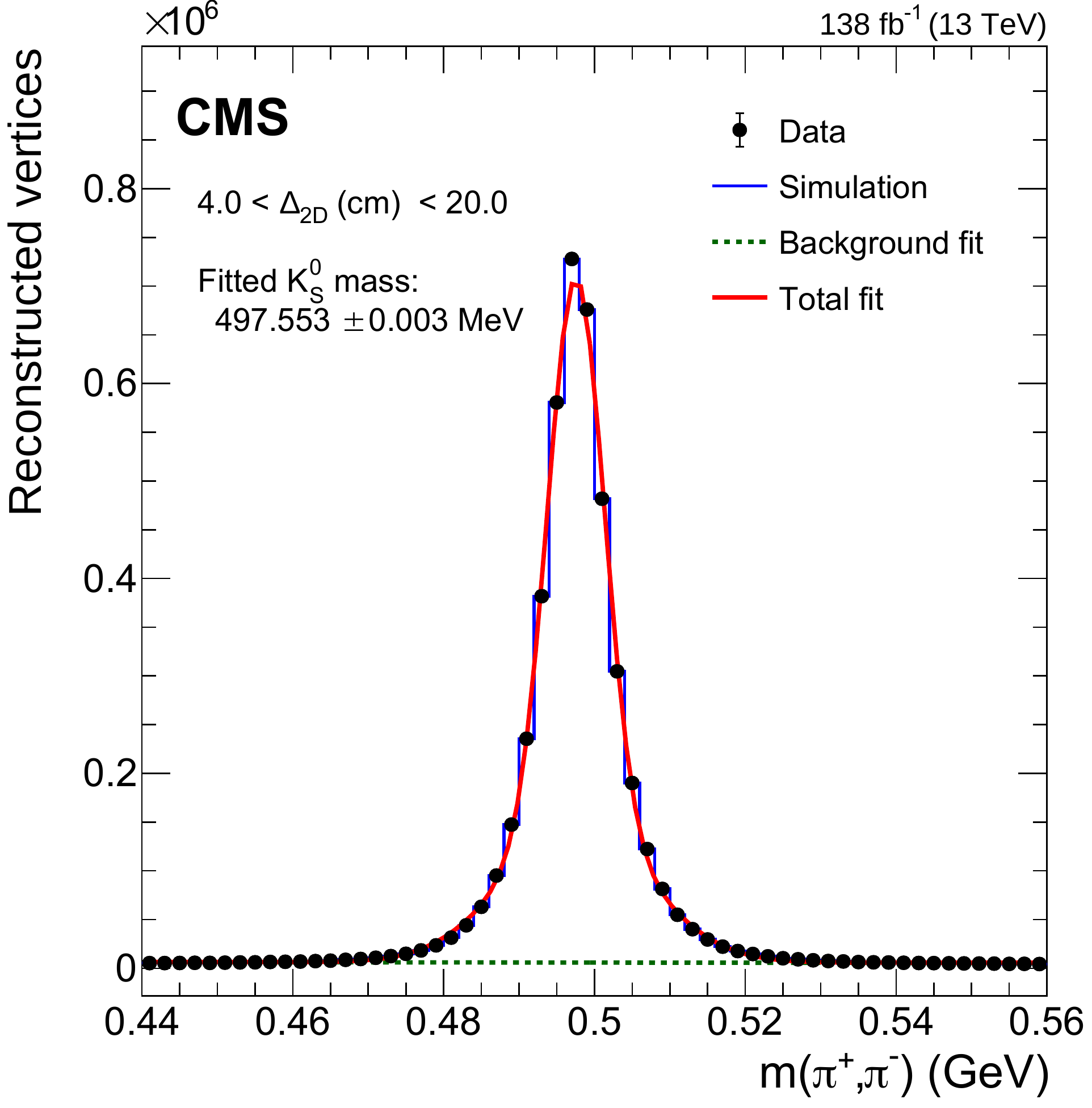} \\
\includegraphics[width=0.42\textwidth]{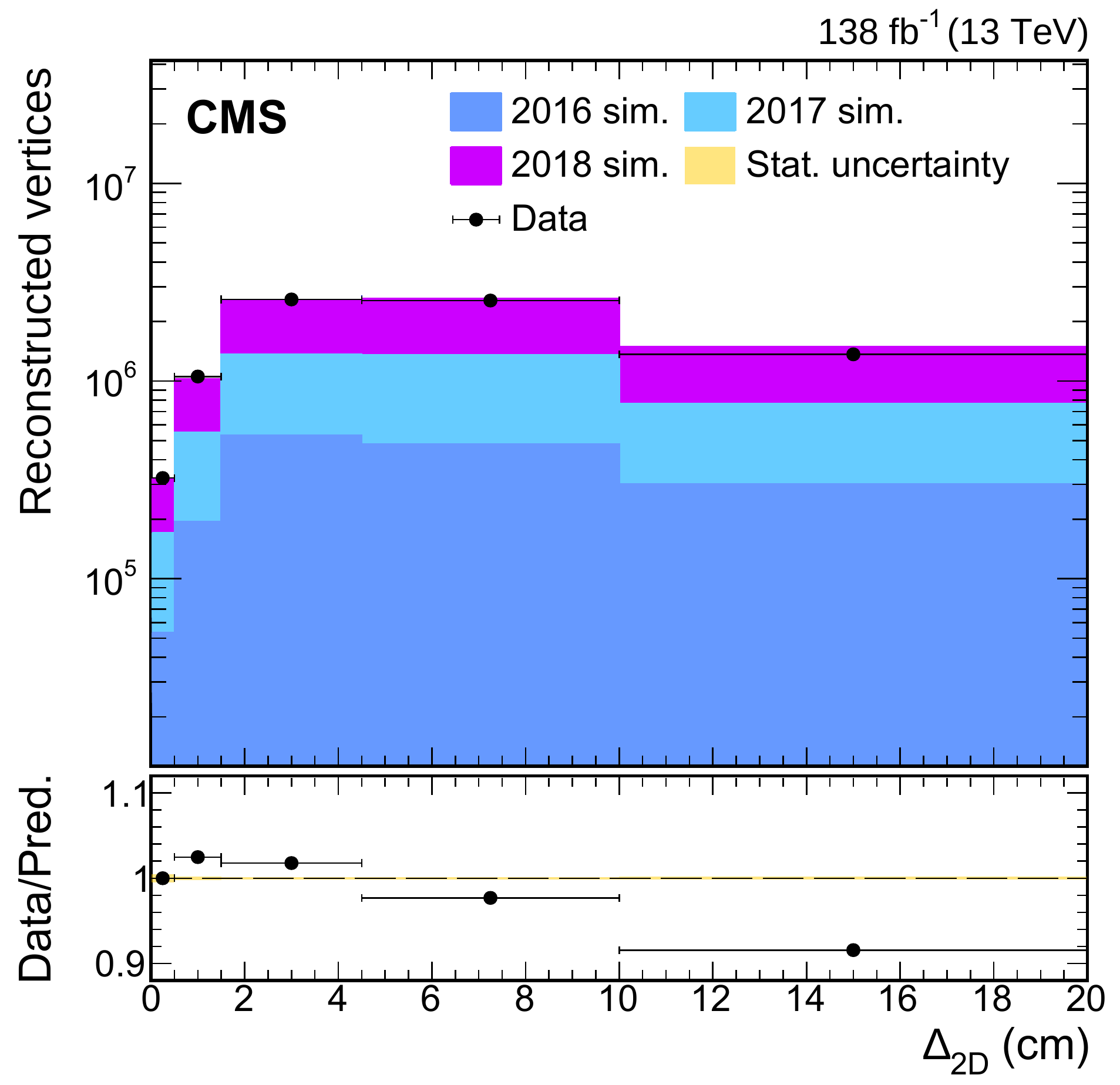}
\caption{\label{fig:validation:kshort}
    The invariant mass distribution of the \PKzS candidates reconstructed using
    the \PiPi tracks for various displacement regions.
    The fitted \PKzS candidate mass in data in each region is also shown.
    The lowermost plot shows the \PKzS candidate yield after subtracting the
    background in data and in simulation, as well as their ratio, as a function
    of radial distance of the \PKzS vertex to the PV.
    The simulated yields are scaled to match the data in the lowest
    displacement bin.
}
\end{figure}

In order to validate the efficiency of the displaced-track reconstruction for
muons in simulation, we study \PKzS mesons decaying into two charged pions.
To allow for the comparison of the efficiency in data and simulated events, this
validation is performed in events with $\PZ\to\MM$ decays.
Events are selected with exactly two muons with $\pt>30$ (25)\GeV and with the
invariant mass of the muon pairs within 10\GeV of the \PZ~boson mass.
Events containing at least one \PQb~jet are vetoed to further suppress the top
quark background.
In these events, \PKzS~mesons are reconstructed using pairs of tracks.
The number of \PKzS~mesons in data and simulation are extracted through a
template fit to the invariant mass distribution of the \PiPi pairs, with the
combinatorial background subtracted.
The results of this validation are presented in bins of radial displacement in
Fig.~\ref{fig:validation:kshort}.

The event yield in simulation is normalized to the corresponding yield in data
at small radial distances (0--0.5\cm), representing an overall normalization
correction for possible MC mismodeling of the \PKzS production in \PZ~boson
events.
The correction scale factors for displaced-track reconstruction efficiencies are
measured in bins of the radial distance from the PV, as well as in bins of the
\pt of the reconstructed \PKzS for each data-taking period.
In Fig.~\ref{fig:validation:kshort}, the comparison of data to simulation for
the full data set and all \pt ranges is shown as a function of radial
displacement.
In events with either two displaced muons or one displaced muon, these results
are used to extract correction factors for MC signal event acceptances in bins
of displacement, as well as the corresponding systematic uncertainties for
displaced-track reconstruction.
The per-event correction factors deviate from unity by less than 10\% for a wide
range of displacements and \pt.
Larger deviations, due to a known inefficiency in the tracker readout electronics~\cite{CMS-DP-2020-035}, are found in the 2016 data set.
These only occur for displacements above 4\cm, and result in correction factors
deviating from unity by up to 30\%.

\section{Systematic uncertainties}
\label{sec:systematic}

Several sources of systematic uncertainties affect the expectations for the
number of events in the different signal region bins.
The main uncertainties are described below.

The MC generation of HNL signal events is done at LO precision, resulting in
large theoretical uncertainties in the cross section of up to 15\% when
accounting for QCD scale variations and the choice of PDF.
To reduce the uncertainty, we apply the $K$-factor evaluated for
$\PW\to\Pell\PGn$ production to HNL signal events.
The predicted cross section for the signal process is therefore estimated with
NNLO precision, and we obtain a total uncertainty of 3.9\% for the combined
effects of QCD scale variations and PDF choice.

The integrated luminosity is measured with a precision between 1.2 and 2.5\%
separately for the three years of data taking~\cite{CMS:2021xjt, CMS:2018elu,
CMS:2019jhq}.
Because of correlations between some systematic sources in the luminosity
calibration, the integrated luminosity of the combined data set is measured with
a precision of 1.6\%.
The uncertainty in the integrated luminosity affects the yields of the signal
and of the simulation-based background estimations.

The uncertainty in the modeling of pileup interactions is evaluated by varying
the total \pp inelastic cross section in simulation by
$\pm$5\%~\cite{CMS:2018mlc}.
This results in an uncertainty in the signal yields of 1--3\%, depending on the
signal region bin.

To estimate the uncertainties associated with the prompt-lepton identification
and isolation efficiency scale factors, the total yields in each signal region
bin are recomputed with the scale factors varied up and down by the
tag-and-probe fit uncertainties, treating all bins as correlated.
Uncertainties of 2--4 (1--3)\% in the signal yields are found in events with a
prompt electron (muon).

Possible discrepancies in the reconstruction, identification, and isolation
efficiencies of displaced leptons between data and simulation are studied using
different strategies for electrons and muons, and are used to assess an
uncertainty in the MC modeling of the signal.
In the different signal region bins, the correction factors applied to the
simulated events are varied within their uncertainties, and the resulting
uncertainties in the final yields for the displacement bin $1.5<\Deltwod<4\cm$
are found to be about 7, 4, and 5\% for events with $(\ltwo,\lthree)=\Cee$,
\Cem, and \Cmm, respectively.

Trigger scale factors are used to correct the simulated event yields.
Uncertainties extracted from the tag-and-probe fits are used to assess
systematic uncertainties.
These are found to be less than 1\% for muon triggers and about 1\% for electron
triggers.

The uncertainties in the jet energy scale, as well as the jet energy resolution
in simulation, affect directly the veto on \PQb~jets with $\pt>25\GeV$.
Their impact is estimated by scaling independently the energy of jets up and
down by one standard deviation.
These systematic effects lead to negligible uncertainties in the analysis.

The efficiency of the \PQb~jet veto is corrected for the difference between data
and simulation.
The uncertainty associated with this correction is used to estimate an
uncertainty in the final event yields.
The resulting uncertainty is found to be 1--5\%.

The contributions from background leptons are estimated using control samples in
data, as described in Section~\ref{sec:backgrounds}.
The general features of the method were validated in simulation as a function of
the search variables.
Based on the closure test results in the background validation region in data
shown in Fig.~\ref{fig:backgrounds:controlregions}, we apply a 30\% systematic
uncertainty in all channels, and include an additional uncorrelated 50\%
systematic uncertainty in the highest displacement and the $\mtwol>4\GeV$ signal
region bins, to cover the observed deviations.
These two uncertainty sources are considered to be correlated over all
data-taking periods and lepton channels.
Additionally, we include three uncertainties to account for the different
background sources and the differences in statistical power in the \ttlDB
measurement regions for the different lepton channels.
These uncertainties are in the range  10--17, 10--20, and 25--40\%, in the \Cmm,
\Cem, and \Cee channels, respectively, and are taken to be uncorrelated among
the channels and years.

The statistical uncertainty in the predicted background lepton contributions in
the majority of signal region bins is largely driven by the limited number of
events in the application regions.
We adopt a Gamma distribution, reflecting the Poisson statistics of the observed
event yields in an application region bin with a mean corresponding to the
background event yields in the signal region bin divided by
\ttl~\cite{Linnemann:2003vw, Cousins:2007yta}, to model the statistical
uncertainty in the predicted background yields.
For each bin, an independent nuisance parameter is assigned accounting for the
68\% confidence level (\CL) interval of the Gamma distribution with the average
measured value of \ttl in that bin.
If no events are observed in a given bin in the application region, the Gamma
distribution is evaluated with the maximum measured value of \ttl.

For the processes estimated from simulation, the available number of events in
the simulated samples limits the precision of the modeling.
The statistical uncertainty in the event yield in each search bin is taken as a
systematic uncertainty in the shape of the distributions used in the analysis.

\section{Results}

\begin{figure}[bp!]
\centering
\includegraphics[width=.48\textwidth]{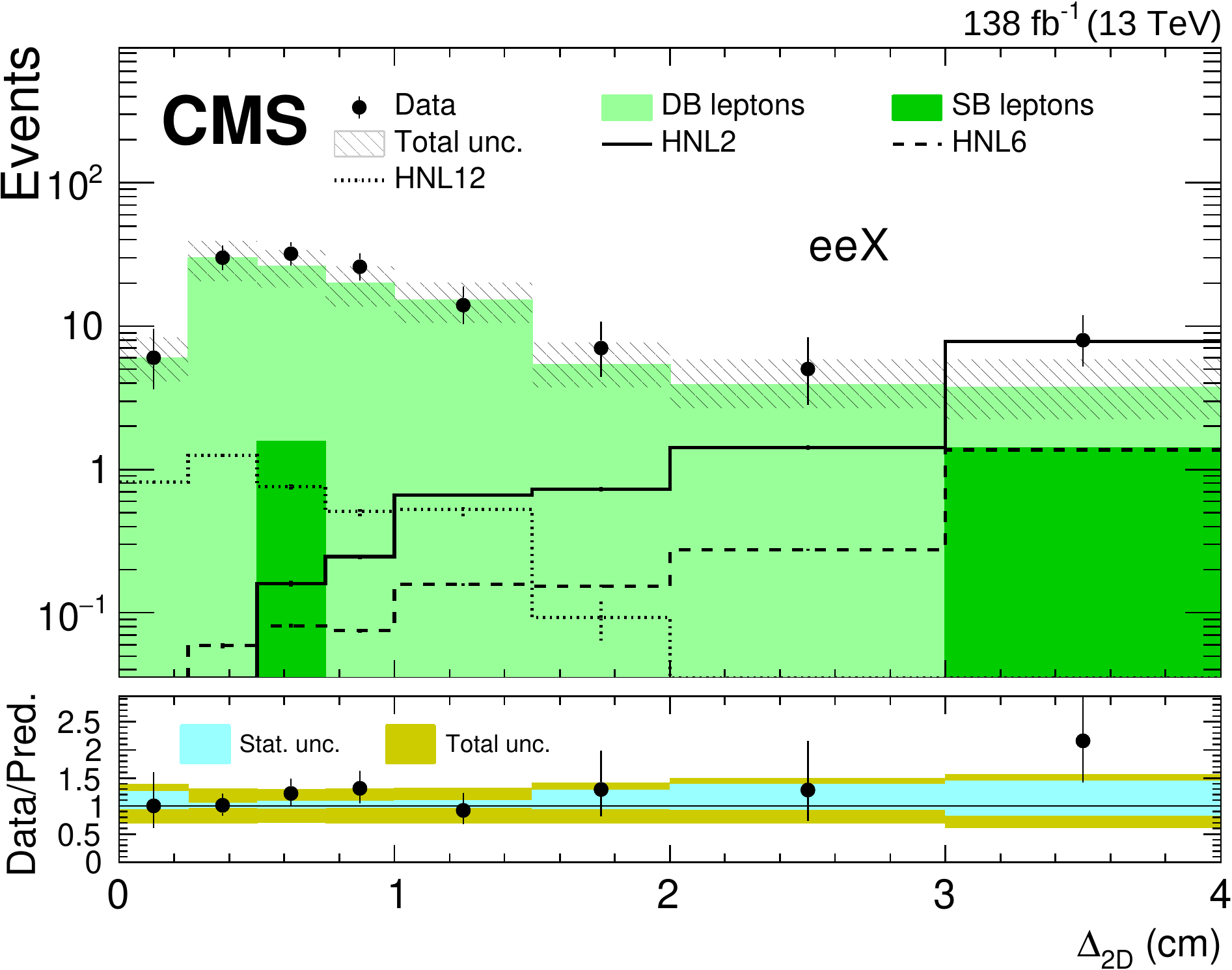}
\hfill
\includegraphics[width=.48\textwidth]{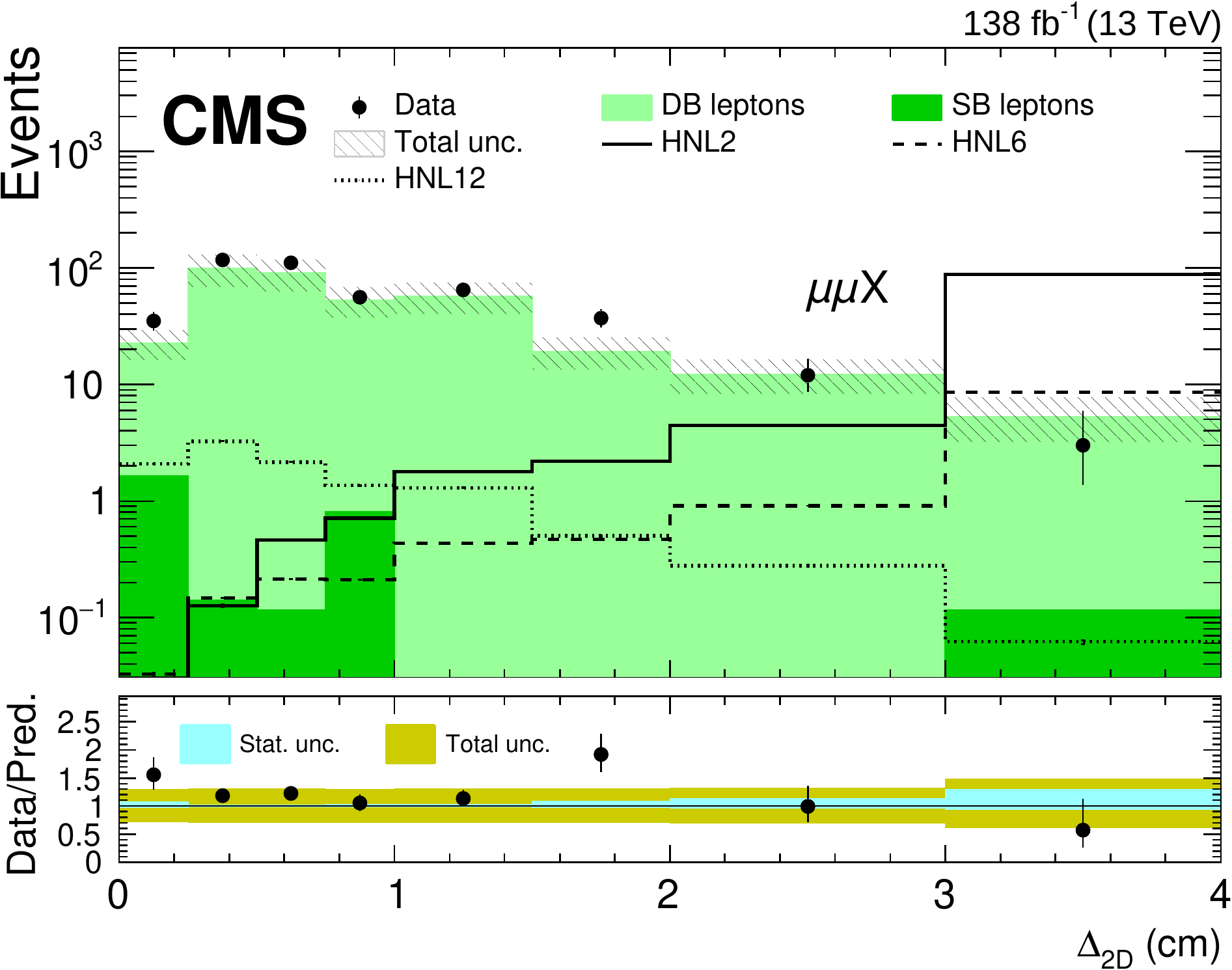}
\caption{\label{fig:results:displacement}
    Comparison between the number of observed events in data and the background
    predictions (shaded histograms, stacked) for the \Deltwod variable in \eex
    (left) and \mmx (right) final states.
    Events in the overflow bin are included in the last bin.
    The hashed band indicates the total systematic and statistical uncertainty
    in the background prediction.
    The lower panels indicate the ratio between data and prediction.
    Predictions for signal events are shown for several benchmark hypotheses for
    Majorana HNL production: \mHNLtwo and \mixZpEtmF (HNL2), \mHNLsix and
    \mixOpTtmS (HNL6), \mHNLtwelve and \mixOpZtmS (HNL12).
}
\end{figure}

\begin{figure}[tp!]
\centering
\includegraphics[width=.48\textwidth]{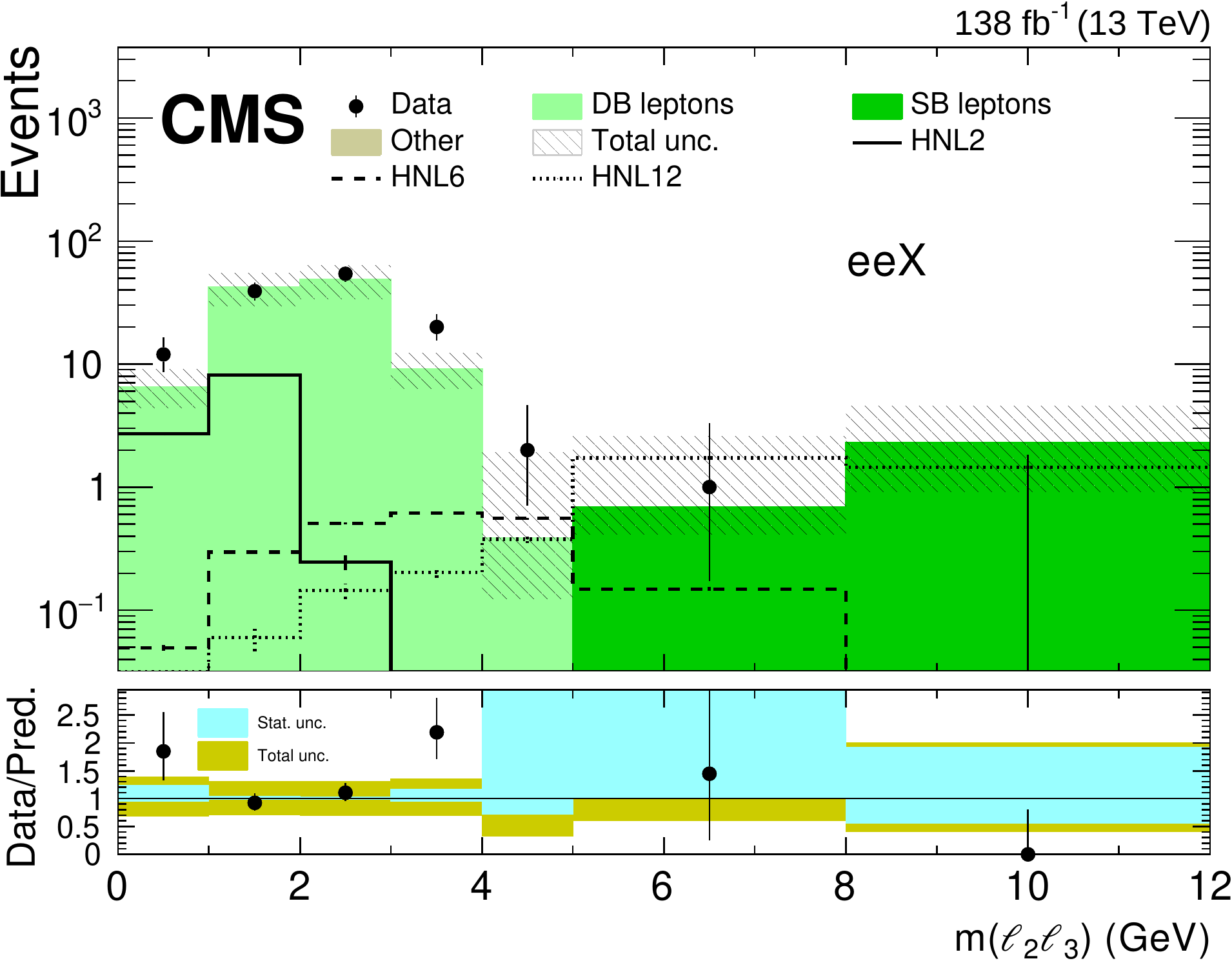}
\hfill
\includegraphics[width=.48\textwidth]{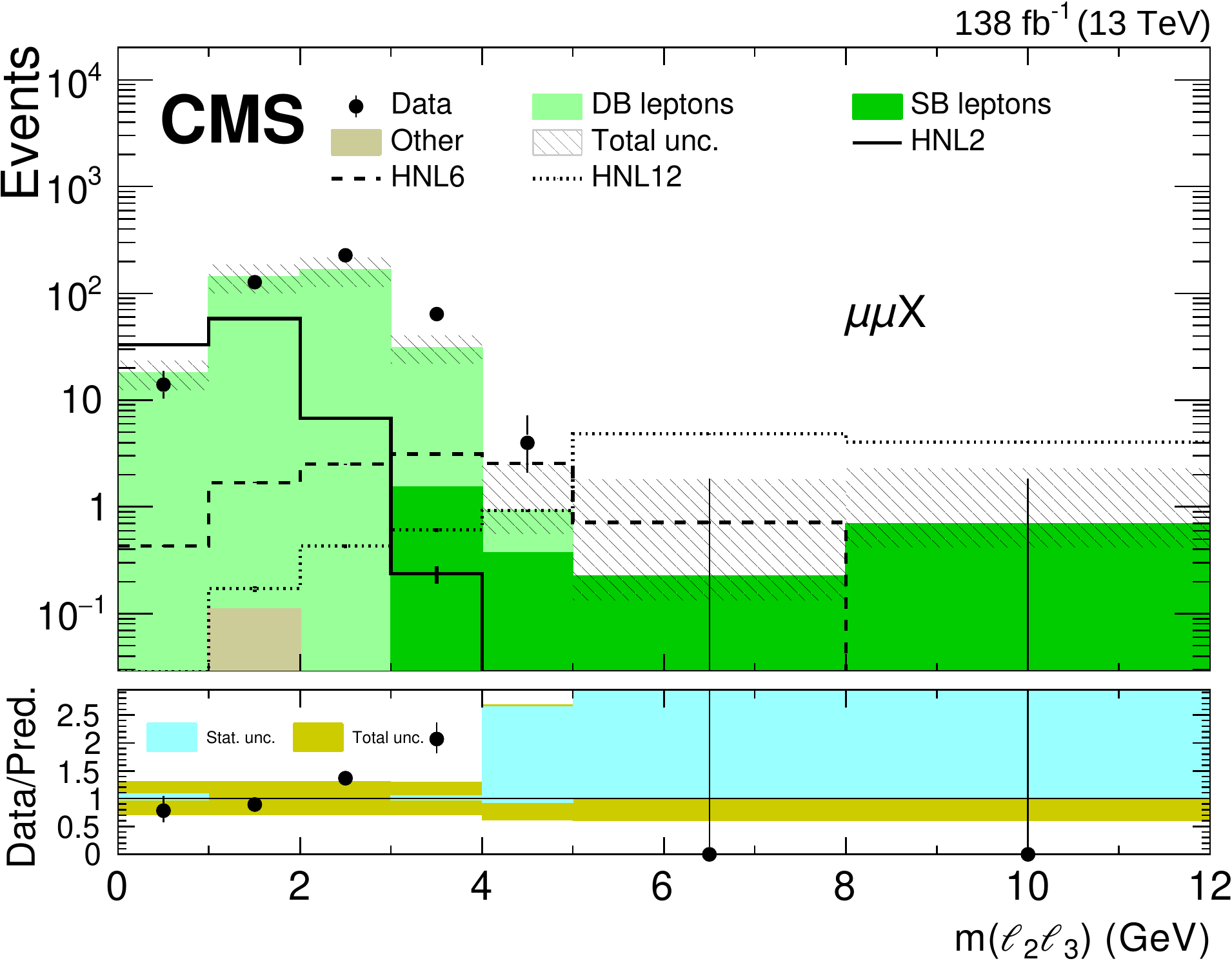}
\caption{\label{fig:results:massl2l3}
    Comparison between the number of observed events in data and the background
    predictions (shaded histograms, stacked) for the \mtwol variable in \eex
    (left) and \mmx (right) final states.
    Events in the overflow bin are included in the last bin.
    The hashed band indicates the total systematic and statistical uncertainty
    in the background prediction.
    The lower panels indicate the ratio between data and prediction, and missing
    points indicate that the ratio lies outside the axis range.
    Predictions for signal events are shown for several benchmark hypotheses for
    Majorana HNL production: \mHNLtwo and \mixZpEtmF (HNL2), \mHNLsix and
    \mixOpTtmS (HNL6), \mHNLtwelve and \mixOpZtmS (HNL12).
    Small contributions from background processes that are estimated from
    simulation are collectively referred to as ``Other''.
}
\end{figure}

\begin{figure}[tp!]
\centering
\includegraphics[width=.85\textwidth]{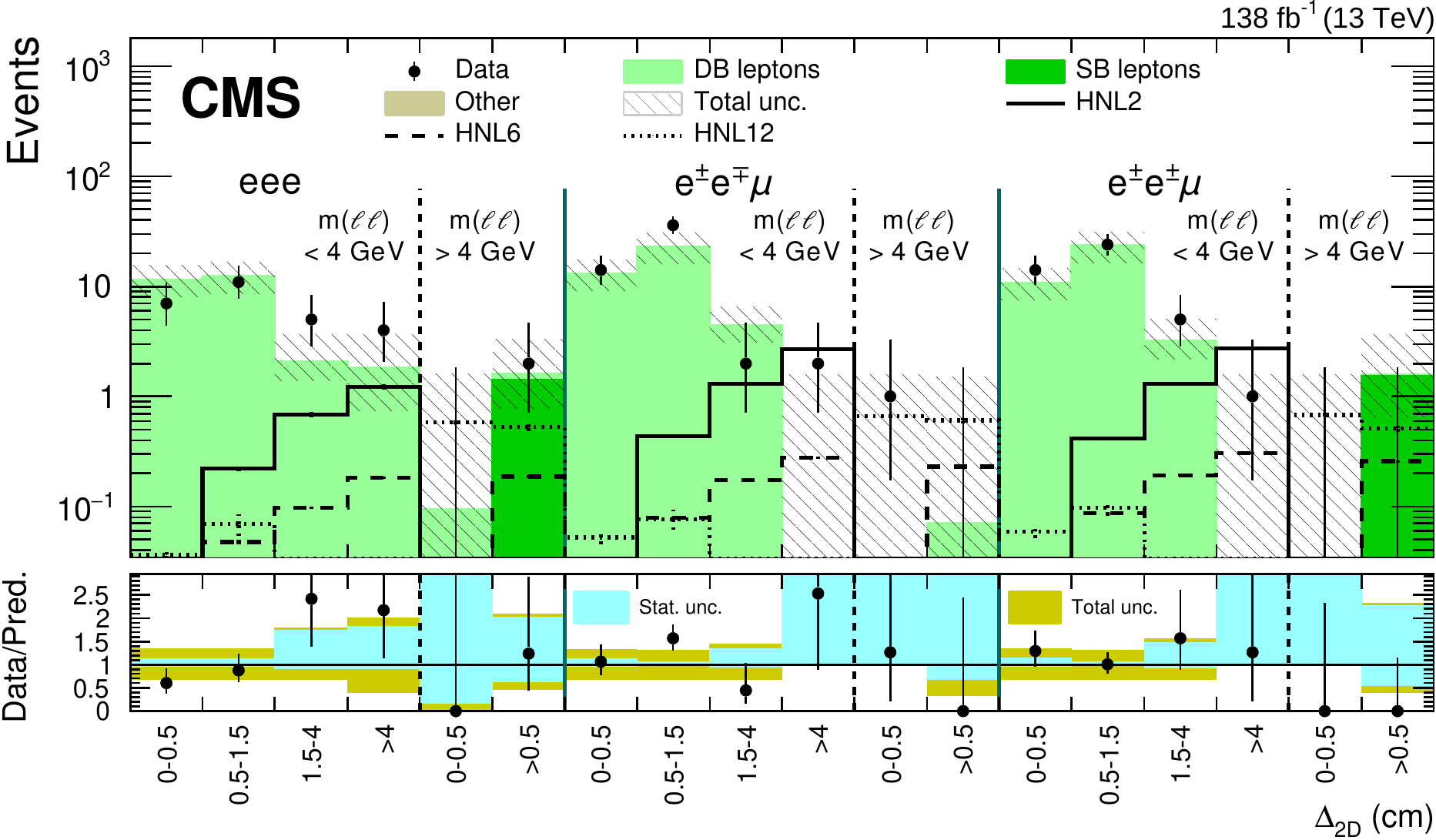} \\[1ex]
\includegraphics[width=.85\textwidth]{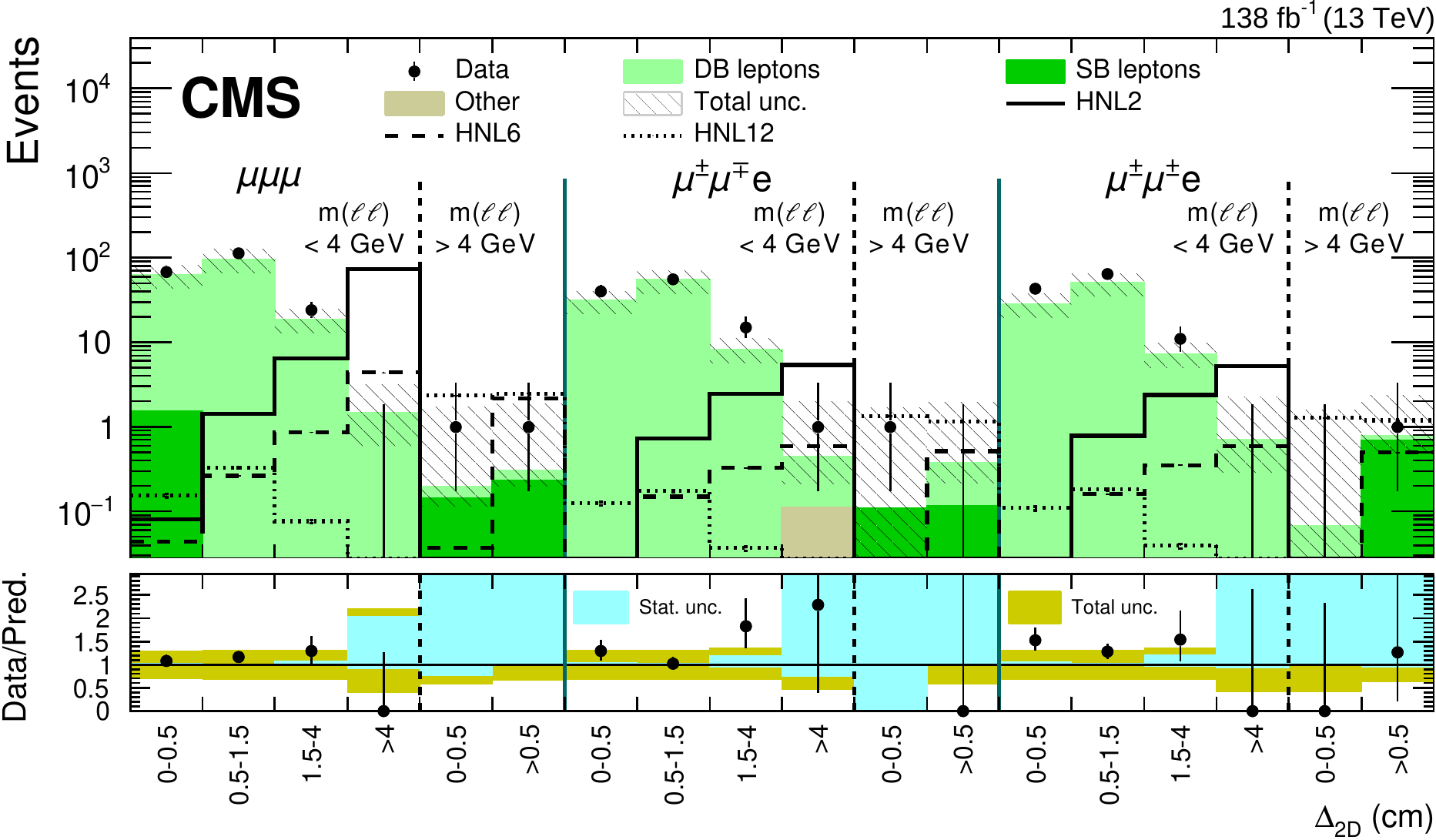}
\caption{\label{fig:results:searchregions}
    Comparison between the number of observed events in data and the background
    predictions (shaded histograms, stacked) in the signal region for \eex
    (upper) and \mmx (lower) final states.
    The hashed band indicates the total systematic and statistical uncertainty
    in the background prediction.
    The lower panels indicate the ratio between data and prediction, and missing
    points indicate that the ratio lies outside the axis range.
    Predictions for signal events are shown for several benchmark hypotheses
    for Majorana HNL production: \mHNLtwo and \mixZpEtmF (HNL2), \mHNLsix and
    \mixOpTtmS (HNL6), \mHNLtwelve and \mixOpZtmS (HNL12).
    The uncertainty band assigned to the background prediction includes
    statistical and systematic contributions.
    Small contributions from background processes that are estimated from
    simulation are collectively referred to as ``Other''.
}
\end{figure}

The two main variables used to categorize the selected events are \Deltwod and
\mtwol.
The comparison between the number of observed events in data and the background
predictions is shown in Figs.~\ref{fig:results:displacement}
and~\ref{fig:results:massl2l3}.
The expected and observed yields in each of the signal region bins are shown in
Fig.~\ref{fig:results:searchregions}, and listed in
Tables~\ref{tab:results:yieldsEE} and~\ref{tab:results:yieldsMM}.
In total, 128 (436) data events are observed in the \eex (\mmx) final states.
The significantly larger event yield in the \mmx final states is a consequence
of the better resolution and higher efficiency in the reconstruction of muons
compared to electrons with the CMS experiment, which correspondingly means that
also the signal efficiency is higher in the \mmx final states.

\begin{table}[!tp]
\renewcommand{\arraystretch}{1.2}
\centering
\topcaption{\label{tab:results:yieldsEE}
    Number of predicted and observed events in the \eex final states.
    The quoted uncertainties include statistical and systematic components.
}
\begin{tabular}{cl@{\cmsColSkip}cccc@{\cmsColSkip}cc}
    \multicolumn{2}{r@{\cmsColSkip}}{\mtwol (\GeVns{})} & \multicolumn{4}{c}{$<$4} & \multicolumn{2}{c}{$>$4} \\
    \multicolumn{2}{r@{\cmsColSkip}}{\Deltwod (cm)} & $<$0.5 & 0.5--1.5 & 1.5--4 & $>$4 & $<$0.5 & $>$0.5 \\
    \hline
    \multirow{2}{*}{\EEE} & Data & 7 & 11 & 5 & 4 & 0 & 2 \\
    & Background & $11.6_{-3.6}^{+3.9}$ & $12.5_{-3.9}^{+4.2}$ & $2.1_{-0.7}^{+1.7}$ & $1.8_{-1.1}^{+1.9}$ & $0.1_{-0.1}^{+1.5}$ & $1.6_{-0.8}^{+1.7}$ \\[\cmsTabSkip]
    \multirow{2}{*}{\EEMos} & Data & 14 & 36 & 2 & 2 & 1 & 0 \\
    & Background & $13.1_{-4.0}^{+4.3}$ & $23.0_{-7.0}^{+7.2}$ & $4.5_{-1.4}^{+2.1}$ & $0_{-0.0}^{+1.6}$ & $0_{-0.0}^{+1.6}$ & $0.1_{-0.0}^{+1.5}$ \\[\cmsTabSkip]
    \multirow{2}{*}{\EEMss} & Data & 14 & 24 & 5 & 1 & 0 & 0 \\
    & Background & $10.8_{-3.3}^{+3.6}$ & $23.8_{-7.3}^{+7.4}$ & $3.2_{-1.0}^{+1.8}$ & $0_{-0.0}^{+1.6}$ & $0_{-0.0}^{+1.6}$ & $1.6_{-0.9}^{+2.1}$ \\
\end{tabular}
\end{table}

\begin{table}[!tp]
\renewcommand{\arraystretch}{1.2}
\centering
\topcaption{\label{tab:results:yieldsMM}
    Number of predicted and observed events in the \mmx final states.
    The quoted uncertainties include statistical and systematic components.
}
\begin{tabular}{cl@{\cmsColSkip}cccc@{\cmsColSkip}cc}
    \multicolumn{2}{r@{\cmsColSkip}}{\mtwol (\GeVns{})} & \multicolumn{4}{c}{$<$4} & \multicolumn{2}{c}{$>$4} \\
    \multicolumn{2}{r@{\cmsColSkip}}{\Deltwod (cm)} & $<$0.5 & 0.5--1.5 & 1.5--4 & $>$4 & $<$0.5 & $>$0.5 \\
    \hline
    \multirow{2}{*}{\MMM} & Data & 67 & 112 & 24 & 0 & 1 & 1 \\
    & Background & $62_{-18}^{+19}$ & $96_{-29}^{+29}$ & $18.6_{-5.6}^{+5.9}$ & $1.4_{-0.9}^{+1.7}$ & $0.2_{-0.1}^{+1.5}$ & $0.3_{-0.1}^{+1.6}$ \\[\cmsTabSkip]
    \multirow{2}{*}{\MMEos} & Data & 40 & 55 & 15 & 1 & 1 & 0 \\
    & Background & $30.8_{-9.3}^{+9.5}$ & $54_{-16}^{+16}$ & $8.2_{-2.6}^{+3.0}$ & $0.4_{-0.2}^{+1.5}$ & $0.1_{-0.1}^{+1.6}$ & $0.4_{-0.2}^{+1.6}$ \\[\cmsTabSkip]
    \multirow{2}{*}{\MMEss} & Data & 43 & 64 & 11 & 0 & 0 & 1 \\
    & Background & $28.2_{-8.5}^{+8.7}$ & $50_{-15}^{+15}$ & $7.2_{-2.2}^{+2.7}$ & $0.7_{-0.4}^{+1.6}$ & $0.1_{-0.0}^{+1.6}$ & $0.8_{-0.3}^{+1.5}$ \\
\end{tabular}
\end{table}

The number of observed events in data is in good agreement with the SM
background expectations within the statistical and systematic uncertainties and
no significant excess is found, for all final states and in all signal region
bins.
The numbers of expected HNL signal events for a selection of representative
signal scenarios are listed in Tables~\ref{tab:results:signalEE}
and~\ref{tab:results:signalMM}.
As discussed in Section~\ref{sec:introduction}, only scenarios where the HNL
mixes with a single neutrino flavor are considered in this search, \ie, these
are scenarios where only one of \mixpare and \mixparm is nonzero.
The \eex~channels (\EEE, \EEMos, and \EEMss) are sensitive to the \mixpare
mixing parameter, while the \mmx~channels (\MMM, \MMEos, and \MMEss) are
sensitive to \mixparm.

\begin{table}[!tp]
\renewcommand{\arraystretch}{1.2}
\centering
\topcaption{\label{tab:results:signalEE}
    Number of predicted signal events in the \eex final states.
    The results are presented for several benchmark signal hypotheses for
    Majorana HNL production: \mHNLtwo and \mixZpEtmF (HNL2), \mHNLsix and
    \mixOpTtmS (HNL6), \mHNLtwelve and \mixOpZtmS (HNL12).
    The quoted uncertainties include statistical and systematic components.
}
\cmsTable{\begin{tabular}{cl@{\cmsColSkip}cccc@{\cmsColSkip}cc}
    \multicolumn{2}{r@{\cmsColSkip}}{\mtwol (\GeVns{})} & \multicolumn{4}{c}{$<$4} & \multicolumn{2}{c}{$>$4} \\
    \multicolumn{2}{r@{\cmsColSkip}}{\Deltwod (cm)} & $<$0.5 & 0.5--1.5 & 1.5--4 & $>$4 & $<$0.5 & $>$0.5 \\
    \hline
    & \EEE & $0.000\pm0.002$ & $0.22\pm0.02$ & $0.69\pm0.06$ & $1.2\pm0.1$ & 0 & 0 \\
    HNL2 & \EEMos & $0.013\pm0.002$ & $0.43\pm0.03$ & $1.3\pm0.1$ & $2.7\pm0.3$ & 0 & 0 \\
    & \EEMss & $0.013\pm0.002$ & $0.41\pm0.03$ & $1.3\pm0.1$ & $2.7\pm0.3$ & 0 & 0 \\[\cmsTabSkip]
    & \EEE & 0 & $0.048\pm0.004$ & $0.096\pm0.008$ & $0.18\pm0.01$ & $0.000\pm0.001$ & $0.19\pm0.01$ \\
    HNL6 & \EEMos & $0.014\pm0.001$ & $0.078\pm0.005$ & $0.17\pm0.01$ & $0.28\pm0.02$ & $0.000\pm0.001$ & $0.23\pm0.02$ \\
    & \EEMss & $0.017\pm0.002$ & $0.086\pm0.006$ & $0.19\pm0.01$ & $0.30\pm0.03$ & $0.011\pm0.001$ & $0.26\pm0.02$ \\[\cmsTabSkip]
    & \EEE & $0.036\pm0.007$ & $0.07\pm0.02$ & $0.00\pm0.01$ & 0 & $0.58\pm0.04$ & $0.53\pm0.06$ \\
    HNL12 & \EEMos & $0.05\pm0.01$ & $0.08\pm0.02$ & $0.00\pm0.01$ & 0 & $0.66\pm0.05$ & $0.60\pm0.06$ \\
    & \EEMss & $0.06\pm0.01$ & $0.10\pm0.02$ & $0.000\pm0.001$ & 0 & $0.68\pm0.05$ & $0.51\pm0.05$ \\
\end{tabular}}
\end{table}

\begin{table}[!tp]
\renewcommand{\arraystretch}{1.2}
\centering
\topcaption{\label{tab:results:signalMM}
    Number of predicted signal events in the \mmx final states.
    The results are presented for several benchmark signal hypotheses for
    Majorana HNL production: \mHNLtwo and \mixZpEtmF (HNL2), \mHNLsix and
    \mixOpTtmS (HNL6), \mHNLtwelve and \mixOpZtmS (HNL12).
    The quoted uncertainties include statistical and systematic components.
}
\cmsTable{\begin{tabular}{cl@{\cmsColSkip}cccc@{\cmsColSkip}cc}
    \multicolumn{2}{r@{\cmsColSkip}}{\mtwol (\GeVns{})} & \multicolumn{4}{c}{$<$4} & \multicolumn{2}{c}{$>$4} \\
    \multicolumn{2}{r@{\cmsColSkip}}{\Deltwod (cm)} & $<$0.5 & 0.5--1.5 & 1.5--4 & $>$4 & $<$0.5 & $>$0.5 \\
    \hline
    & \MMM & $0.081\pm0.008$ & $1.40\pm0.09$ & $6.5\pm0.4$ & $73\pm7$ & 0 & 0 \\
    HNL2 & \MMEos & $0.024\pm0.003$ & $0.73\pm0.05$ & $2.5\pm0.2$ & $5.4\pm0.5$ & 0 & $0.000\pm0.004$ \\
    & \MMEss & $0.024\pm0.003$ & $0.78\pm0.05$ & $2.4\pm0.2$ & $5.2\pm0.5$ & 0 & 0 \\[\cmsTabSkip]
    & \MMM & $0.044\pm0.004$ & $0.27\pm0.02$ & $0.87\pm0.05$ & $4.4\pm0.5$ & $0.037\pm0.003$ & $2.2\pm0.2$ \\
    HNL6 & \MMEos & $0.027\pm0.002$ & $0.15\pm0.01$ & $0.33\pm0.02$ & $0.58\pm0.05$ & $0.024\pm0.002$ & $0.51\pm0.03$ \\
    & \MMEss & $0.03\pm0.00$ & $0.16\pm0.01$ & $0.35\pm0.02$ & $0.58\pm0.04$ & $0.02\pm0.00$ & $0.50\pm0.03$ \\[\cmsTabSkip]
    & \MMM & $0.15\pm0.02$ & $0.33\pm0.02$ & $0.08\pm0.01$ & 0 & $2.3\pm0.2$ & $2.5\pm0.1$ \\
    HNL12 & \MMEos & $0.13\pm0.01$ & $0.18\pm0.02$ & $0.037\pm0.005$ & 0 & $1.30\pm0.08$ & $1.10\pm0.07$ \\
    & \MMEss & $0.11\pm0.01$ & $0.18\pm0.02$ & $0.04\pm0.01$ & 0 & $1.30\pm0.08$ & $1.20\pm0.08$ \\
\end{tabular}}
\end{table}

To derive exclusion limits on HNL signal scenarios, the modified frequentist
\CLs approach~\cite{Junk:1999kv, Read:2002hq, ATLAS:2011tau, Cowan:2010js} is
applied.
For each signal scenario, a binned likelihood function
$\likeli(\sigstr,\nuisan)$ is constructed from the product of Poisson
probabilities to obtain the observed yields given the HNL signal scenario scaled
with a signal strength~\sigstr and the background estimates, using the bins
defined in Table~\ref{tab:selection:binning}.
Additional terms are included to account for all sources of systematic
uncertainty, and \nuisan denotes the full set of corresponding nuisance
parameters.
Based on the profile likelihood method, the LHC test statistic
$q(\sigstr)=-2\ln\likeli(\sigstr,\nuisanr)/\likeli(\sigstrgl,\nuisangl)$ is
used, where \nuisanr is the maximum likelihood estimator of \nuisan for a given
\sigstr, and \sigstrgl and \nuisangl are the estimators corresponding to the
global maximum of \likeli~\cite{ATLAS:2011tau}.
The \CLs value is calculated in the asymptotic
approximation~\cite{Cowan:2010js}, and we exclude the signal scenario if the
signal strength value of $\sigstr=1$ is incompatible with the observed data at
95\% \CL.

\begin{figure}[!ht]
\centering
\includegraphics[width=.48\textwidth]{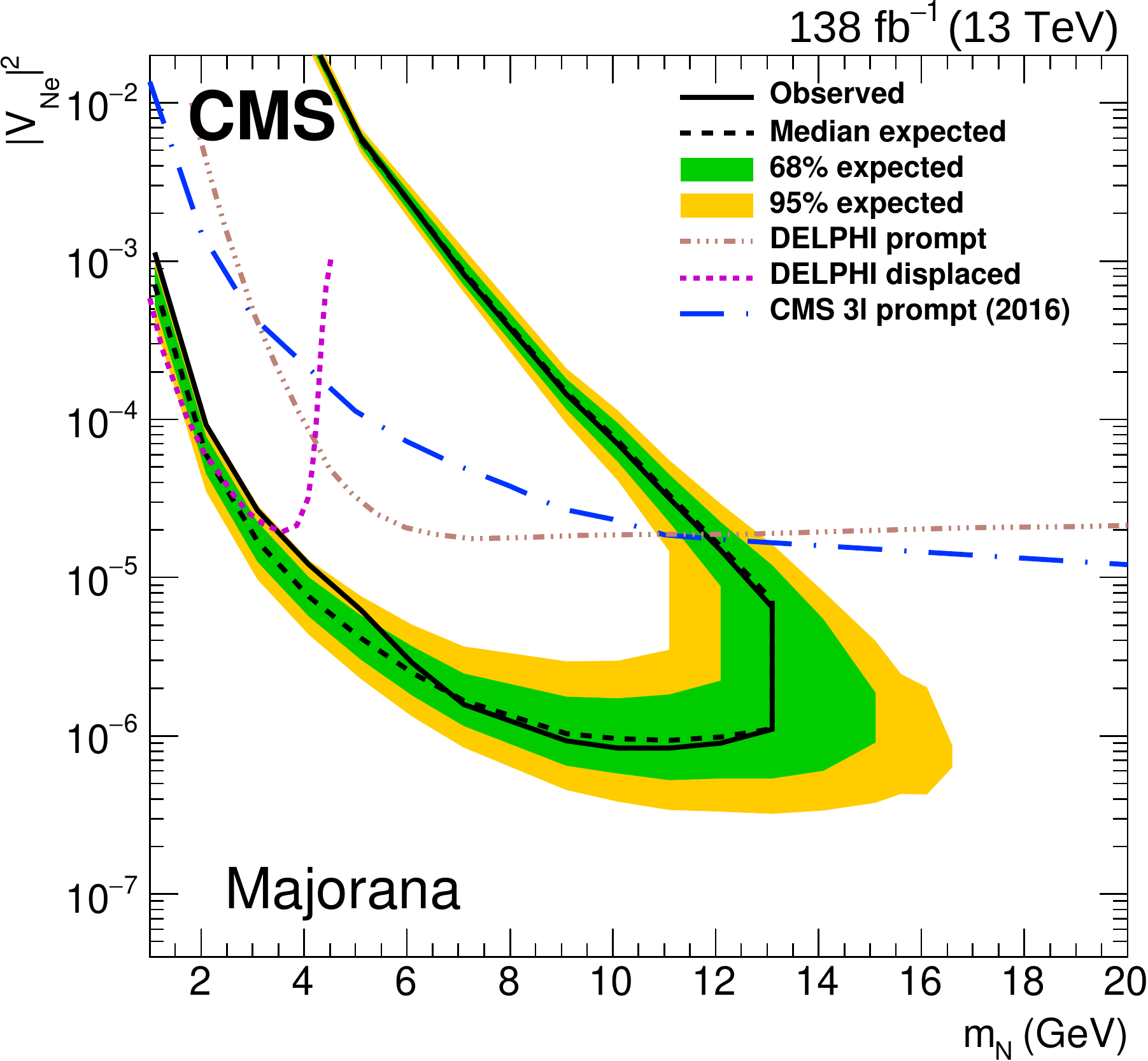}
\hfill
\includegraphics[width=.48\textwidth]{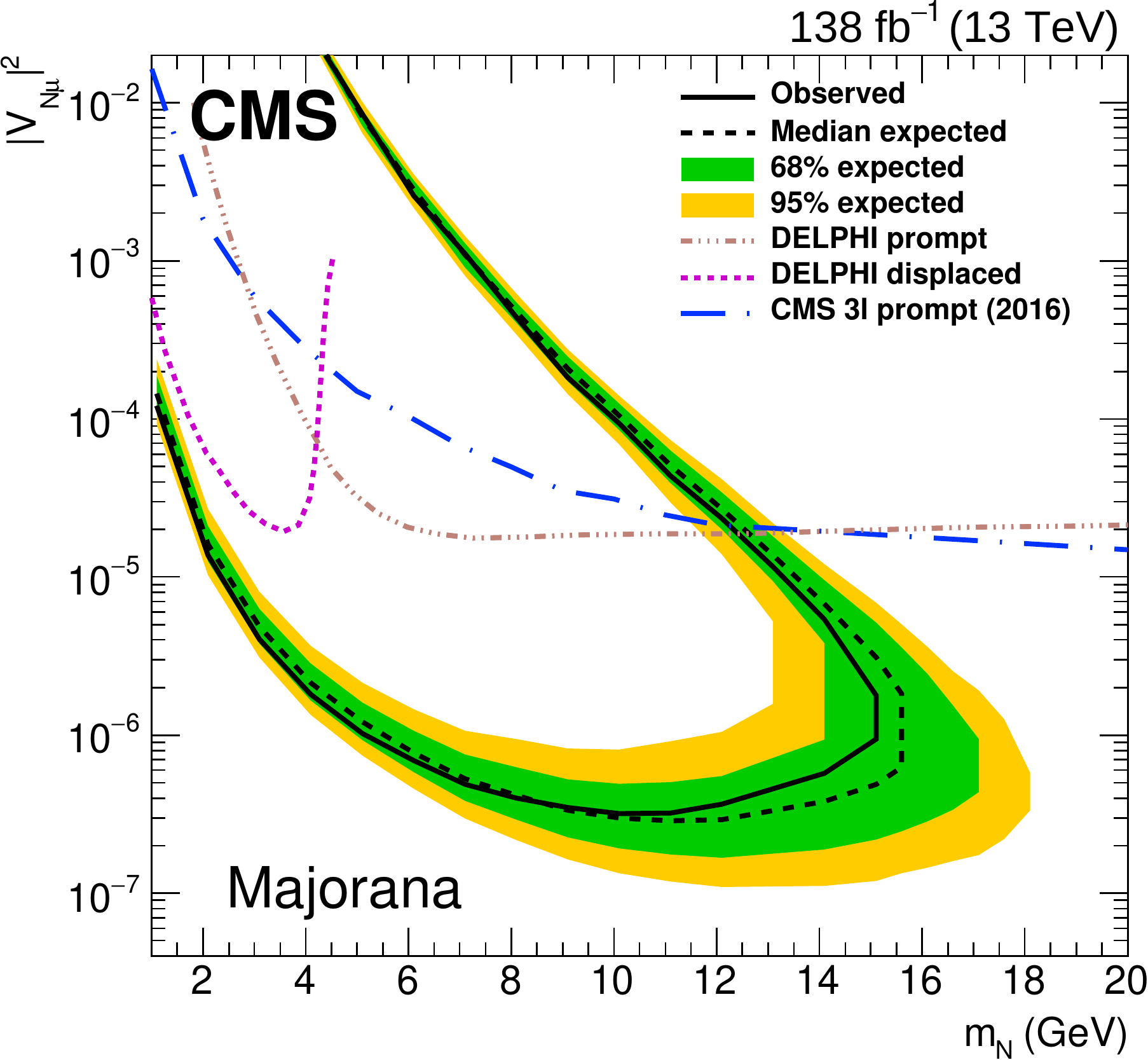}
\caption{\label{fig:results:limitsMaj}
    The 95\% \CL limits on \mixpare (left) and \mixparm (right) as functions of
    \mhnl for a Majorana HNL.
    The area inside the solid (dashed) black curve indicates the observed
    (expected) exclusion region.
    Results from the DELPHI~\cite{DELPHI:1996qcc} and the CMS~\cite{CMS:2018jxx,
    CMS:2018iaf} Collaborations are shown for reference.
}
\end{figure}

\begin{figure}[!ht]
\centering
\includegraphics[width=.48\textwidth]{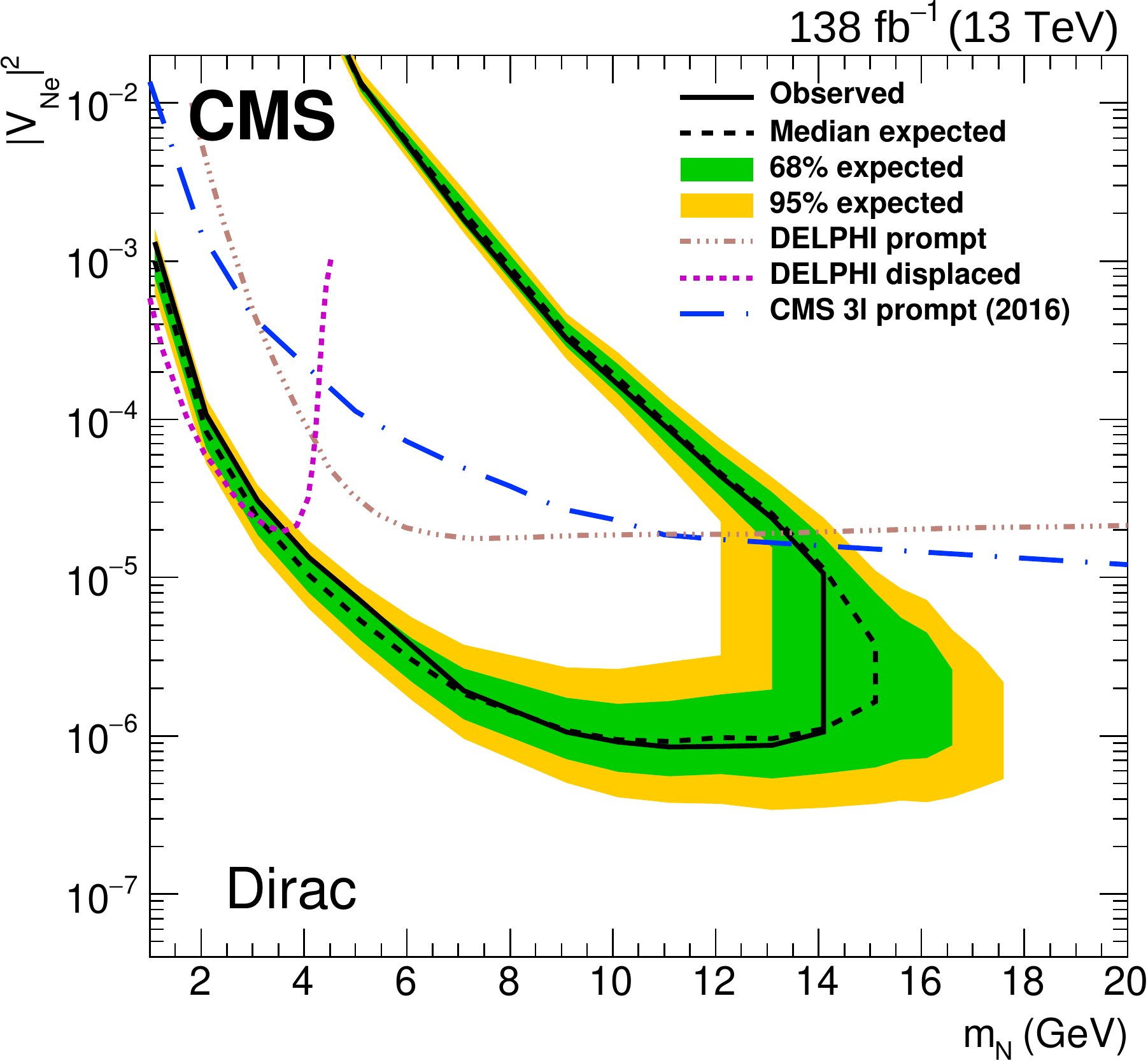}
\hfill
\includegraphics[width=.48\textwidth]{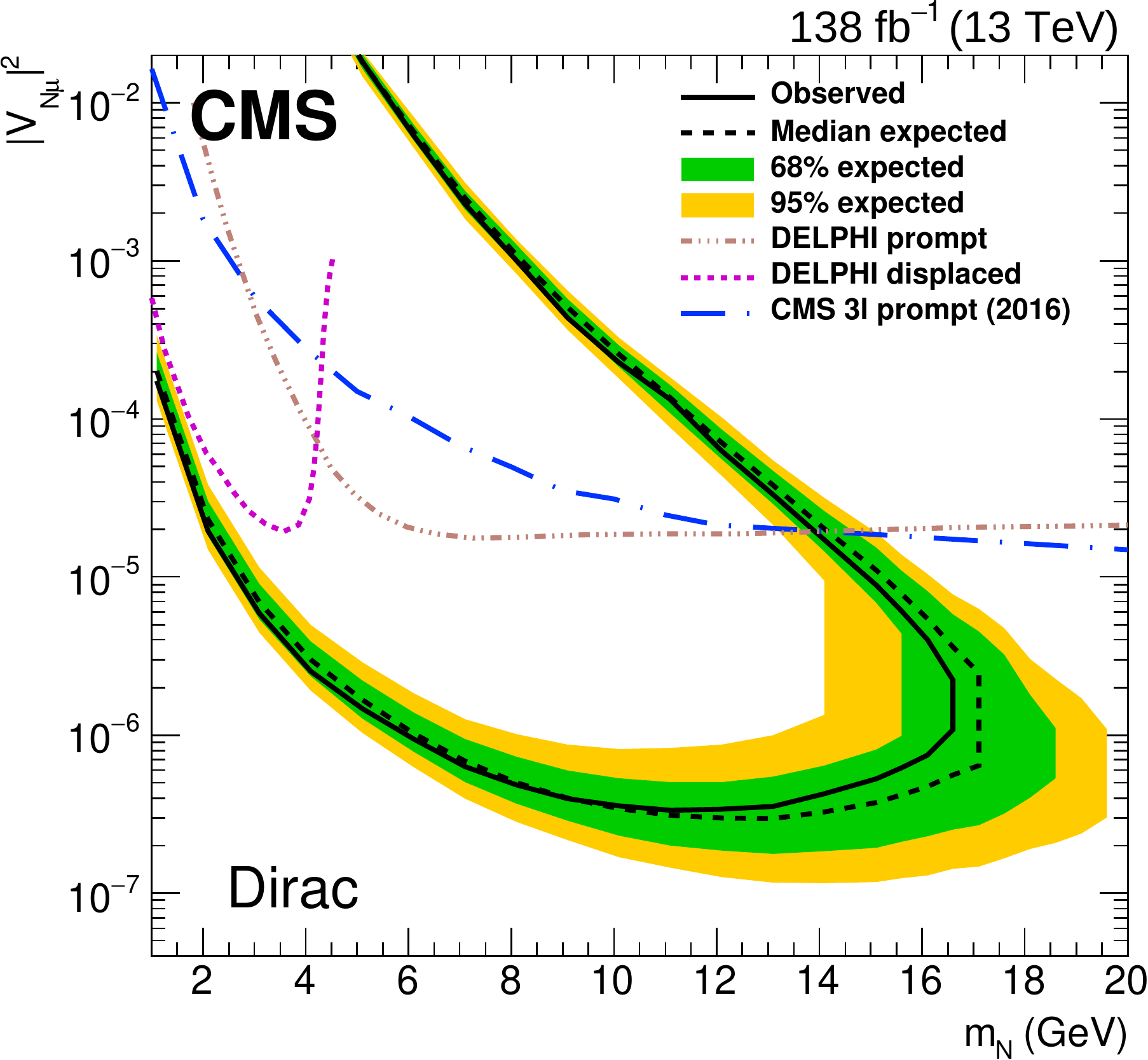}
\caption{\label{fig:results:limitsDir}
    The 95\% \CL limits on \mixpare (left) and \mixparm (right) as functions of
    \mhnl for a Dirac HNL.
    The area inside the solid (dashed) black curve indicates the observed
    (expected) exclusion region.
    Results from the DELPHI~\cite{DELPHI:1996qcc} and the CMS~\cite{CMS:2018jxx,
    CMS:2018iaf} Collaborations are shown for reference.
}
\end{figure}

We derive exclusion limits on \mixpare and \mixparm as a function of \mhnl,
separately for the cases of Majorana and Dirac HNLs, using a grid of points in
the (\mhnl,\mixpar) parameter space.
For \mhnl, we consider values between 1 and 20\GeV, with a step size of 0.5 or
1\GeV.
For each mass point, per-event reweighting is used to interpolate to the full
range of interest in \mixpar starting from the available MC samples.
The results are shown in Fig.~\ref{fig:results:limitsMaj} for a Majorana HNL and
in Fig.~\ref{fig:results:limitsDir} for a Dirac HNL.
The simulation of the Dirac HNL scenarios is obtained from the same MC samples
used for the Majorana HNL interpretation, by applying proper lifetime
corrections and classifying all selected events as LNC, as explained in
Section~\ref{sec:simulation}.
The obtained limits are connected with straight lines between neighboring mass
points.
For the last mass point for which any coupling values are excluded, the upper
and lower limits are connected with a vertical line.

In the Dirac HNL scenario, we exclude squared values of electron (muon)
couplings of \Ten{1.1}{-4} (\Ten{2.0}{-5}) and higher for an HNL mass of 2\GeV,
and between \Ten{9.1}{-7} and \Ten{1.7}{-4} (\Ten{3.6}{-7} and \Ten{2.3}{-4}) for an HNL mass of 10\GeV.
The highest excluded HNL masses are 14 (16.5)\GeV, with \mixpar values between
\Ten{1.1}{-6} and \Ten{1.1}{-5} (\Ten{1.1}{-6} and \Ten{2.2}{-6}).
Compared to the previous CMS result targeting prompt
signatures~\cite{CMS:2018iaf}, the limits are improved by more than one order of
magnitude in the probed mass range, which is a result of the dedicated
reconstruction of the displaced decay vertex in this analysis.
Compared to the results of the DELPHI Collaboration~\cite{DELPHI:1996qcc}, the
limits are similarly improved in the probed mass range except for the case of
the electron coupling and masses below 3.5\GeV, for which the DELPHI exclusion
limits extend to slightly smaller coupling values.
Compared to the results of the ATLAS Collaboration~\cite{ATLAS:2019kpx}, which
only probe the muon coupling in displaced signatures, we exclude both lower
couplings and lower masses.
In the Majorana HNL scenario, the exclusion limits are slightly less stringent,
with the highest excluded HNL masses being 13 (15)\GeV and \mixpar values range
between \Ten{1.1}{-6} and \Ten{6.5}{-6} (\Ten{9.5}{-7} and \Ten{1.8}{-6}) for
electron (muon) couplings.

The largest explicitly simulated coupling strength is $\mixpar=0.5$, which is
available for the HNL masses of 1, 2, and 3\GeV, and corresponds to a mean
lifetime $c\tauhnl=1.3$, 0.039, and 0.0045\unit{cm}, respectively.
We find that this coupling strength is excluded for masses of 1--2 (1--3)\GeV in
case of electron (muon) couplings.

\section{Summary}

A search for heavy neutral leptons (HNLs) has been performed in the decays of
\PW~bosons produced in proton-proton collisions at \sqrts and collected by the
CMS experiment at the LHC.
The analysis uses a data set corresponding to an integrated luminosity of
138\fbinv.
Events with three charged leptons are selected, and dedicated methods are
applied to identify two displaced leptons consistent with the decay of a
long-lived HNL in the mass range 1--20\GeV.
Novel methods have been developed to estimate relevant background contributions
from control samples in data, addressing one of the important challenges in this
type of search at the LHC.

No significant deviation from the standard model predictions is observed.
Constraints are derived for models with a single HNL generation of Majorana or
Dirac nature, coupled exclusively to electrons or muons.
Exclusion limits are evaluated at 95\% confidence level on the coupling
strengths of HNLs to standard model neutrinos as functions of the HNL mass,
covering HNL masses from 1 up to 16.5\GeV and squared mixing parameters as low
as \Ten{3.2}{-7}, depending on the scenario.
These results exceed previous experimental constraints in the mass range 3--14
(1--16.5)\GeV for HNLs coupling to electrons (muons) and provide the most
stringent limits to date.

\begin{acknowledgments}
    We congratulate our colleagues in the CERN accelerator departments for the excellent performance of the LHC and thank the technical and administrative staffs at CERN and at other CMS institutes for their contributions to the success of the CMS effort. In addition, we gratefully acknowledge the computing centers and personnel of the Worldwide LHC Computing Grid and other centers for delivering so effectively the computing infrastructure essential to our analyses. Finally, we acknowledge the enduring support for the construction and operation of the LHC, the CMS detector, and the supporting computing infrastructure provided by the following funding agencies: BMBWF and FWF (Austria); FNRS and FWO (Belgium); CNPq, CAPES, FAPERJ, FAPERGS, and FAPESP (Brazil); MES and BNSF (Bulgaria); CERN; CAS, MoST, and NSFC (China); MINCIENCIAS (Colombia); MSES and CSF (Croatia); RIF (Cyprus); SENESCYT (Ecuador); MoER, ERC PUT and ERDF (Estonia); Academy of Finland, MEC, and HIP (Finland); CEA and CNRS/IN2P3 (France); BMBF, DFG, and HGF (Germany); GSRI (Greece); NKFIA (Hungary); DAE and DST (India); IPM (Iran); SFI (Ireland); INFN (Italy); MSIP and NRF (Republic of Korea); MES (Latvia); LAS (Lithuania); MOE and UM (Malaysia); BUAP, CINVESTAV, CONACYT, LNS, SEP, and UASLP-FAI (Mexico); MOS (Montenegro); MBIE (New Zealand); PAEC (Pakistan); MSHE and NSC (Poland); FCT (Portugal); JINR (Dubna); MON, RosAtom, RAS, RFBR, and NRC KI (Russia); MESTD (Serbia); MCIN/AEI and PCTI (Spain); MOSTR (Sri Lanka); Swiss Funding Agencies (Switzerland); MST (Taipei); ThEPCenter, IPST, STAR, and NSTDA (Thailand); TUBITAK and TAEK (Turkey); NASU (Ukraine); STFC (United Kingdom); DOE and NSF (USA).

\hyphenation{Rachada-pisek} Individuals have received support from the Marie-Curie program and the European Research Council and Horizon 2020 Grant, contract Nos.\ 675440, 724704, 752730, 758316, 765710, 824093, 884104, and COST Action CA16108 (European Union); the Leventis Foundation; the Alfred P.\ Sloan Foundation; the Alexander von Humboldt Foundation; the Belgian Federal Science Policy Office; the Fonds pour la Formation \`a la Recherche dans l'Industrie et dans l'Agriculture (FRIA-Belgium); the Agentschap voor Innovatie door Wetenschap en Technologie (IWT-Belgium); the F.R.S.-FNRS and FWO (Belgium) under the ``Excellence of Science -- EOS" -- be.h project n.\ 30820817; the Beijing Municipal Science \& Technology Commission, No. Z191100007219010; the Ministry of Education, Youth and Sports (MEYS) of the Czech Republic; the Deutsche Forschungsgemeinschaft (DFG), under Germany's Excellence Strategy -- EXC 2121 ``Quantum Universe" -- 390833306, and under project number 400140256 - GRK2497; the Lend\"ulet (``Momentum") Program and the J\'anos Bolyai Research Scholarship of the Hungarian Academy of Sciences, the New National Excellence Program \'UNKP, the NKFIA research grants 123842, 123959, 124845, 124850, 125105, 128713, 128786, and 129058 (Hungary); the Council of Science and Industrial Research, India; the Latvian Council of Science; the Ministry of Science and Higher Education and the National Science Center, contracts Opus 2014/15/B/ST2/03998 and 2015/19/B/ST2/02861 (Poland); the Funda\c{c}\~ao para a Ci\^encia e a Tecnologia, grant CEECIND/01334/2018 (Portugal); the National Priorities Research Program by Qatar National Research Fund; the Ministry of Science and Higher Education, projects no. 0723-2020-0041 and no. FSWW-2020-0008, and the Russian Foundation for Basic Research, project No.19-42-703014 (Russia); MCIN/AEI/10.13039/501100011033, ERDF ``a way of making Europe", and the Programa Estatal de Fomento de la Investigaci{\'o}n Cient{\'i}fica y T{\'e}cnica de Excelencia Mar\'{\i}a de Maeztu, grant MDM-2017-0765 and Programa Severo Ochoa del Principado de Asturias (Spain); the Stavros Niarchos Foundation (Greece); the Rachadapisek Sompot Fund for Postdoctoral Fellowship, Chulalongkorn University and the Chulalongkorn Academic into Its 2nd Century Project Advancement Project (Thailand); the Kavli Foundation; the Nvidia Corporation; the SuperMicro Corporation; the Welch Foundation, contract C-1845; and the Weston Havens Foundation (USA).
\end{acknowledgments}

\bibliography{auto_generated}

\providecommand{\href}[2]{#2}\begingroup\raggedright\begin{thebibliography}{10}%
\makeatletter
\providecommand{\hrefCMSnoop }[0]{\@secondoftwo}%
\makeatother
\providecommand{\doi}{\texttt{doi:}\begingroup \urlstyle{tt}\Url}

\bibitem{Super-Kamiokande:1998kpq}
\hrefCMSnoop {}{{Super-Kamiokande} Collaboration, ``Evidence for oscillation of
  atmospheric neutrinos'',} \textit{ Phys. Rev. Lett.} \textbf{ 81} (1998)
  1562,
  \href{http://dx.doi.org/10.1103/PhysRevLett.81.1562}{\doi{10.1103/PhysRevLett.81.1562}},
  \href{http://www.arXiv.org/abs/hep-ex/9807003}{\texttt{arXiv:hep-ex/9807003}}.

\bibitem{SNO:2002tuh}
\hrefCMSnoop {}{{SNO} Collaboration, ``Direct evidence for neutrino flavor
  transformation from neutral-current interactions in the {Sudbury Neutrino
  Observatory}'',} \textit{ Phys. Rev. Lett.} \textbf{ 89} (2002) 011301,
  \href{http://dx.doi.org/10.1103/PhysRevLett.89.011301}{\doi{10.1103/PhysRevLett.89.011301}},
  \href{http://www.arXiv.org/abs/nucl-ex/0204008}{\texttt{arXiv:nucl-ex/0204008}}.

\bibitem{KamLAND:2002uet}
\hrefCMSnoop {}{{KamLAND} Collaboration, ``First results from {KamLAND}:
  Evidence for reactor antineutrino disappearance'',} \textit{ Phys. Rev.
  Lett.} \textbf{ 90} (2003) 021802,
  \href{http://dx.doi.org/10.1103/PhysRevLett.90.021802}{\doi{10.1103/PhysRevLett.90.021802}},
  \href{http://www.arXiv.org/abs/hep-ex/0212021}{\texttt{arXiv:hep-ex/0212021}}.

\bibitem{Bilenky:2016pep}
\hrefCMSnoop {}{S.~Bilenky, ``Neutrino oscillations: From a historical
  perspective to the present status'',} \textit{ Nucl. Phys. B} \textbf{ 908}
  (2016) 2,
  \href{http://dx.doi.org/10.1016/j.nuclphysb.2016.01.025}{\doi{10.1016/j.nuclphysb.2016.01.025}},
  \href{http://www.arXiv.org/abs/1602.00170}{\texttt{arXiv:1602.00170}}.

\bibitem{RoyChoudhury:2019hls}
\hrefCMSnoop {}{S.~Roy~Choudhury and S.~Hannestad, ``Updated results on
  neutrino mass and mass hierarchy from cosmology with {Planck} 2018
  likelihoods'',} \textit{ JCAP} \textbf{ 07} (2020) 037,
  \href{http://dx.doi.org/10.1088/1475-7516/2020/07/037}{\doi{10.1088/1475-7516/2020/07/037}},
  \href{http://www.arXiv.org/abs/1907.12598}{\texttt{arXiv:1907.12598}}.

\bibitem{Ivanov:2019hqk}
\hrefCMSnoop {}{M.~Ivanov, M.~Simonovi{\'c}, and M.~Zaldarriaga, ``Cosmological
  parameters and neutrino masses from the final {Planck} and full-shape {BOSS}
  data'',} \textit{ Phys. Rev. D} \textbf{ 101} (2020) 083504,
  \href{http://dx.doi.org/10.1103/PhysRevD.101.083504}{\doi{10.1103/PhysRevD.101.083504}},
  \href{http://www.arXiv.org/abs/1912.08208}{\texttt{arXiv:1912.08208}}.

\bibitem{Formaggio:2021nfz}
\hrefCMSnoop {}{J.~Formaggio, A.~de~Gouv{\^e}a, and R.~Robertson, ``Direct
  measurements of neutrino mass'',} \textit{ Phys. Rept.} \textbf{ 914} (2021)
  1,
  \href{http://dx.doi.org/10.1016/j.physrep.2021.02.002}{\doi{10.1016/j.physrep.2021.02.002}},
  \href{http://www.arXiv.org/abs/2102.00594}{\texttt{arXiv:2102.00594}}.

\bibitem{Minkowski:1977sc}
\hrefCMSnoop {}{P.~Minkowski, ``${\PGm\to\Pe\PGg}$ at a rate of one out of
  {\TEN{9}} muon decays?'',} \textit{ Phys. Lett. B} \textbf{ 67} (1977) 421,
  \href{http://dx.doi.org/10.1016/0370-2693(77)90435-X}{\doi{10.1016/0370-2693(77)90435-X}}.

\bibitem{Yanagida:1979as}
\hrefCMSnoop {}{T.~Yanagida, ``Horizontal gauge symmetry and masses of
  neutrinos'',} in \textit{ Proc. Workshop on the Unified Theories and the
  Baryon Number in the Universe: Tsukuba, Japan, February 13--14, 1979}.
\newblock 1979.
\newblock [Conf. Proc. C 7902131 (1979) 95].

\bibitem{Gell-Mann:1979vob}
\hrefCMSnoop {}{M.~Gell-Mann, P.~Ramond, and R.~Slansky, ``Complex spinors and
  unified theories'',} in \textit{ Supergravity}, p.~315.
\newblock North Holland Publishing, 1979.
\newblock \href{http://www.arXiv.org/abs/1306.4669}{\texttt{arXiv:1306.4669}}.

\bibitem{Glashow:1979nm}
\hrefCMSnoop {}{S.~Glashow, ``The future of elementary particle physics'',}
  \textit{ NATO Sci. Ser. B} \textbf{ 61} (1980) 687,
  \href{http://dx.doi.org/10.1007/978-1-4684-7197-7_15}{\doi{10.1007/978-1-4684-7197-7_15}}.

\bibitem{Mohapatra:1979ia}
\hrefCMSnoop {}{R.~Mohapatra and G.~Senjanovi{\'c}, ``Neutrino mass and
  spontaneous parity nonconservation'',} \textit{ Phys. Rev. Lett.} \textbf{
  44} (1980) 912,
  \href{http://dx.doi.org/10.1103/PhysRevLett.44.912}{\doi{10.1103/PhysRevLett.44.912}}.

\bibitem{Schechter:1980gr}
\hrefCMSnoop {}{J.~Schechter and J.~Valle, ``Neutrino masses in
  $\mathrm{SU}(2)\otimes\mathrm{U}(1)$ theories'',} \textit{ Phys. Rev. D}
  \textbf{ 22} (1980) 2227,
  \href{http://dx.doi.org/10.1103/PhysRevD.22.2227}{\doi{10.1103/PhysRevD.22.2227}}.

\bibitem{Shrock:1980ct}
\hrefCMSnoop {}{R.~Shrock, ``General theory of weak leptonic and semileptonic
  decays. {I}. leptonic pseudoscalar meson decays, with associated tests for,
  and bounds on, neutrino masses and lepton mixing'',} \textit{ Phys. Rev. D}
  \textbf{ 24} (1981) 1232,
  \href{http://dx.doi.org/10.1103/PhysRevD.24.1232}{\doi{10.1103/PhysRevD.24.1232}}.

\bibitem{Cai:2017mow}
\hrefCMSnoop {}{Y.~Cai, T.~Han, T.~Li, and R.~Ruiz, ``Lepton number violation:
  Seesaw models and their collider tests'',} \textit{ Front. in Phys.} \textbf{
  6} (2018) 40,
  \href{http://dx.doi.org/10.3389/fphy.2018.00040}{\doi{10.3389/fphy.2018.00040}},
  \href{http://www.arXiv.org/abs/1711.02180}{\texttt{arXiv:1711.02180}}.

\bibitem{Maki:1962mu}
\hrefCMSnoop {}{Z.~Maki, M.~Nakagawa, and S.~Sakata, ``Remarks on the unified
  model of elementary particles'',} \textit{ Prog. Theor. Phys.} \textbf{ 28}
  (1962) 870,
  \href{http://dx.doi.org/10.1143/PTP.28.870}{\doi{10.1143/PTP.28.870}}.

\bibitem{Pontecorvo:1967fh}
\hrefCMSnoop {}{B.~Pontecorvo, ``Neutrino experiments and the problem of
  conservation of leptonic charge'',} \textit{ Zh. Eksp. Teor. Fiz.} \textbf{
  53} (1967) 1717.

\bibitem{ParticleDataGroup:2020ssz}
\hrefCMSnoop {}{{Particle Data Group}, P.~A. Zyla {et~al.}, ``Review of
  particle physics'',} \textit{ Prog. Theor. Exp. Phys.} \textbf{ 2020} (2020)
  083C01,
  \href{http://dx.doi.org/10.1093/ptep/ptaa104}{\doi{10.1093/ptep/ptaa104}}.

\bibitem{Dodelson:1993je}
\hrefCMSnoop {}{S.~Dodelson and L.~Widrow, ``Sterile neutrinos as dark
  matter'',} \textit{ Phys. Rev. Lett.} \textbf{ 72} (1994) 17,
  \href{http://dx.doi.org/10.1103/PhysRevLett.72.17}{\doi{10.1103/PhysRevLett.72.17}},
  \href{http://www.arXiv.org/abs/hep-ph/9303287}{\texttt{arXiv:hep-ph/9303287}}.

\bibitem{Boyarsky:2018tvu}
A.~Boyarsky\hrefCMSnoop {}{ {et~al.}, ``Sterile neutrino dark matter'',}
  \textit{ Prog. Part. Nucl. Phys.} \textbf{ 104} (2019) 1,
  \href{http://dx.doi.org/10.1016/j.ppnp.2018.07.004}{\doi{10.1016/j.ppnp.2018.07.004}},
  \href{http://www.arXiv.org/abs/1807.07938}{\texttt{arXiv:1807.07938}}.

\bibitem{Fukugita:1986hr}
\hrefCMSnoop {}{M.~Fukugita and T.~Yanagida, ``Baryogenesis without grand
  unification'',} \textit{ Phys. Lett. B} \textbf{ 174} (1986) 45,
  \href{http://dx.doi.org/10.1016/0370-2693(86)91126-3}{\doi{10.1016/0370-2693(86)91126-3}}.

\bibitem{Chun:2017spz}
E.~Chun\hrefCMSnoop {}{ {et~al.}, ``Probing leptogenesis'',} \textit{ Int. J.
  Mod. Phys. A} \textbf{ 33} (2018) 1842005,
  \href{http://dx.doi.org/10.1142/S0217751X18420058}{\doi{10.1142/S0217751X18420058}},
  \href{http://www.arXiv.org/abs/1711.02865}{\texttt{arXiv:1711.02865}}.

\bibitem{Deppisch:2015qwa}
\hrefCMSnoop {}{F.~Deppisch, P.~Bhupal~Dev, and A.~Pilaftsis, ``Neutrinos and
  collider physics'',} \textit{ New J. Phys.} \textbf{ 17} (2015) 075019,
  \href{http://dx.doi.org/10.1088/1367-2630/17/7/075019}{\doi{10.1088/1367-2630/17/7/075019}},
  \href{http://www.arXiv.org/abs/1502.06541}{\texttt{arXiv:1502.06541}}.

\bibitem{DELPHI:1996qcc}
\hrefCMSnoop {}{{DELPHI} Collaboration, ``Search for neutral heavy leptons
  produced in {\PZ} decays'',} \textit{ Z. Phys. C} \textbf{ 74} (1997) 57,
  \href{http://dx.doi.org/10.1007/s002880050370}{\doi{10.1007/s002880050370}}.
  [Erratum: \DOI{10.1007/BF03546181}].

\bibitem{CMS:2012wqj}
\hrefCMSnoop {}{{CMS Collaboration}, ``Search for heavy {Majorana} neutrinos in
  ${\PGmpm\PGmpm}$+jets and ${\Pepm\Pepm}$+jets events in \pp collisions at
  \sqrts[7]'',} \textit{ Phys. Lett. B} \textbf{ 717} (2012) 109,
  \href{http://dx.doi.org/10.1016/j.physletb.2012.09.012}{\doi{10.1016/j.physletb.2012.09.012}},
  \href{http://www.arXiv.org/abs/1207.6079}{\texttt{arXiv:1207.6079}}.

\bibitem{CMS:2015qur}
\hrefCMSnoop {}{{CMS Collaboration}, ``Search for heavy {Majorana} neutrinos in
  ${\PGmpm\PGmpm}$+jets events in proton-proton collisions at \sqrts[8]'',}
  \textit{ Phys. Lett. B} \textbf{ 748} (2015) 144,
  \href{http://dx.doi.org/10.1016/j.physletb.2015.06.070}{\doi{10.1016/j.physletb.2015.06.070}},
  \href{http://www.arXiv.org/abs/1501.05566}{\texttt{arXiv:1501.05566}}.

\bibitem{ATLAS:2015gtp}
\hrefCMSnoop {}{{ATLAS Collaboration}, ``Search for heavy {Majorana} neutrinos
  with the {ATLAS} detector in \pp collisions at \sqrts[8]'',} \textit{ JHEP}
  \textbf{ 07} (2015) 162,
  \href{http://dx.doi.org/10.1007/JHEP07(2015)162}{\doi{10.1007/JHEP07(2015)162}},
  \href{http://www.arXiv.org/abs/1506.06020}{\texttt{arXiv:1506.06020}}.

\bibitem{CMS:2016aro}
\hrefCMSnoop {}{{CMS Collaboration}, ``Search for heavy {Majorana} neutrinos in
  ${\Pepm\Pepm}$+jets and ${\Pepm\PGmpm}$+jets events in proton-proton
  collisions at \sqrts[8]'',} \textit{ JHEP} \textbf{ 04} (2016) 169,
  \href{http://dx.doi.org/10.1007/JHEP04(2016)169}{\doi{10.1007/JHEP04(2016)169}},
  \href{http://www.arXiv.org/abs/1603.02248}{\texttt{arXiv:1603.02248}}.

\bibitem{CMS:2018iaf}
\hrefCMSnoop {}{{CMS Collaboration}, ``Search for heavy neutral leptons in
  events with three charged leptons in proton-proton collisions at \sqrts'',}
  \textit{ Phys. Rev. Lett.} \textbf{ 120} (2018) 221801,
  \href{http://dx.doi.org/10.1103/PhysRevLett.120.221801}{\doi{10.1103/PhysRevLett.120.221801}},
  \href{http://www.arXiv.org/abs/1802.02965}{\texttt{arXiv:1802.02965}}.

\bibitem{CMS:2018jxx}
\hrefCMSnoop {}{{CMS Collaboration}, ``Search for heavy {Majorana} neutrinos in
  same-sign dilepton channels in proton-proton collisions at \sqrts'',}
  \textit{ JHEP} \textbf{ 01} (2019) 122,
  \href{http://dx.doi.org/10.1007/JHEP01(2019)122}{\doi{10.1007/JHEP01(2019)122}},
  \href{http://www.arXiv.org/abs/1806.10905}{\texttt{arXiv:1806.10905}}.

\bibitem{ATLAS:2019kpx}
\hrefCMSnoop {}{{ATLAS Collaboration}, ``Search for heavy neutral leptons in
  decays of {\PW}~bosons produced in {13\TeV} \pp collisions using prompt and
  displaced signatures with the {ATLAS} detector'',} \textit{ JHEP} \textbf{
  10} (2019) 265,
  \href{http://dx.doi.org/10.1007/JHEP10(2019)265}{\doi{10.1007/JHEP10(2019)265}},
  \href{http://www.arXiv.org/abs/1905.09787}{\texttt{arXiv:1905.09787}}.

\bibitem{LHCb:2020wxx}
\hrefCMSnoop {}{{LHCb Collaboration}, ``Search for heavy neutral leptons in
  ${\PWp\to\PGmp\PGmpm\,\text{jet}}$ decays'',} \textit{ Eur. Phys. J. C}
  \textbf{ 81} (2021) 248,
  \href{http://dx.doi.org/10.1140/epjc/s10052-021-08973-5}{\doi{10.1140/epjc/s10052-021-08973-5}},
  \href{http://www.arXiv.org/abs/2011.05263}{\texttt{arXiv:2011.05263}}.

\bibitem{Asaka:2005an}
\hrefCMSnoop {}{T.~Asaka, S.~Blanchet, and M.~Shaposhnikov, ``The {\PGn}{MSM},
  dark matter and neutrino masses'',} \textit{ Phys. Lett. B} \textbf{ 631}
  (2005) 151,
  \href{http://dx.doi.org/10.1016/j.physletb.2005.09.070}{\doi{10.1016/j.physletb.2005.09.070}},
  \href{http://www.arXiv.org/abs/hep-ph/0503065}{\texttt{arXiv:hep-ph/0503065}}.

\bibitem{Drewes:2021nqr}
\hrefCMSnoop {}{M.~Drewes, Y.~Georis, and J.~Klari{\'c}, ``Mapping the viable
  parameter space for testable leptogenesis'',} \textit{ Phys. Rev. Lett.}
  \textbf{ 128} (2022) 051801,
  \href{http://dx.doi.org/10.1103/PhysRevLett.128.051801}{\doi{10.1103/PhysRevLett.128.051801}},
  \href{http://www.arXiv.org/abs/2106.16226}{\texttt{arXiv:2106.16226}}.

\bibitem{Abada:2018sfh}
\hrefCMSnoop {}{A.~Abada, N.~Bernal, M.~Losada, and X.~Marcano, ``Inclusive
  displaced vertex searches for heavy neutral leptons at the {LHC}'',} \textit{
  JHEP} \textbf{ 01} (2019) 093,
  \href{http://dx.doi.org/10.1007/JHEP01(2019)093}{\doi{10.1007/JHEP01(2019)093}},
  \href{http://www.arXiv.org/abs/1807.10024}{\texttt{arXiv:1807.10024}}.

\bibitem{Tastet:2021vwp}
\hrefCMSnoop {}{J.-L. Tastet, O.~Ruchayskiy, and I.~Timiryasov,
  ``Reinterpreting the {ATLAS} bounds on heavy neutral leptons in a realistic
  neutrino oscillation model'',} \textit{ JHEP} \textbf{ 12} (2021) 182,
  \href{http://dx.doi.org/10.1007/JHEP12(2021)182}{\doi{10.1007/JHEP12(2021)182}},
  \href{http://www.arXiv.org/abs/2107.12980}{\texttt{arXiv:2107.12980}}.

\bibitem{Boiarska:2021yho}
\hrefCMSnoop {}{I.~Boiarska, A.~Boyarsky, O.~Mikulenko, and M.~Ovchynnikov,
  ``Constraints from the {CHARM} experiment on heavy neutral leptons with tau
  mixing'',} \textit{ Phys. Rev. D} \textbf{ 104} (2021) 095019,
  \href{http://dx.doi.org/10.1103/PhysRevD.104.095019}{\doi{10.1103/PhysRevD.104.095019}},
  \href{http://www.arXiv.org/abs/2107.14685}{\texttt{arXiv:2107.14685}}.

\bibitem{Degrande:2016aje}
\hrefCMSnoop {}{C.~Degrande, O.~Mattelaer, R.~Ruiz, and J.~Turner, ``Fully
  automated precision predictions for heavy neutrino production mechanisms at
  hadron colliders'',} \textit{ Phys. Rev. D} \textbf{ 94} (2016) 053002,
  \href{http://dx.doi.org/10.1103/PhysRevD.94.053002}{\doi{10.1103/PhysRevD.94.053002}},
  \href{http://www.arXiv.org/abs/1602.06957}{\texttt{arXiv:1602.06957}}.

\bibitem{Das:2016hof}
\hrefCMSnoop {}{A.~Das, P.~Konar, and S.~Majhi, ``Production of heavy neutrino
  in next-to-leading order {QCD} at the {LHC} and beyond'',} \textit{ JHEP}
  \textbf{ 06} (2016) 019,
  \href{http://dx.doi.org/10.1007/JHEP06(2016)019}{\doi{10.1007/JHEP06(2016)019}},
  \href{http://www.arXiv.org/abs/1604.00608}{\texttt{arXiv:1604.00608}}.

\bibitem{hepdata}
\hrefCMSnoop {}{}{HEPD}ata record for this analysis, 2022.
\newblock
  \href{http://dx.doi.org/10.17182/hepdata.115355}{\doi{10.17182/hepdata.115355}}.

\bibitem{CMS:2008xjf}
\hrefCMSnoop {}{{CMS Collaboration}, ``The {CMS} experiment at the {CERN}
  {LHC}'',} \textit{ JINST} \textbf{ 3} (2008) S08004,
  \href{http://dx.doi.org/10.1088/1748-0221/3/08/S08004}{\doi{10.1088/1748-0221/3/08/S08004}}.

\bibitem{CMS:2020cmk}
\hrefCMSnoop {}{{CMS Collaboration}, ``Performance of the {CMS} {Level-1}
  trigger in proton-proton collisions at \sqrts'',} \textit{ JINST} \textbf{
  15} (2020) P10017,
  \href{http://dx.doi.org/10.1088/1748-0221/15/10/P10017}{\doi{10.1088/1748-0221/15/10/P10017}},
  \href{http://www.arXiv.org/abs/2006.10165}{\texttt{arXiv:2006.10165}}.

\bibitem{CMS:2016ngn}
\hrefCMSnoop {}{{CMS Collaboration}, ``The {CMS} trigger system'',} \textit{
  JINST} \textbf{ 12} (2017) P01020,
  \href{http://dx.doi.org/10.1088/1748-0221/12/01/P01020}{\doi{10.1088/1748-0221/12/01/P01020}},
  \href{http://www.arXiv.org/abs/1609.02366}{\texttt{arXiv:1609.02366}}.

\bibitem{CMS:2014pgm}
\hrefCMSnoop {}{{CMS Collaboration}, ``Description and performance of track and
  primary-vertex reconstruction with the {CMS} tracker'',} \textit{ JINST}
  \textbf{ 9} (2014) P10009,
  \href{http://dx.doi.org/10.1088/1748-0221/9/10/P10009}{\doi{10.1088/1748-0221/9/10/P10009}},
  \href{http://www.arXiv.org/abs/1405.6569}{\texttt{arXiv:1405.6569}}.

\bibitem{CMS:DP-2017-015}
\href {https://cds.cern.ch/record/2290524}{{CMS Collaboration}, ``2017 tracking
  performance plots'',} CMS Detector Performance Summary CMS-DP-2017-015, 2017.

\bibitem{CMSTrackerGroup:2020edz}
\hrefCMSnoop {}{{CMS Tracker Group}, W.~Adam {et~al.}, ``The {CMS} {Phase-1}
  pixel detector upgrade'',} \textit{ JINST} \textbf{ 16} (2021) P02027,
  \href{http://dx.doi.org/10.1088/1748-0221/16/02/P02027}{\doi{10.1088/1748-0221/16/02/P02027}},
  \href{http://www.arXiv.org/abs/2012.14304}{\texttt{arXiv:2012.14304}}.

\bibitem{CMS:DP-2020-049}
\href {https://cds.cern.ch/record/2743740}{{CMS Collaboration}, ``Track impact
  parameter resolution for the full pseudo rapidity coverage in the 2017
  dataset with the {CMS} {Phase-1} pixel detector'',} CMS Detector Performance
  Summary CMS-DP-2020-049, 2020.

\bibitem{Cacciari:2008gp}
\hrefCMSnoop {}{M.~Cacciari, G.~Salam, and G.~Soyez, ``The anti-\kt jet
  clustering algorithm'',} \textit{ JHEP} \textbf{ 04} (2008) 063,
  \href{http://dx.doi.org/10.1088/1126-6708/2008/04/063}{\doi{10.1088/1126-6708/2008/04/063}},
  \href{http://www.arXiv.org/abs/0802.1189}{\texttt{arXiv:0802.1189}}.

\bibitem{Cacciari:2011ma}
\hrefCMSnoop {}{M.~Cacciari, G.~Salam, and G.~Soyez, ``{\FASTJET} user
  manual'',} \textit{ Eur. Phys. J. C} \textbf{ 72} (2012) 1896,
  \href{http://dx.doi.org/10.1140/epjc/s10052-012-1896-2}{\doi{10.1140/epjc/s10052-012-1896-2}},
  \href{http://www.arXiv.org/abs/1111.6097}{\texttt{arXiv:1111.6097}}.

\bibitem{CMS:2017yfk}
\hrefCMSnoop {}{{CMS Collaboration}, ``Particle-flow reconstruction and global
  event description with the {CMS} detector'',} \textit{ JINST} \textbf{ 12}
  (2017) P10003,
  \href{http://dx.doi.org/10.1088/1748-0221/12/10/P10003}{\doi{10.1088/1748-0221/12/10/P10003}},
  \href{http://www.arXiv.org/abs/1706.04965}{\texttt{arXiv:1706.04965}}.

\bibitem{CMS:2020uim}
\hrefCMSnoop {}{{CMS Collaboration}, ``Electron and photon reconstruction and
  identification with the {CMS} experiment at the {CERN} {LHC}'',} \textit{
  JINST} \textbf{ 16} (2021) P05014,
  \href{http://dx.doi.org/10.1088/1748-0221/16/05/P05014}{\doi{10.1088/1748-0221/16/05/P05014}},
  \href{http://www.arXiv.org/abs/2012.06888}{\texttt{arXiv:2012.06888}}.

\bibitem{CMS:2018rym}
\hrefCMSnoop {}{{CMS Collaboration}, ``Performance of the {CMS} muon detector
  and muon reconstruction with proton-proton collisions at \sqrts'',} \textit{
  JINST} \textbf{ 13} (2018) P06015,
  \href{http://dx.doi.org/10.1088/1748-0221/13/06/P06015}{\doi{10.1088/1748-0221/13/06/P06015}},
  \href{http://www.arXiv.org/abs/1804.04528}{\texttt{arXiv:1804.04528}}.

\bibitem{CMS:2019ctu}
\hrefCMSnoop {}{{CMS Collaboration}, ``Performance of missing transverse
  momentum reconstruction in proton-proton collisions at \sqrts using the {CMS}
  detector'',} \textit{ JINST} \textbf{ 14} (2019) P07004,
  \href{http://dx.doi.org/10.1088/1748-0221/14/07/P07004}{\doi{10.1088/1748-0221/14/07/P07004}},
  \href{http://www.arXiv.org/abs/1903.06078}{\texttt{arXiv:1903.06078}}.

\bibitem{CMS:2018mlc}
\hrefCMSnoop {}{{CMS Collaboration}, ``Measurement of the inelastic
  proton-proton cross section at \sqrts'',} \textit{ JHEP} \textbf{ 07} (2018)
  161,
  \href{http://dx.doi.org/10.1007/JHEP07(2018)161}{\doi{10.1007/JHEP07(2018)161}},
  \href{http://www.arXiv.org/abs/1802.02613}{\texttt{arXiv:1802.02613}}.

\bibitem{GEANT4:2002zbu}
\hrefCMSnoop {}{{GEANT4} Collaboration, ``{\GEANTfour}---a simulation
  toolkit'',} \textit{ Nucl. Instrum. Meth. A} \textbf{ 506} (2003) 250,
  \href{http://dx.doi.org/10.1016/S0168-9002(03)01368-8}{\doi{10.1016/S0168-9002(03)01368-8}}.

\bibitem{Alwall:2014hca}
J.~Alwall\hrefCMSnoop {}{ {et~al.}, ``The automated computation of tree-level
  and next-to-leading order differential cross sections, and their matching to
  parton shower simulations'',} \textit{ JHEP} \textbf{ 07} (2014) 079,
  \href{http://dx.doi.org/10.1007/JHEP07(2014)079}{\doi{10.1007/JHEP07(2014)079}},
  \href{http://www.arXiv.org/abs/1405.0301}{\texttt{arXiv:1405.0301}}.

\bibitem{Artoisenet:2012st}
\hrefCMSnoop {}{P.~Artoisenet, R.~Frederix, O.~Mattelaer, and R.~Rietkerk,
  ``Automatic spin-entangled decays of heavy resonances in {Monte Carlo}
  simulations'',} \textit{ JHEP} \textbf{ 03} (2013) 015,
  \href{http://dx.doi.org/10.1007/JHEP03(2013)015}{\doi{10.1007/JHEP03(2013)015}},
  \href{http://www.arXiv.org/abs/1212.3460}{\texttt{arXiv:1212.3460}}.

\bibitem{Atre:2009rg}
\hrefCMSnoop {}{A.~Atre, T.~Han, S.~Pascoli, and B.~Zhang, ``The search for
  heavy {Majorana} neutrinos'',} \textit{ JHEP} \textbf{ 05} (2009) 030,
  \href{http://dx.doi.org/10.1088/1126-6708/2009/05/030}{\doi{10.1088/1126-6708/2009/05/030}},
  \href{http://www.arXiv.org/abs/0901.3589}{\texttt{arXiv:0901.3589}}.

\bibitem{Alva:2014gxa}
\hrefCMSnoop {}{D.~Alva, T.~Han, and R.~Ruiz, ``Heavy {Majorana} neutrinos from
  ${\PW\PGg}$ fusion at hadron colliders'',} \textit{ JHEP} \textbf{ 02} (2015)
  072,
  \href{http://dx.doi.org/10.1007/JHEP02(2015)072}{\doi{10.1007/JHEP02(2015)072}},
  \href{http://www.arXiv.org/abs/1411.7305}{\texttt{arXiv:1411.7305}}.

\bibitem{Pascoli:2018heg}
\hrefCMSnoop {}{S.~Pascoli, R.~Ruiz, and C.~Weiland, ``Heavy neutrinos with
  dynamic jet vetoes:\ multilepton searches at $\sqrt{s}=14$, 27, and
  {100\TeV}'',} \textit{ JHEP} \textbf{ 06} (2019) 049,
  \href{http://dx.doi.org/10.1007/JHEP06(2019)049}{\doi{10.1007/JHEP06(2019)049}},
  \href{http://www.arXiv.org/abs/1812.08750}{\texttt{arXiv:1812.08750}}.

\bibitem{Melnikov:2006kv}
\hrefCMSnoop {}{K.~Melnikov and F.~Petriello, ``Electroweak gauge boson
  production at hadron colliders through $\mathcal{O}({\alpS^2})$'',} \textit{
  Phys. Rev. D} \textbf{ 74} (2006) 114017,
  \href{http://dx.doi.org/10.1103/PhysRevD.74.114017}{\doi{10.1103/PhysRevD.74.114017}},
  \href{http://www.arXiv.org/abs/hep-ph/0609070}{\texttt{arXiv:hep-ph/0609070}}.

\bibitem{Gavin:2010az}
\hrefCMSnoop {}{R.~Gavin, Y.~Li, F.~Petriello, and S.~Quackenbush,
  ``{\FEWZ2.0}: A code for hadronic {\PZ} production at next-to-next-to-leading
  order'',} \textit{ Comput. Phys. Commun.} \textbf{ 182} (2011) 2388,
  \href{http://dx.doi.org/10.1016/j.cpc.2011.06.008}{\doi{10.1016/j.cpc.2011.06.008}},
  \href{http://www.arXiv.org/abs/1011.3540}{\texttt{arXiv:1011.3540}}.

\bibitem{Gavin:2012sy}
\hrefCMSnoop {}{R.~Gavin, Y.~Li, F.~Petriello, and S.~Quackenbush, ``{\PW}
  physics at the {LHC} with {\FEWZ2.1}'',} \textit{ Comput. Phys. Commun.}
  \textbf{ 184} (2013) 208,
  \href{http://dx.doi.org/10.1016/j.cpc.2012.09.005}{\doi{10.1016/j.cpc.2012.09.005}},
  \href{http://www.arXiv.org/abs/1201.5896}{\texttt{arXiv:1201.5896}}.

\bibitem{Li:2012wna}
\hrefCMSnoop {}{Y.~Li and F.~Petriello, ``Combining {QCD} and electroweak
  corrections to dilepton production in {\FEWZ}'',} \textit{ Phys. Rev. D}
  \textbf{ 86} (2012) 094034,
  \href{http://dx.doi.org/10.1103/PhysRevD.86.094034}{\doi{10.1103/PhysRevD.86.094034}},
  \href{http://www.arXiv.org/abs/1208.5967}{\texttt{arXiv:1208.5967}}.

\bibitem{Frederix:2012ps}
\hrefCMSnoop {}{R.~Frederix and S.~Frixione, ``Merging meets matching in
  {\MCATNLO}'',} \textit{ JHEP} \textbf{ 12} (2012) 061,
  \href{http://dx.doi.org/10.1007/JHEP12(2012)061}{\doi{10.1007/JHEP12(2012)061}},
  \href{http://www.arXiv.org/abs/1209.6215}{\texttt{arXiv:1209.6215}}.

\bibitem{Alwall:2007fs}
J.~Alwall\hrefCMSnoop {}{ {et~al.}, ``Comparative study of various algorithms
  for the merging of parton showers and matrix elements in hadronic
  collisions'',} \textit{ Eur. Phys. J. C} \textbf{ 53} (2008) 473,
  \href{http://dx.doi.org/10.1140/epjc/s10052-007-0490-5}{\doi{10.1140/epjc/s10052-007-0490-5}},
  \href{http://www.arXiv.org/abs/0706.2569}{\texttt{arXiv:0706.2569}}.

\bibitem{Nason:2004rx}
\hrefCMSnoop {}{P.~Nason, ``A new method for combining {NLO} {QCD} with shower
  {Monte Carlo} algorithms'',} \textit{ JHEP} \textbf{ 11} (2004) 040,
  \href{http://dx.doi.org/10.1088/1126-6708/2004/11/040}{\doi{10.1088/1126-6708/2004/11/040}},
  \href{http://www.arXiv.org/abs/hep-ph/0409146}{\texttt{arXiv:hep-ph/0409146}}.

\bibitem{Frixione:2007nw}
\hrefCMSnoop {}{S.~Frixione, P.~Nason, and G.~Ridolfi, ``A positive-weight
  next-to-leading-order {Monte Carlo} for heavy flavour hadroproduction'',}
  \textit{ JHEP} \textbf{ 09} (2007) 126,
  \href{http://dx.doi.org/10.1088/1126-6708/2007/09/126}{\doi{10.1088/1126-6708/2007/09/126}},
  \href{http://www.arXiv.org/abs/0707.3088}{\texttt{arXiv:0707.3088}}.

\bibitem{Frixione:2007vw}
\hrefCMSnoop {}{S.~Frixione, P.~Nason, and C.~Oleari, ``Matching {NLO} {QCD}
  computations with parton shower simulations: the {\POWHEG} method'',}
  \textit{ JHEP} \textbf{ 11} (2007) 070,
  \href{http://dx.doi.org/10.1088/1126-6708/2007/11/070}{\doi{10.1088/1126-6708/2007/11/070}},
  \href{http://www.arXiv.org/abs/0709.2092}{\texttt{arXiv:0709.2092}}.

\bibitem{Alioli:2009je}
\hrefCMSnoop {}{S.~Alioli, P.~Nason, C.~Oleari, and E.~Re, ``{NLO} single-top
  production matched with shower in {\POWHEG}: $s$- and $t$-channel
  contributions'',} \textit{ JHEP} \textbf{ 09} (2009) 111,
  \href{http://dx.doi.org/10.1088/1126-6708/2009/09/111}{\doi{10.1088/1126-6708/2009/09/111}},
  \href{http://www.arXiv.org/abs/0907.4076}{\texttt{arXiv:0907.4076}}.
  [Erratum: \DOI{10.1007/JHEP02(2010)011}].

\bibitem{Alioli:2010xd}
\hrefCMSnoop {}{S.~Alioli, P.~Nason, C.~Oleari, and E.~Re, ``A general
  framework for implementing {NLO} calculations in shower {Monte Carlo}
  programs: the {\POWHEG} \textsc{box}'',} \textit{ JHEP} \textbf{ 06} (2010)
  043,
  \href{http://dx.doi.org/10.1007/JHEP06(2010)043}{\doi{10.1007/JHEP06(2010)043}},
  \href{http://www.arXiv.org/abs/1002.2581}{\texttt{arXiv:1002.2581}}.

\bibitem{Re:2010bp}
\hrefCMSnoop {}{E.~Re, ``Single-top ${\PW\PQt}$-channel production matched with
  parton showers using the {\POWHEG} method'',} \textit{ Eur. Phys. J. C}
  \textbf{ 71} (2011) 1547,
  \href{http://dx.doi.org/10.1140/epjc/s10052-011-1547-z}{\doi{10.1140/epjc/s10052-011-1547-z}},
  \href{http://www.arXiv.org/abs/1009.2450}{\texttt{arXiv:1009.2450}}.

\bibitem{Melia:2011tj}
\hrefCMSnoop {}{T.~Melia, P.~Nason, R.~R{\"o}ntsch, and G.~Zanderighi,
  ``${\PWp\PWm}$, {\WZ} and {\ZZ} production in the {\POWHEG} \textsc{box}'',}
  \textit{ JHEP} \textbf{ 11} (2011) 078,
  \href{http://dx.doi.org/10.1007/JHEP11(2011)078}{\doi{10.1007/JHEP11(2011)078}},
  \href{http://www.arXiv.org/abs/1107.5051}{\texttt{arXiv:1107.5051}}.

\bibitem{Nason:2013ydw}
\hrefCMSnoop {}{P.~Nason and G.~Zanderighi, ``${\PWp\PWm}$, {\WZ} and {\ZZ}
  production in the {\POWHEG}\textsc{-box-v2}'',} \textit{ Eur. Phys. J. C}
  \textbf{ 74} (2014) 2702,
  \href{http://dx.doi.org/10.1140/epjc/s10052-013-2702-5}{\doi{10.1140/epjc/s10052-013-2702-5}},
  \href{http://www.arXiv.org/abs/1311.1365}{\texttt{arXiv:1311.1365}}.

\bibitem{NNPDF:2014otw}
\hrefCMSnoop {}{{NNPDF} Collaboration, ``Parton distributions for the {LHC Run
  II}'',} \textit{ JHEP} \textbf{ 04} (2015) 040,
  \href{http://dx.doi.org/10.1007/JHEP04(2015)040}{\doi{10.1007/JHEP04(2015)040}},
  \href{http://www.arXiv.org/abs/1410.8849}{\texttt{arXiv:1410.8849}}.

\bibitem{NNPDF:2017mvq}
\hrefCMSnoop {}{{NNPDF} Collaboration, ``Parton distributions from
  high-precision collider data'',} \textit{ Eur. Phys. J. C} \textbf{ 77}
  (2017) 663,
  \href{http://dx.doi.org/10.1140/epjc/s10052-017-5199-5}{\doi{10.1140/epjc/s10052-017-5199-5}},
  \href{http://www.arXiv.org/abs/1706.00428}{\texttt{arXiv:1706.00428}}.

\bibitem{Sjostrand:2014zea}
T.~Sj{\"o}strand\hrefCMSnoop {}{ {et~al.}, ``An introduction to
  {\PYTHIA8.2}'',} \textit{ Comput. Phys. Commun.} \textbf{ 191} (2015) 159,
  \href{http://dx.doi.org/10.1016/j.cpc.2015.01.024}{\doi{10.1016/j.cpc.2015.01.024}},
  \href{http://www.arXiv.org/abs/1410.3012}{\texttt{arXiv:1410.3012}}.

\bibitem{CMS:2016kle}
\href {http://cds.cern.ch/record/2235192}{{CMS Collaboration}, ``Investigations
  of the impact of the parton shower tuning in {\PYTHIA8} in the modelling of
  \ttbar at $\sqrt{s}=8$ and {13\TeV}'',} CMS Physics Analysis Summary
  CMS-PAS-TOP-16-021, 2016.

\bibitem{CMS:2019csb}
\hrefCMSnoop {}{{CMS Collaboration}, ``Extraction and validation of a new set
  of {CMS} {\PYTHIA8} tunes from underlying-event measurements'',} \textit{
  Eur. Phys. J. C} \textbf{ 80} (2020) 4,
  \href{http://dx.doi.org/10.1140/epjc/s10052-019-7499-4}{\doi{10.1140/epjc/s10052-019-7499-4}},
  \href{http://www.arXiv.org/abs/1903.12179}{\texttt{arXiv:1903.12179}}.

\bibitem{Cacciari:2007fd}
\hrefCMSnoop {}{M.~Cacciari and G.~Salam, ``Pileup subtraction using jet
  areas'',} \textit{ Phys. Lett. B} \textbf{ 659} (2008) 119,
  \href{http://dx.doi.org/10.1016/j.physletb.2007.09.077}{\doi{10.1016/j.physletb.2007.09.077}},
  \href{http://www.arXiv.org/abs/0707.1378}{\texttt{arXiv:0707.1378}}.

\bibitem{CMS:2016lmd}
\hrefCMSnoop {}{{CMS Collaboration}, ``Jet energy scale and resolution in the
  {CMS} experiment in pp collisions at {8\TeV}'',} \textit{ JINST} \textbf{ 12}
  (2017) P02014,
  \href{http://dx.doi.org/10.1088/1748-0221/12/02/P02014}{\doi{10.1088/1748-0221/12/02/P02014}},
  \href{http://www.arXiv.org/abs/1607.03663}{\texttt{arXiv:1607.03663}}.

\bibitem{CMS:2017wtu}
\hrefCMSnoop {}{{CMS Collaboration}, ``Identification of heavy-flavour jets
  with the {CMS} detector in pp collisions at {13\TeV}'',} \textit{ JINST}
  \textbf{ 13} (2018) P05011,
  \href{http://dx.doi.org/10.1088/1748-0221/13/05/P05011}{\doi{10.1088/1748-0221/13/05/P05011}},
  \href{http://www.arXiv.org/abs/1712.07158}{\texttt{arXiv:1712.07158}}.

\bibitem{Fruhwirth:1987fm}
\hrefCMSnoop {}{R.~Fr{\"u}hwirth, ``Application of {Kalman} filtering to track
  and vertex fitting'',} \textit{ Nucl. Instrum. Meth. A} \textbf{ 262} (1987)
  444,
  \href{http://dx.doi.org/10.1016/0168-9002(87)90887-4}{\doi{10.1016/0168-9002(87)90887-4}}.

\bibitem{CMS-DP-2020-035}
\href {https://cds.cern.ch/record/2724492}{{CMS Collaboration}, ``Muon tracking
  performance in the {CMS Run-2 Legacy} data using the tag-and-probe
  technique'',} CMS Detector Performance Note CMS-DP-2020-035, 2020.

\bibitem{CMS:2021xjt}
\hrefCMSnoop {}{{CMS Collaboration}, ``Precision luminosity measurement in
  proton-proton collisions at \sqrts in 2015 and 2016 at {CMS}'',} \textit{
  Eur. Phys. J. C} \textbf{ 81} (2021) 800,
  \href{http://dx.doi.org/10.1140/epjc/s10052-021-09538-2}{\doi{10.1140/epjc/s10052-021-09538-2}},
  \href{http://www.arXiv.org/abs/2104.01927}{\texttt{arXiv:2104.01927}}.

\bibitem{CMS:2018elu}
\href {http://cds.cern.ch/record/2621960}{{CMS Collaboration}, ``{CMS}
  luminosity measurement for the 2017 data-taking period at \sqrts'',} CMS
  Physics Analysis Summary CMS-PAS-LUM-17-004, 2018.

\bibitem{CMS:2019jhq}
\href {http://cds.cern.ch/record/2676164}{{CMS Collaboration}, ``{CMS}
  luminosity measurement for the 2018 data-taking period at \sqrts'',} CMS
  Physics Analysis Summary CMS-PAS-LUM-18-002, 2019.

\bibitem{Linnemann:2003vw}
\hrefCMSnoop {}{J.~Linnemann, ``Measures of significance in {HEP} and
  astrophysics'',} in \textit{ Proc. Statistical Problems in Particle Physics,
  Astrophysics, and Cosmology (PHYSTAT2003): Menlo Park CA, United States,
  September 8--11, 2003}.
\newblock 2003.
\newblock
  \href{http://www.arXiv.org/abs/physics/0312059}{\texttt{arXiv:physics/0312059}}.
\newblock [eConf C030908 (2003) MOBT001].

\bibitem{Cousins:2007yta}
\hrefCMSnoop {}{R.~Cousins, J.~Linnemann, and J.~Tucker, ``Evaluation of three
  methods for calculating statistical significance when incorporating a
  systematic uncertainty into a test of the background-only hypothesis for a
  poisson process'',} \textit{ Nucl. Instrum. Meth. A} \textbf{ 595} (2008)
  480,
  \href{http://dx.doi.org/10.1016/j.nima.2008.07.086}{\doi{10.1016/j.nima.2008.07.086}},
  \href{http://www.arXiv.org/abs/physics/0702156}{\texttt{arXiv:physics/0702156}}.

\bibitem{Junk:1999kv}
\hrefCMSnoop {}{T.~Junk, ``Confidence level computation for combining searches
  with small statistics'',} \textit{ Nucl. Instrum. Meth. A} \textbf{ 434}
  (1999) 435,
  \href{http://dx.doi.org/10.1016/S0168-9002(99)00498-2}{\doi{10.1016/S0168-9002(99)00498-2}},
  \href{http://www.arXiv.org/abs/hep-ex/9902006}{\texttt{arXiv:hep-ex/9902006}}.

\bibitem{Read:2002hq}
\hrefCMSnoop {}{A.~Read, ``Presentation of search results: The {\CLs}
  technique'',} \textit{ J. Phys. G} \textbf{ 28} (2002) 2693,
  \href{http://dx.doi.org/10.1088/0954-3899/28/10/313}{\doi{10.1088/0954-3899/28/10/313}}.

\bibitem{ATLAS:2011tau}
\href {https://cds.cern.ch/record/1379837}{{ATLAS and CMS Collaborations, and
  LHC Higgs Combination Group}, ``Procedure for the {LHC} {Higgs} boson search
  combination in {Summer} 2011'',} Technical Report CMS-NOTE-2011-005,
  ATL-PHYS-PUB-2011-11, 2011.

\bibitem{Cowan:2010js}
\hrefCMSnoop {}{G.~Cowan, K.~Cranmer, E.~Gross, and O.~Vitells, ``Asymptotic
  formulae for likelihood-based tests of new physics'',} \textit{ Eur. Phys. J.
  C} \textbf{ 71} (2011) 1554,
  \href{http://dx.doi.org/10.1140/epjc/s10052-011-1554-0}{\doi{10.1140/epjc/s10052-011-1554-0}},
  \href{http://www.arXiv.org/abs/1007.1727}{\texttt{arXiv:1007.1727}}.
  [Erratum: \DOI{10.1140/epjc/s10052-013-2501-z}].

\end{thebibliography}\endgroup
\cleardoublepage \appendix\section{The CMS Collaboration \label{app:collab}}\begin{sloppypar}\hyphenpenalty=5000\widowpenalty=500\clubpenalty=5000\cmsinstitute{Yerevan~Physics~Institute, Yerevan, Armenia}
A.~Tumasyan
\cmsinstitute{Institut~f\"{u}r~Hochenergiephysik, Vienna, Austria}
W.~Adam\cmsorcid{0000-0001-9099-4341}, J.W.~Andrejkovic, T.~Bergauer\cmsorcid{0000-0002-5786-0293}, S.~Chatterjee\cmsorcid{0000-0003-2660-0349}, K.~Damanakis, M.~Dragicevic\cmsorcid{0000-0003-1967-6783}, A.~Escalante~Del~Valle\cmsorcid{0000-0002-9702-6359}, R.~Fr\"{u}hwirth\cmsAuthorMark{1}, M.~Jeitler\cmsAuthorMark{1}\cmsorcid{0000-0002-5141-9560}, N.~Krammer, L.~Lechner\cmsorcid{0000-0002-3065-1141}, D.~Liko, I.~Mikulec, P.~Paulitsch, F.M.~Pitters, J.~Schieck\cmsAuthorMark{1}\cmsorcid{0000-0002-1058-8093}, R.~Sch\"{o}fbeck\cmsorcid{0000-0002-2332-8784}, D.~Schwarz, S.~Templ\cmsorcid{0000-0003-3137-5692}, W.~Waltenberger\cmsorcid{0000-0002-6215-7228}, C.-E.~Wulz\cmsAuthorMark{1}\cmsorcid{0000-0001-9226-5812}
% ...
\vskip\cmsinstskip
% ...
1:~Also at TU Wien, Wien, Austria\\
% ...
\end{sloppypar}
\end{document}